\documentclass[pra, aps, notitlepage, nofootinbib, 12pt, superscriptaddress]{revtex4-1}

\usepackage[braket, qm]{qcircuit}
\usepackage{amsmath}
\usepackage{amsfonts}
\usepackage{amssymb}
\usepackage{braket}
\usepackage[multiple]{footmisc}
\usepackage{stmaryrd}
\usepackage{graphicx}
\usepackage[hidelinks]{hyperref}
\usepackage{dsfont}
\usepackage{caption}
\usepackage{subcaption}
\usepackage{nicefrac}
\usepackage{enumerate}
\usepackage{multirow}
\usepackage{stmaryrd}
\usepackage{parskip}
\usepackage{tikz}
\usepackage{varwidth}

\usepackage[capitalise, noabbrev]{cleveref}

\newcommand{\id}{\mathds{1}}

\providecommand{\cnot}{\textsc{cnot}}

\begin{document}
\title{Code Deformation and Lattice Surgery Are Gauge Fixing}
\author{Christophe Vuillot\footnote{These authors have made equal contributions to the research discussed herein.}}
\author{Lingling Lao\footnotemark[1]}
\affiliation{QuTech, TU Delft, The Netherlands}
\author{Ben Criger}
\affiliation{QuTech, TU Delft, The Netherlands}
\affiliation{Institute for Globally Distributed Open Research and Education (IGDORE)}
\author{Carmen Garc\'{i}a Almud\'{e}ver}
\author{Koen Bertels}
\affiliation{QuTech, TU Delft, The Netherlands}
\author{Barbara M. Terhal}
\affiliation{QuTech, TU Delft, The Netherlands}
\affiliation{JARA Institute for Quantum Information (PGI-11), Forschungszentrum Juelich, D-52425, Juelich, Germany}
\date{\today}
\begin{abstract}
The large-scale execution of quantum algorithms requires basic quantum operations to be implemented fault-tolerantly.
The most popular technique for accomplishing this, using the devices that can be realised in the near term, uses stabilizer codes which can be embedded in a planar layout.
The set of fault-tolerant operations which can be executed in these systems using unitary gates is typically very limited. 
This has driven the development of measurement-based schemes for performing logical operations in these codes, known as lattice surgery and code deformation.
In parallel, gauge fixing has emerged as a measurement-based method for performing universal gate sets in subsystem stabilizer codes.
In this work, we show that lattice surgery and code deformation can be expressed as special cases of gauge fixing, permitting a simple and rigorous test for fault-tolerance together with simple guiding principles for the implementation of these operations.
We demonstrate the accuracy of this method numerically with examples based on the surface code, some of which are novel.
\end{abstract}
\maketitle

\section{Introduction}
Quantum computers can implement algorithms which are much faster than their classical counterparts, with exponential speedup for problems such as prime factorisation \cite{shor1994algorithms}, and polynomial speedup for many others \cite{jordan2011quantum}.
The main obstacle to constructing a large-scale quantum computer is decoherence, which partially randomizes quantum states and operations.
Although state-of-the-art coherence times are now appreciably longer than gate times \cite{riste2015detecting, kelly2015state}, they remain too short for useful quantum computation.

To counter the effect of decoherence on quantum states which are stored or manipulated imperfectly, we can encode logical qubit states into several physical qubits, and perform non-destructive multi-qubit measurements of the resulting system to extract information about which errors have occurred, called the \emph{syndrome}. 
The spaces of multi-qubit states used to encode these logical states are called \emph{quantum error-correcting codes}, and their ability to correct errors is measured by the \emph{distance} $d$, which is the number of independent errors (or error \emph{weight}) necessary to alter the state of the logical qubits without being detected.
In order to use one of these codes in practice, it is also necessary to account for the effect of decoherence on operations. 
For example, a syndrome measurement may involve a sequence of entangling gates, and the error caused by a faulty gate on a small set of qubits in the beginning of the circuit may propagate onto many qubits, producing a high-weight error, increasing the likelihood of a logical error.
Measurement results can also be corrupted by decoherence, so syndrome extraction often has to be repeated.
In order to prevent error propagation during repeated measurement, syndrome extraction circuits must be designed such that a small number of faults (from imperfect gates or memory errors on data qubits) will result in a small number of errors on the physical qubits, which can be corrected using noisy syndromes.
Given a family of codes of different distances, we can determine a \emph{threshold error rate}, the rate beneath which codes with higher distance produce lower logical error probabilities.

Several such families of quantum error-correcting codes have been developed, including concatenated codes \cite{steane1996error, knill1996concatenated}, subsystem codes such as Bacon-Shor codes \cite{bacon2006operator}, and 2D topological codes. 
The most prominent 2D topological codes are surface codes \cite{fowler2012surface} derived from Kitaev's toric code \cite{kitaev2003fault}, which we will focus on in the remainder of this manuscript.
2D topological codes can be implemented using entangling gates which are local in two dimensions, allowing fault-tolerance in near-term devices which have limited connectivity. 
In addition, 2D topological codes generally have high fault-tolerant memory thresholds, with the surface code having the highest at $\sim 1\%$ \cite{wang2011surface}.

These advantages come at a cost, however.
While other 2D topological codes permit certain single-qubit logical operations to be implemented \emph{transversally},
the surface code does not. 
In addition, the constraint that computation be carried out in a single plane does not permit two-qubit physical gates to be carried out between physical qubits in different code blocks, precluding the two-qubit gates which, in principle, can be carried out transversally.

These two restrictions have led to the design of measurement-based protocols for performing single- and two-qubit logical gates by making gradual changes to the underlying stabilizer code.
Measurement-based protocols that implement single-qubit gates are typically called \emph{code deformation} \cite{bombin2009quantum}, and protocols that involve multiple logical qubits are usually called \emph{lattice surgery} \cite{horsman2012surface}.
A separate measurement-based technique, called \emph{gauge fixing} \cite{paetznick2013universal}, can be applied to \emph{subsystem codes}, which have operators which can be added to or removed from the stabilizer group as desired, the so-called \emph{gauge operators}.
During gauge fixing, the stabilizer generators of the subsystem code remain unchanged, and can be used to detect and correct errors; so decoding is unaffected by gauge fixing.
This is in contrast to code deformation and lattice surgery, where it is not \emph{a priori} clear which measurement results to incorporate into decoding, or how to process them.
Recently, many different code deformation and lattice surgery techniques have been devised, most of which use tailor-made analysis or decoding techniques, see e.g. \cite{Bombin2011Cliff,landahl2014quantum,Bravyi16deform,Nautrup2016FTi,Brown2017poking,Litinski2018latticesurgery,Fowler2018lowoverhead,Vasmer20183DSurf}.
 
In this paper, we phrase existing lattice surgery and code deformation protocols as special cases of gauge fixing, showing that the underlying subsystem code dictates the fault-tolerance properties of the protocol.
This perspective can simplify the analysis of new measurement-based protocols, provided that they are based on stabilizer codes whose distances can be easily calculated. 
Also, knowing the stabilizer of the underlying subsystem code results in clear guidelines for decoding using the measurement results produced by such a protocol.

The remainder of this paper is organised as follows.
In \cref{sec:LSCD}, we review the ideas behind code deformation and lattice surgery.
In \cref{sec:GF}, we review the formalism of gauge fixing.
Following this, in \cref{sec:Unification}, we formulate lattice surgery and code deformation operations as gauge fixing, demonstrating that fault-tolerant code deformation protocols are in fact based on high-distance subsystem codes.
We also show this explicitly using both well-known and novel protocols.
In \cref{sec:Numerics}, we numerically determine the performance of these protocols.
We conclude and discuss potential future research in \cref{sec:Conclusion}.

In all figures in this paper, qubits are located on the vertices of the drawn lattice. We refer to the local generators of the stabilizer group of the surface code as \emph{stabilizers} or \emph{checks}. In the figures, black regions signify $X$-stabilizers and light grey regions $Z$-stabilizers, with no stabilizers measured on white plaquettes.

\section{Code Deformation and Lattice surgery}
\label{sec:LSCD}

\subsection{Code Deformation}

Code deformation is a technique to convert one code into another by making a series of changes to the set of stabilizer generators to be measured in each round of error correction.

Typically, these protocols use ancillae prepared in entangled and/or encoded states as a resource.
Also, a typical code deformation sequence proceeds gradually, first expanding the code into a large intermediate code by entangling the original code block with the ancillae, then disentangling some of the qubits (which may include some or all of the original data qubits), producing a final code which can then be used for further computation.
The initial and final code may differ in their logical operators, in which case the deformation performs a logical operation. 
Also, the initial and final code may differ in their position or orientation within a larger quantum computer. 

\begin{figure}[tb]
	\centering
	\begin{subfigure}[h]{0.32\textwidth}
		\caption{}
		\includegraphics[width=0.9\textwidth]{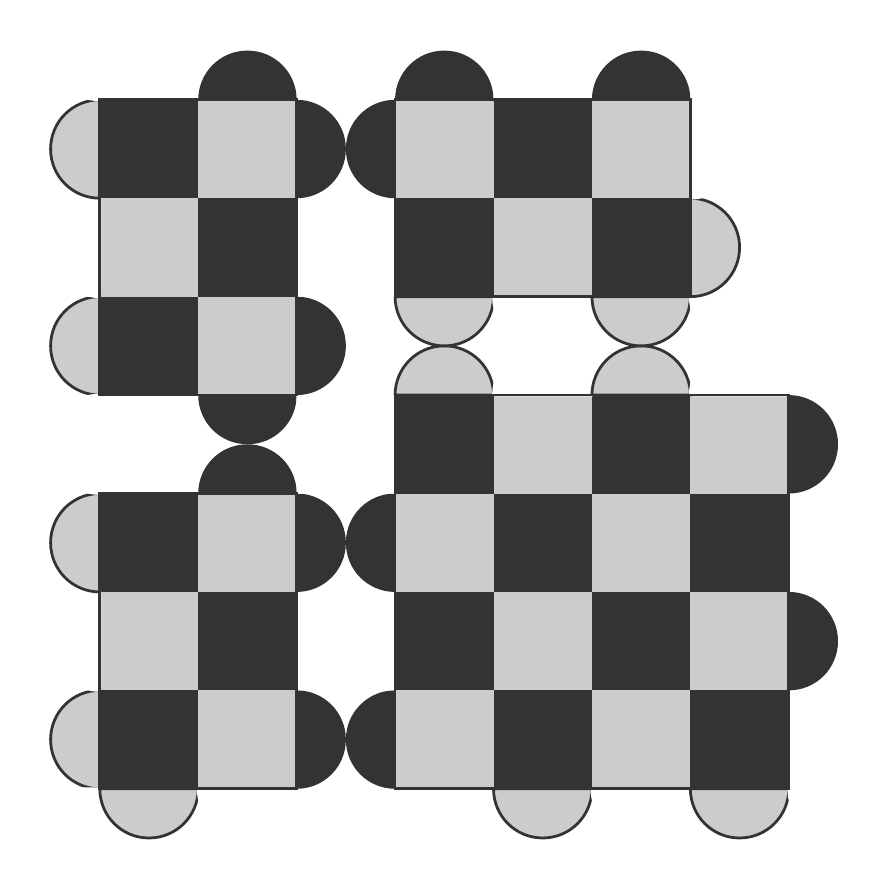}
		\label{fig:layout_r1}
	\end{subfigure}
	\begin{subfigure}[h]{0.32\textwidth}
		\caption{}
		\includegraphics[width=0.9\textwidth]{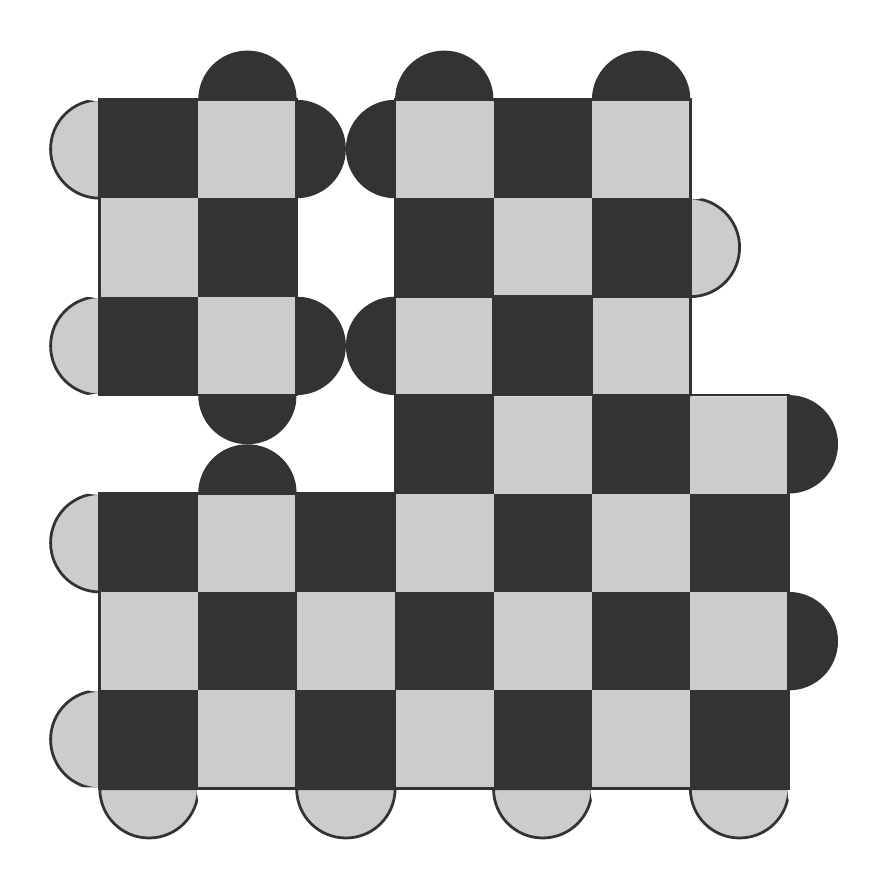}
		\label{fig:layout_r2}
	\end{subfigure}
	\begin{subfigure}[h]{0.32\textwidth}
		\caption{}
		\includegraphics[width=0.9\textwidth]{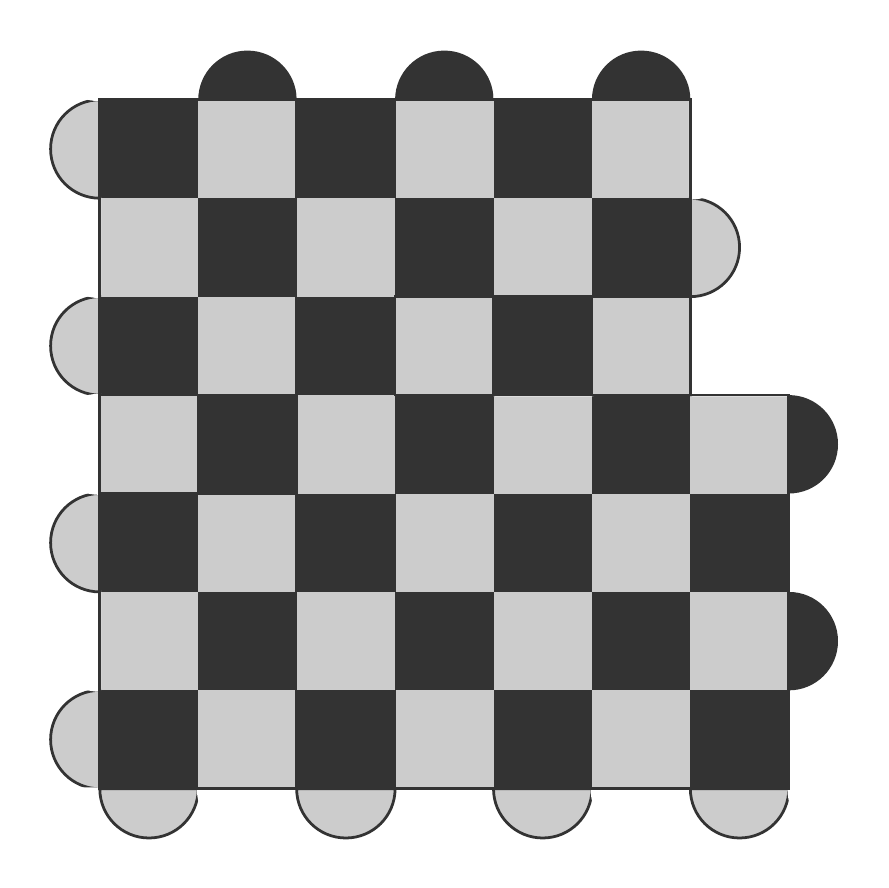}
		\label{fig:layout_r3}
	\end{subfigure}
	\begin{subfigure}[h]{0.32\textwidth}
		\caption{}
		\includegraphics[width=0.9\textwidth]{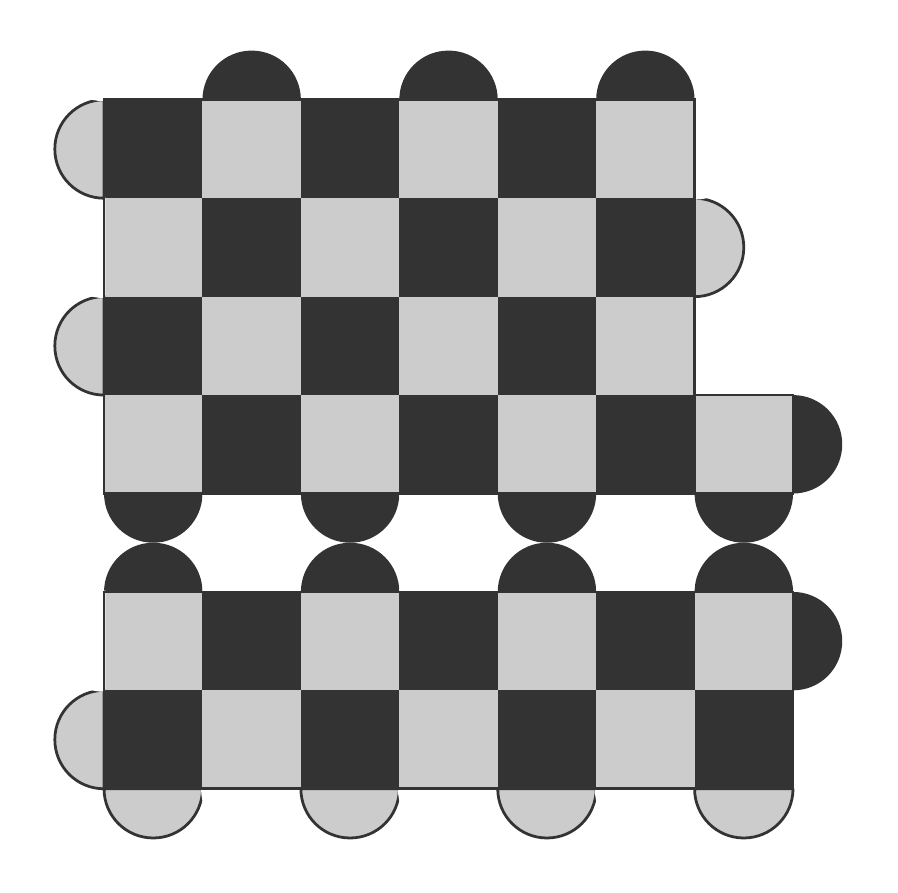}
		\label{fig:layout_r4}
	\end{subfigure}
	\begin{subfigure}[h]{0.32\textwidth}
		\caption{}
		\includegraphics[width=0.9\textwidth]{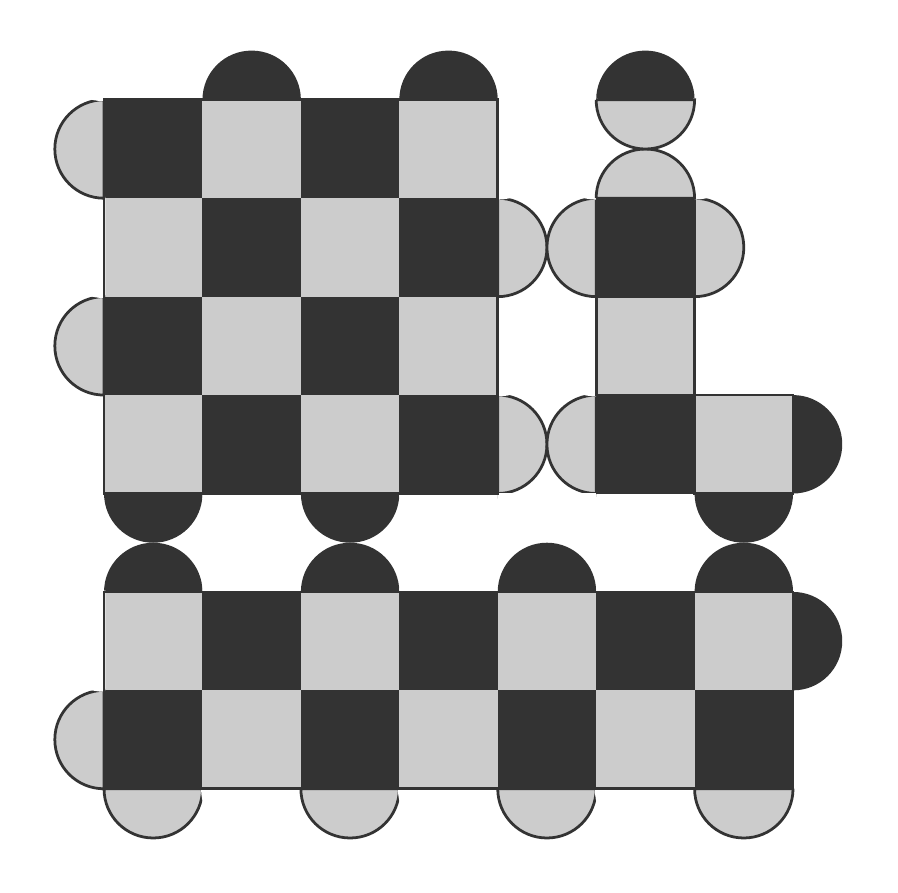}
		\label{fig:layout_r5}
	\end{subfigure}
	\begin{subfigure}[h]{0.32\textwidth}
		\caption{}
		\includegraphics[width=0.9\textwidth]{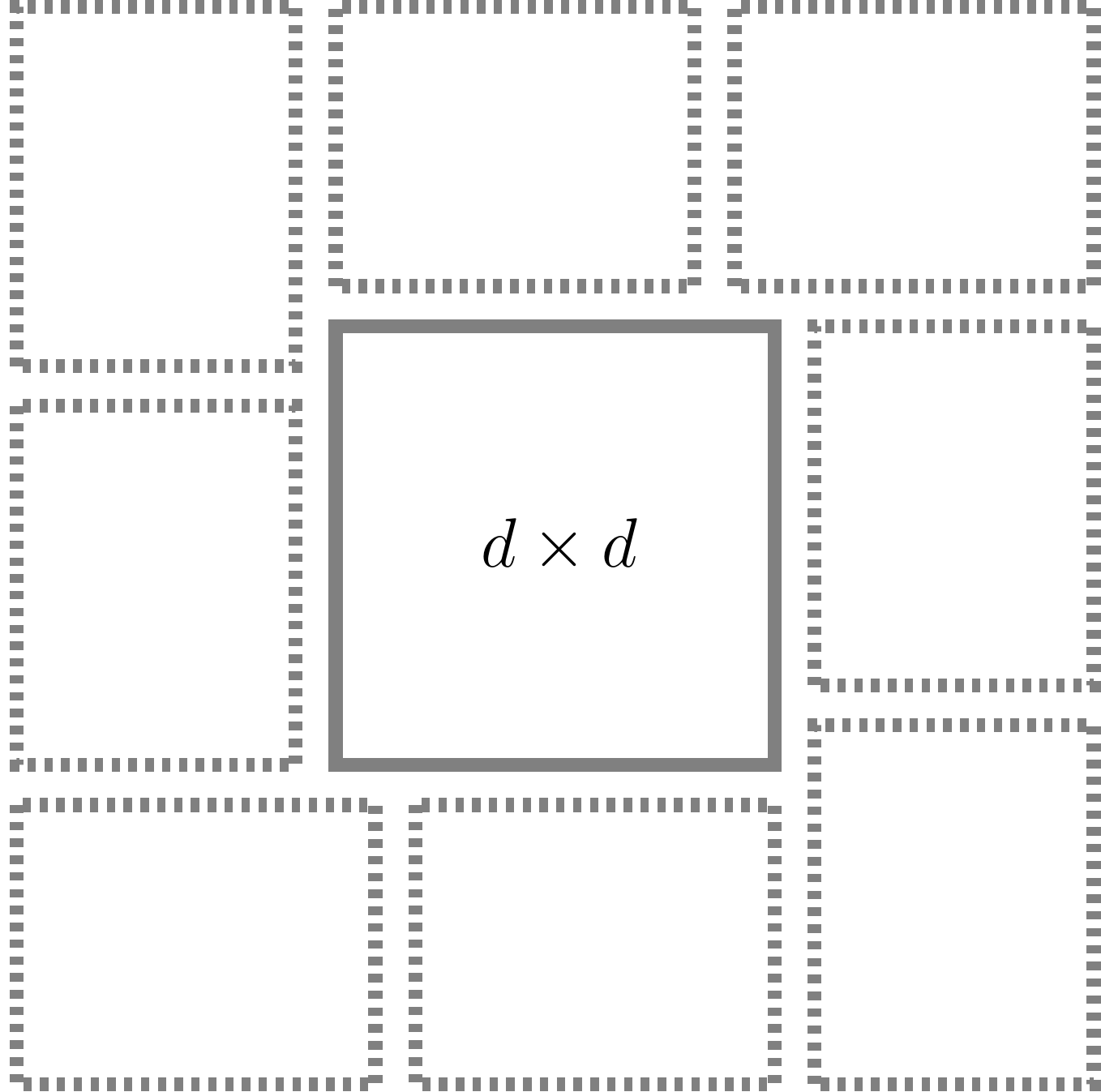}
		\label{fig:layout_enlarge}
	\end{subfigure}
	\caption{Fault-tolerant procedure for rotating a surface code by $90^{\circ}$ and reflecting it about the $x$ axis (see \cite[Figure 10]{horsman2012surface} for the corresponding protocol using smooth/rough boundaries).
		(a) Initial layout where the $5\times5$ lattice is to be rotated, the three $3 \times 4$ patches are ancillas in fixed states, fully specified by the stabilizers shown. 
		(b) Intermediate lattice, this step is required to expand the lattice fault-tolerantly.
		(c) Fully expanded lattice.
		(d) and (e) Splitting operations performed to shrink the lattice.
		(f) By using the two steps from (a) to (c) at the same time on all corners, one can grow a lattice from distance $d$ to $3d-4$.
		The surrounding ancillary patches have $(d-2)\times(d-1)$ qubits each.
	}
	\label{fig:h_layout}
\end{figure}

For example, consider the proposed fault-tolerant procedure for lattice rotation of surface codes shown in \autoref{fig:h_layout}, similar to the one presented in \cite{dennis2002topological}.
One can see five steps which gradually modify the surface code patch starting at the bottom right of \autoref{fig:layout_r1} and ending at the top left of \autoref{fig:layout_r5} in a different orientation. 
First, three ancillary patches are prepared in fixed states, and placed near the upper left corner of the target patch.
Then, the patch undergoes a two-step growing operation, followed by a two-step shrinking operation.
Advancing one step is done by measuring the operators corresponding to the new stabilizers, some of which anti-commute with the old ones. 
Measurement of these new stabilizers will return $\pm 1$ values at random.
This means that additional corrections, unrelated to errors that may have occurred, are needed in order to enter the new code space (the mutual $+1$-eigenspace of all new stabilizers).
Moreover, to account for noisy operations, one must simultaneously perform error correction.
After one is confident that the encoded state is in the new code space, one can proceed to the next step.

\begin{figure}[tb]
	\centering
	\begin{subfigure}[h]{0.2\textwidth}
		\caption{}
		\includegraphics[width=0.9\textwidth]{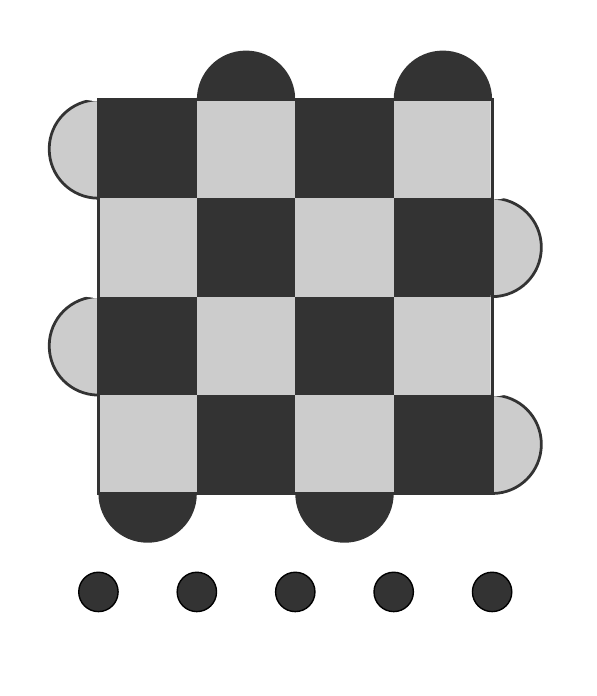}
		\label{fig:layout_f1}
	\end{subfigure}\qquad
	\begin{subfigure}[h]{0.2\textwidth}
		\caption{}
		\includegraphics[width=0.9\textwidth]{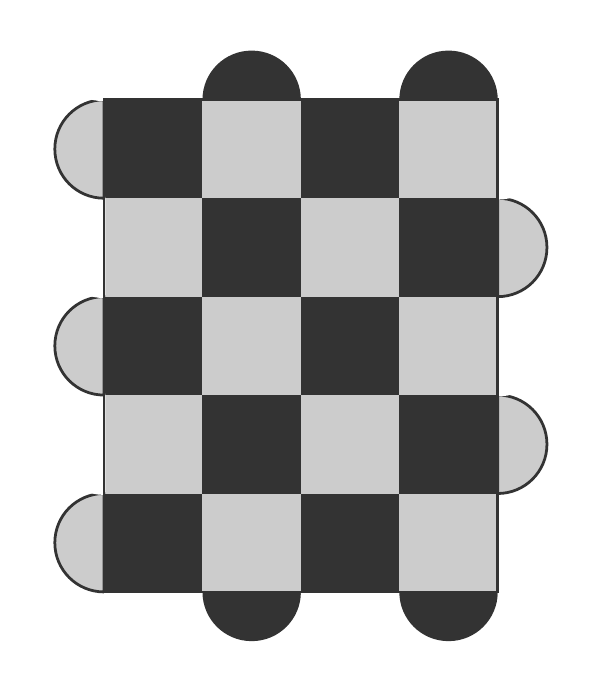}
		\label{fig:layout_f2}
	\end{subfigure}\qquad
	\begin{subfigure}[h]{0.2\textwidth}
		\caption{}
		\includegraphics[width=0.9\textwidth]{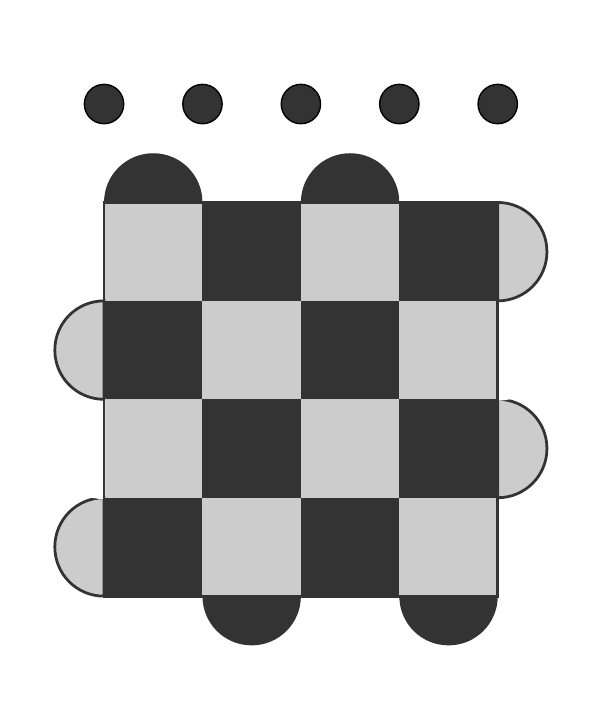}
		\label{fig:layout_f3}
	\end{subfigure}
	\caption{A procedure to flip a lattice using code deformation. 
		(a) The lattice to be flipped, and the physical qubits prepared in $\ket{+}$ states.
		(b) The flip operation is realised by merging the original lattice with the physical qubits below.
		(c) Subsequently measuring the physical qubits at the top in the $X$ basis finishes the flip operation. }
	\label{fig:h_flip}
\end{figure}

In Section \ref{sec:Unification}, we will demonstrate that, following these five steps, one can fault-tolerantly protect the logical information at all times with a distance-5 code.
We also show that the distance would be reduced to 3 if one were to omit step (\subref{fig:layout_f2}), going directly from (\subref{fig:layout_f1}) to (\subref{fig:layout_f3}), as one would do when directly adapting the established surface code rotation method from \cite{horsman2012surface} to rotated surface codes.

This lattice rotation followed by the lattice flip in \autoref{fig:h_flip} are useful for performing a transversal Hadamard gate.
The transversal Hadamard gate on a surface code patch, performed by applying a Hadamard gate on each qubit, interchanges $X$ and $Z$ plaquettes.
This code transformation can be undone by a lattice rotation, followed by a lattice flip. 
Moreover, part of this rotation procedure can be used to grow a code with distance $d$ to a code with distance $(3d-4)$ in two steps by simultaneously growing all corners, see \autoref{fig:layout_enlarge}.

This type of code deformation does not, in itself, perform logical operations, but can be used to move patches of code or to convert between codes where different gates are transversal \cite{Bravyi16deform}.
Other code deformation procedures such as moving holes or twists do perform unitary logical Clifford operations \cite{Rauss2007holebraiding, Bombin2010twists, Brown2017poking}.
In the next section, we present another similar procedure which executes a logical measurement.

\subsection{Lattice Surgery}
Lattice surgery is a particular measurement-based procedure that acts non-trivially on logical information.
By going through two steps of deformation, it implements a joint measurement of logical operators,
typically $\overline{X}_1\overline{X}_2$ or $\overline{Z}_1\overline{Z}_2$, where $\overline{X}_j$ and $\overline{Z}_j$ denote the logical operators of the logical qubit $j$.
We will focus on the $\overline{Z}_1\overline{Z}_2$ measurement and review the protocol used for the surface code \cite{horsman2012surface, landahl2014quantum}.

\begin{figure}[!ht]
\centering
\begin{subfigure}[h]{0.2\textwidth}
\caption{}
\includegraphics[width = .9\textwidth]{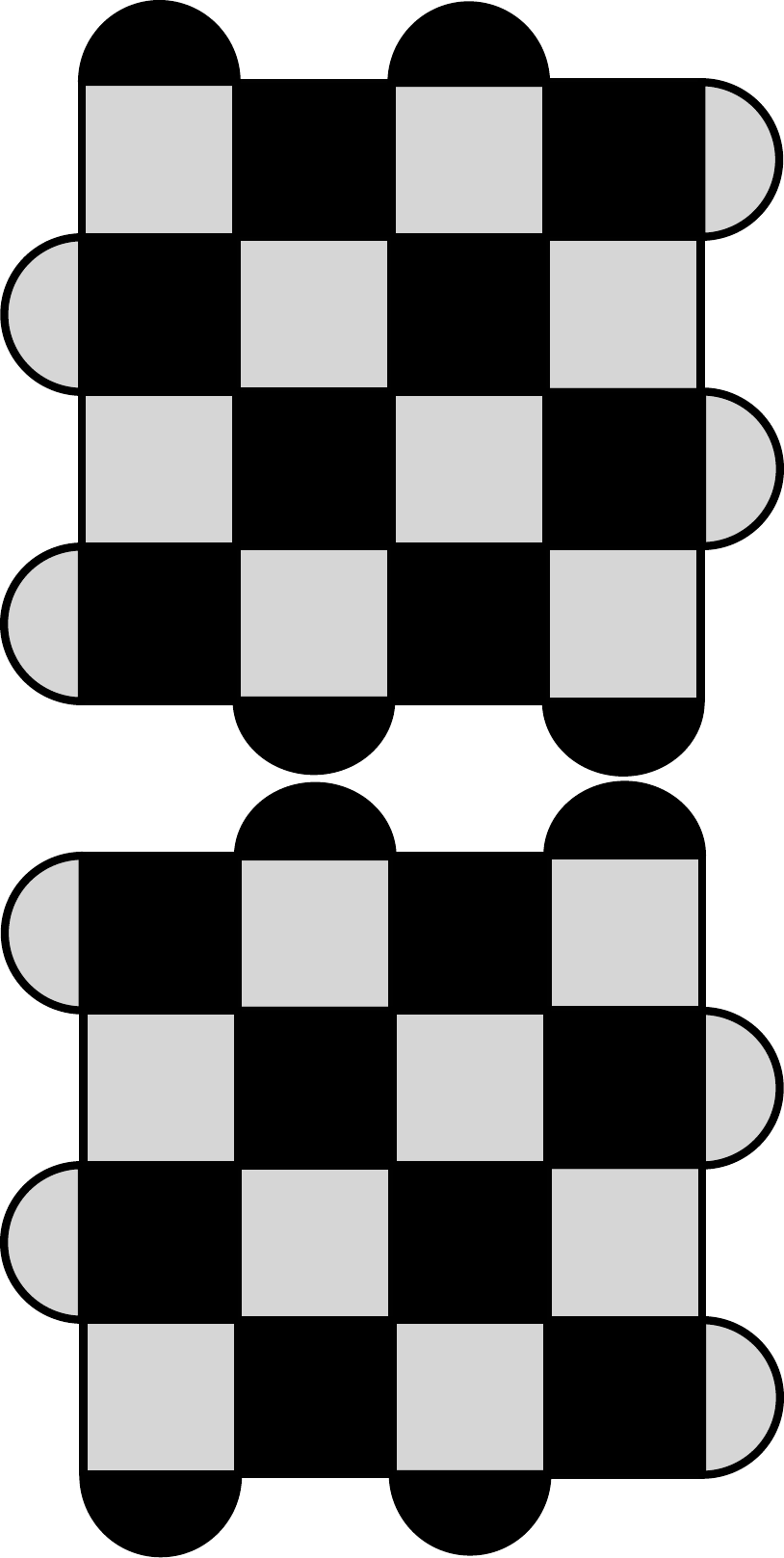}
\label{fig:ssclayout}
\end{subfigure}\scalebox{1}{$\;\xrightarrow{\text{Merge}}\;$}
\begin{subfigure}[h]{0.2\textwidth}
\caption{}
\includegraphics[width = .9\textwidth]{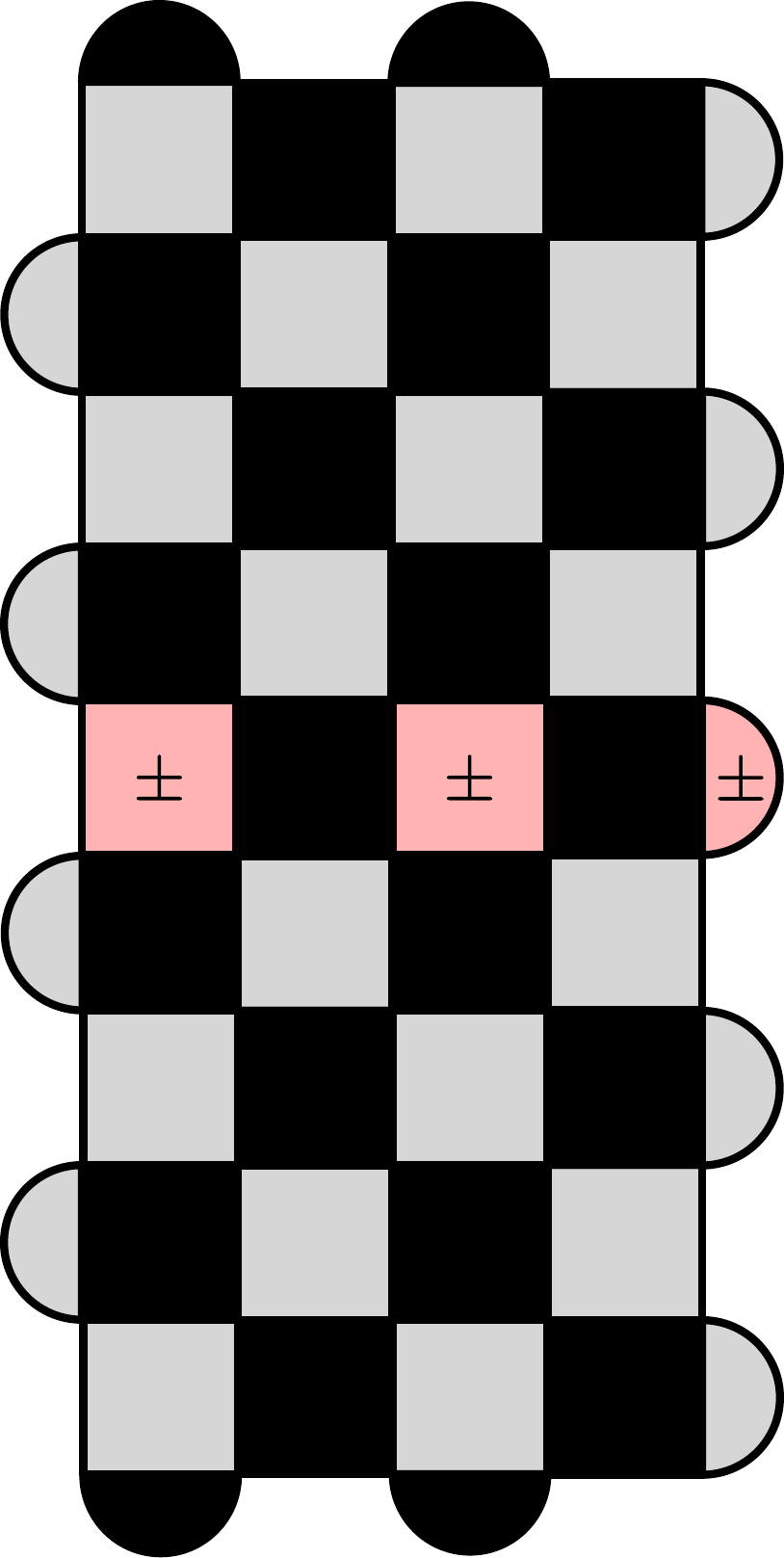}
\label{fig:sscmerged}
\end{subfigure}\scalebox{1}{$\;\xrightarrow{\text{Split}}\;$}
\begin{subfigure}[h]{0.2\textwidth}
\caption{}
\includegraphics[width = .9\textwidth]{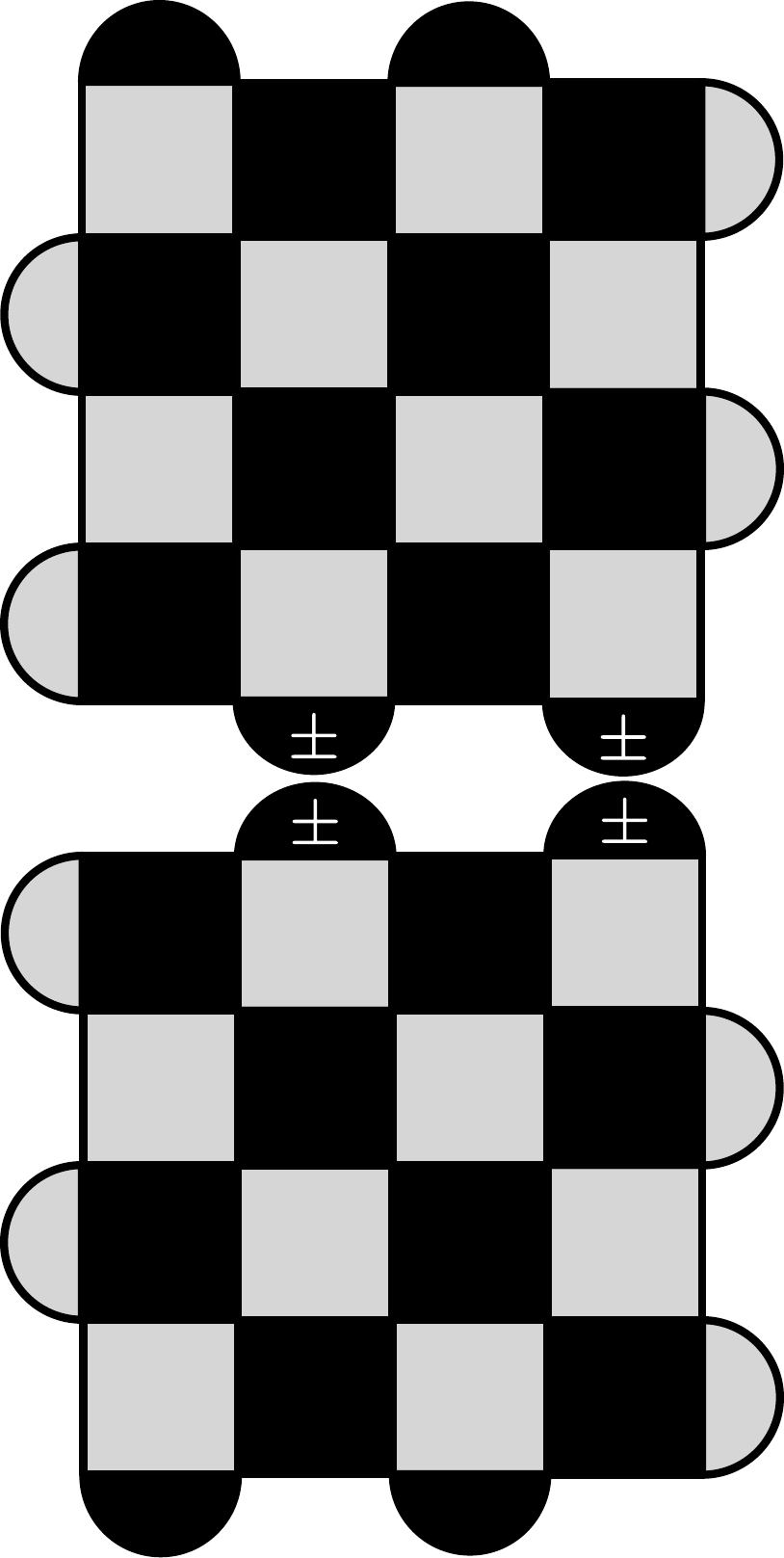}
\label{fig:sscsplit}
\end{subfigure}
\caption{Lattice surgery for the rotated surface code.
Grey plaquettes show $Z$-stabilizers, black plaquettes represent $X$-stabilizers.
A `$\pm$' label indicates a random sign for the corresponding plaquette in the stabilizer group.
(\subref{fig:ssclayout}) Initial layout, two rotated surface codes.
(\subref{fig:sscmerged}) The merged lattice, which is a surface code with random $\pm$ signs on the newly-measured (red) plaquettes.
(\subref{fig:sscsplit}) The \emph{split} lattices, in which the original stabilizers are measured again.
Random $\pm$ signs are produced on the boundary $X$-stabilizers.}
\label{fig:surgerysteps}
\end{figure}

Consider two patches of $L\times L$ rotated surface code, as in \autoref{fig:ssclayout}.
Each has a $\overline{Z}$ along the boundary which faces the other patch.
In the \emph{merge} step, one measures the intermediary $Z$-plaquettes (in pink in \autoref{fig:sscmerged}).
These plaquettes are such that the product of all outcomes is the outcome of the $\overline{Z}_1 \overline{Z}_2$ measurement, but any subset of these outcomes produces a random result when multiplied together. This ensures that the only non-stabilizer operator whose eigenvalue can be inferred from these measurements is $\overline{Z}_1 \overline{Z}_2$.
These measurements do not commute with the weight-2 $X$ stabilizers at the joint boundary (in \autoref{fig:ssclayout}). The Gottesman-Knill theorem \cite{gottesman1998heisenberg} prescribes how to update the stabilizer after such measurements, namely we only retain elements in the original stabilizer group which do commute with the newly measured stabilizers. 
This implies that the code becomes a $2L \times L$ patch of surface code, apart from some minus signs on the newly-measured $Z$-checks.
This merge step is very similar to the rotation presented before, except that some logical information is gained in the process and the additional corrections which fix the state into the new code space may involve one of the original logical operators (when the number of intermediary plaquettes with $-1$ eigenvalues is odd).
To finish the protocol, the original code space must be restored by performing a \emph{splitting} operation, measuring the original stabilizers of the two separate patches instead of the intermediary $Z$-plaquettes.
Those $Z$-plaquettes, as in the merge step, anticommute with the boundary $X$-stabilizers, and will be removed from the stabilizer group. Their product, equal to $\overline{Z}_1\overline{Z}_2$, does commute, and will remain as a stabilizer of the final state.
In addition, the boundary $X$-plaquettes will have random $\pm$ signs which are perfectly correlated between facing pairs.
Therefore, one can eliminate these $\pm$ signs by applying some of the former stabilizers (those supported on the  intermediary $Z$-plaquettes).

One can check (see the algebraic proof in Appendix \ref{sec:proofs}) that depending on the outcome ($\pm 1$) of the logical $\overline{Z}_1\overline{Z}_2$ measurement, the merge and split operations, respectively $M_\pm$ and $S_\pm$ can be expressed as
\begin{alignat}{3}
M_+ &= \ket{\overline{0}}\bra{\overline{0}\overline{0}}+\ket{\overline{1}}\bra{\overline{1}\overline{1}},\qquad &
S_+ &= \ket{\overline{0}\overline{0}}\bra{\overline{0}}+\ket{\overline{1}\overline{1}}\bra{\overline{1}},\label{eq:MSp}\\
M_- &= \ket{\overline{0}}\bra{\overline{0}\overline{1}}+\ket{\overline{1}}\bra{\overline{1}\overline{0}},\qquad &
S_- &= \ket{\overline{0}\overline{1}}\bra{\overline{0}}+\ket{\overline{1}\overline{0}}\bra{\overline{1}}.\label{eq:MSm}
\end{alignat}
They are related to the projections, $P_\pm$, onto the $\pm1$ eigenspace of $\overline{Z}_1\overline{Z}_2$ by composition:
\[P_+ = S_+\circ M_+,\qquad P_- = S_-\circ M_-.\]

\begin{figure}[!ht]
\centering
\begin{subfigure}[h]{0.33\textwidth}
\caption{}
\includegraphics[width = \textwidth]{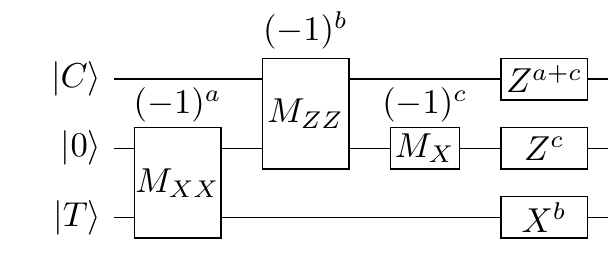}\vspace{5mm}
\includegraphics[width = \textwidth]{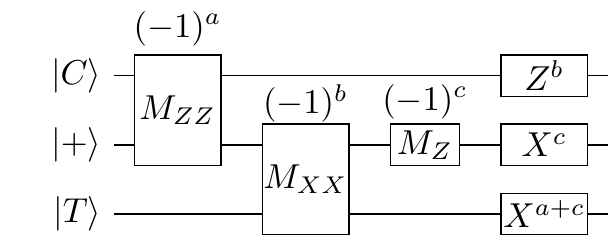}
\label{fig:cnotcircuit}
\end{subfigure}\hspace{5mm}
\begin{subfigure}[h]{0.35\textwidth}
\caption{}
\includegraphics[width = \textwidth]{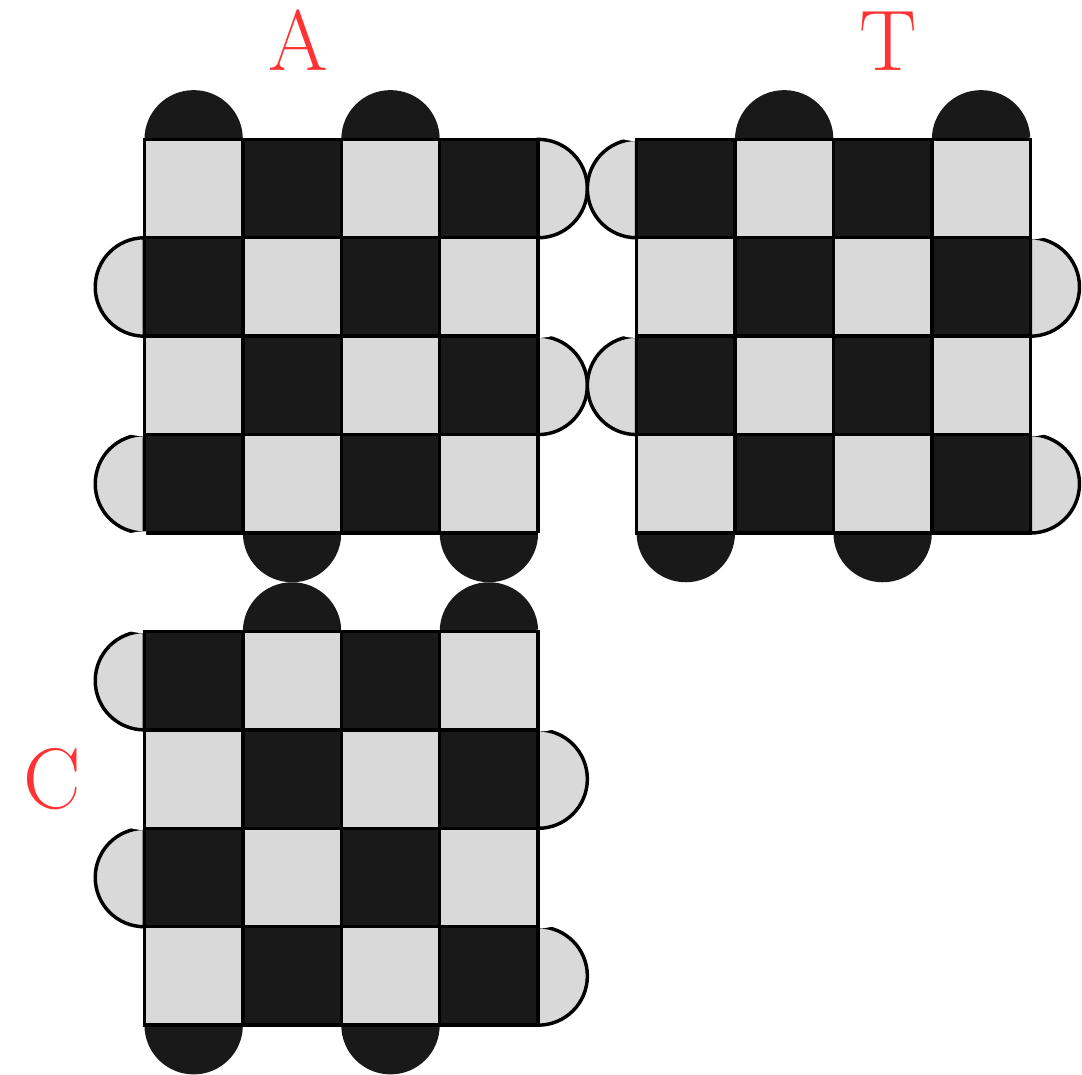}
\label{fig:cnotlayout}
\end{subfigure}\hspace{10mm}
\caption{(a) Two equivalent measurement-based circuits for the \cnot~gate.
(b) The qubit layout for a \cnot~gate between two surface-code qubits. 
\textcolor{red}{C} is the control qubit, \textcolor{red}{T} is the target qubit, and \textcolor{red}{A} is a logical ancilla. }
\label{fig:cnot}
\end{figure}

In particular, lattice surgery allows us to implement the measurement-based \cnot~gate \cite{gottesman1998fault} in a 2D layout with only local operations as shown in \autoref{fig:cnot}.
We note that a more general set of operations which can be implemented by lattice surgery can be constructed using the relation between the merge and split operations considered here and the three-legged nodes of the ZX-calculus \cite{beaudrap2017zx}. 
For the purposes of this work, however, we will limit our discussion to \cnot{} gates.

\section{Gauge Fixing}
\label{sec:GF}

Gauge fixing \cite{paetznick2013universal} is an approach which has been used to implement universal fault-tolerant gate sets in \emph{subsystem codes} \cite{poulin2005stabilizer}.
A subsystem code is equivalent to a stabilizer code in which some of the logical qubits are not used to carry any logical information.
These logical qubits are called \emph{gauge} qubits and they can be acted on or measured without disturbing the states of the other logical qubits, which are used to store and process quantum information.
Then, one way to formally define a subsystem code, $C$, is to define a subgroup of the Pauli group, called the \emph{gauge group} $\mathcal{G}$, containing all the Pauli stabilizers as well as the Pauli operators defining the gauge qubits.
This subgroup is non-Abelian as it contains anti-commuting Pauli operator pairs which represent the gauge qubit logical operators.
The stabilizer group, $\mathcal{S}$, can be derived from $\mathcal{G}$ as its center, denoted ${\rm Z}(\cdot)$, i.e. containing all elements in $\mathcal{G}$ which mutually commute
\begin{equation}
\mathcal{S} = {\rm Z}(\mathcal{G})=\mathcal{C}(\mathcal{G}) \cap \mathcal{G},\label{eq:centerG}
\end{equation}
where $\mathcal{C}(\mathcal{G})$ denotes the centralizer of $\mathcal{G}$ in the Pauli group, i.e. all elements in the Pauli group with commute with all elements in $\mathcal{G}$. Elements in $\mathcal{G}$ which are not in $\mathcal{S}$ are the Pauli operators acting non-trivially on the gauge qubits: this is the set of gauge operators $\mathcal{L}_g$
\begin{equation}
\mathcal{L}_g = \mathcal{G}\setminus\mathcal{S}.\label{eq:gaugeop}
\end{equation}
Following this, one can define operators for the actual logical qubits which by definition are elements in $\mathcal{C}(\mathcal{S})\setminus\mathcal{S}$. If these operators act trivially on the gauge qubits, we call these \emph{bare} logical operators. Bare logical operators can be multiplied by elements in $\mathcal{L}_g$ to become \emph{dressed} logical operators which also act on the gauge qubits. We can write
\begin{equation}
\mathcal{L}_{\rm bare} = \mathcal{C}(\mathcal{G})\setminus\mathcal{G},\qquad\mathcal{L}_{\rm dressed} = \mathcal{C}(\mathcal{S})\setminus\mathcal{G}.\label{eq:baredressed}
\end{equation}
Note that with this definition we have, $\mathcal{L}_{\rm bare} \subset \mathcal{L}_{\rm dressed}$.
The distance of the subsystem code $C$ is the smallest weight of any of its logical operators,
\begin{equation}
d_{C} = \min_{\ell\in \mathcal{L}_{\rm dressed}}{\rm wt}(\ell).\label{eq:gaugedistance}
\end{equation}

One advantage of subsystem codes is that to measure stabilizers, one is free to measure any set of checks in the gauge group as long as this set generates the stabilizer group.
By measuring elements in the full gauge group, one can put the gauge qubits in specific states, permitting different sets of transversal logical gates.
This act of putting the gauge qubits in a specific state is called \emph{gauge fixing}.
The idea is to measure a commuting subset of gauge operators (all the $Z$-type gauge operators, for example), obtaining $\pm 1$ outcomes and applying the anticommuting, or \emph{conjugate} partner operator (an $X$-type gauge operator in the example), wherever a $-1$ outcome has been obtained.
In the example, this would fix all gauge qubits to the $\ket{0}$ state.
While the gauge is fixed in this way, the $Z$-type gauge operators become elements of the stabilizer group, so $\mathcal{S}$ is augmented to some larger Abelian subgroup of $\mathcal{G}$.
\cref{app:conv} shows an example of how code conversion between the $\llbracket 7,\,1,\,3 \rrbracket$ Steane code to the $\llbracket 15,\,7,\,3 \rrbracket$ Reed-Muller code can be viewed as gauge fixing.

\section{Fault-Tolerance Analysis with Gauge Fixing}
\label{sec:Unification}

\begin{figure}[htb]
	\centering
	\begin{subfigure}[p]{.45\textwidth}
		\centering
		\caption{}
		\label{subfig:venn1}
		\includegraphics[width=\textwidth]{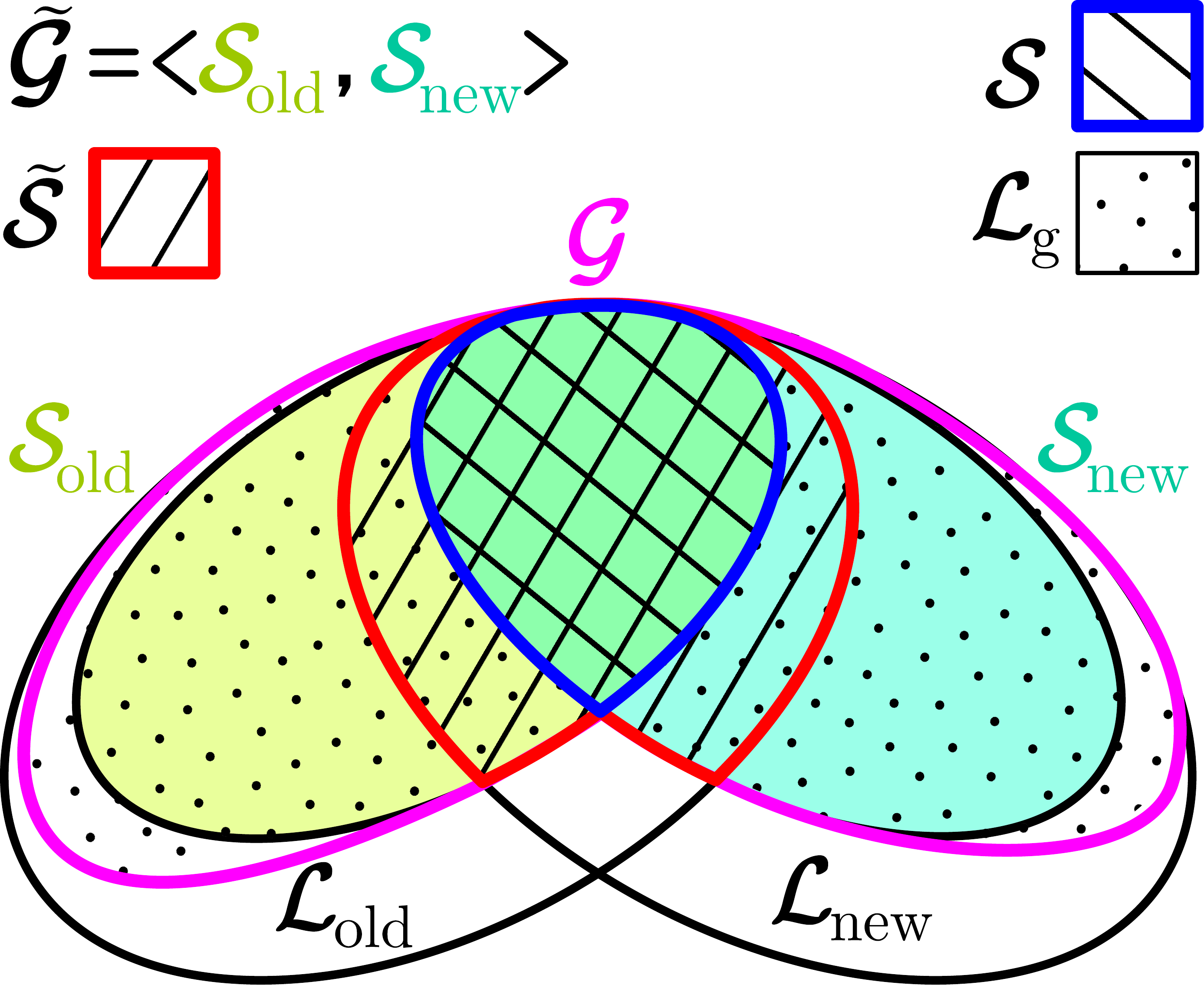}
	\end{subfigure}\qquad
	\begin{subfigure}[p]{.4\textwidth}
		\centering
		\caption{}
		\label{subfig:venn2}
		\includegraphics[width=\textwidth]{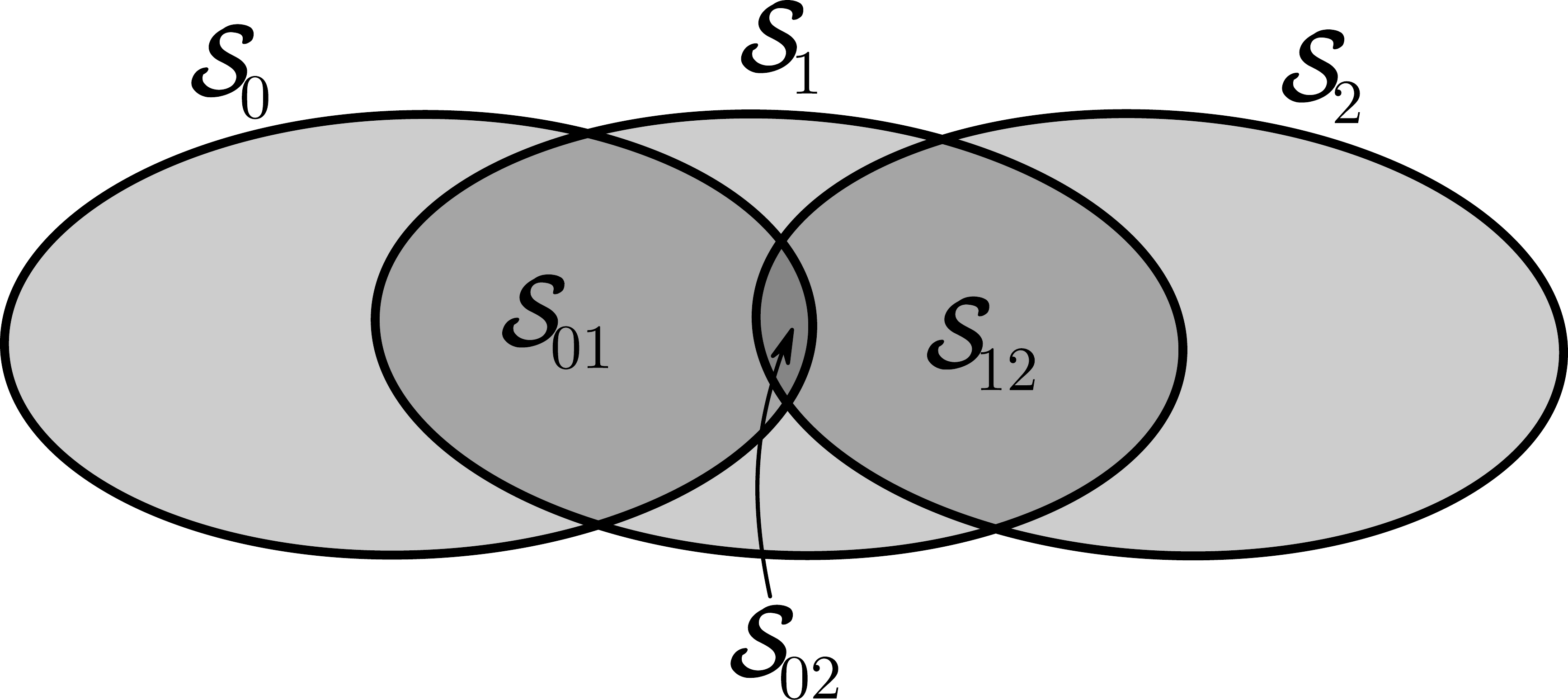}
	\end{subfigure}
	\caption{Venn diagrams depicting the relations between the different sets of Pauli operators concerning the gauge group $\mathcal{G}$ of interest, see main text.
	(\subref{subfig:venn1}) For one step, the yellow-green set represents the old stabilizer group, $\mathcal{S}_{\rm old}$, and the blue-green set the new group, $\mathcal{S}_{\rm new}$. Both are surrounded by the logical operators,  $\mathcal{L}_{\rm old}$ and $\mathcal{L}_{\rm new}$ respectively. The gauge group generated by both, $\tilde{\mathcal{G}} = \left \langle\mathcal{S}_{\rm old}, \mathcal{S}_{\rm new}\right \rangle$, has $\tilde{\mathcal{S}}$ as its center, shown by the down-left-dashed region.
	The gauge group of interest, $\mathcal{G}$, is outlined in purple and has $\mathcal{S}$, in the down-right-dashed region as its center.
	The set of gauge operators defining the gauge qubits, $\mathcal{L}_g$, is the dotted region. When switching from $\mathcal{S}_{\rm old}$ to $\mathcal{S}_{\rm new}$ one fixes the gauge for the elements in the blue-green dotted region $\mathcal{M}_{\rm fix}=\tilde{\mathcal{G}}\setminus \mathcal{S}_{\rm old}$.
	(\subref{subfig:venn2}) One possible scenario for two successive steps of deformation.
	Doing it in two steps, i.e. from $\mathcal{S}_0 \rightarrow \mathcal{S}_1$, and then from $\mathcal{S}_1 \rightarrow \mathcal{S}_2$ permits to use successively the stabilizer groups $\mathcal{S}_{01}$ and then $\mathcal{S}_{12}$ for error correction. Skipping the intermediary steps, one can only use $\mathcal{S}_{02}$ which might offer less protection.}
	\label{fig:venndiag}
\end{figure}
In this section, we show how both code deformation and lattice surgery can be viewed as gauge fixing operations and therefore, one can use gauge fixing to analyze the fault-tolerance of these operations.

We consider the QEC codes before and after an deformation step, denoted as $C_{\rm old}$ and $C_{\rm new}$, with stabilizer groups $\mathcal{S}_{\rm old}$ and $\mathcal{S}_{\rm new}$, respectively. Both codes are fully defined on the same set of qubits.
The logical operators of each code are defined as
\[\mathcal{L}_{\rm old} = \mathcal{C}(\mathcal{S}_{\rm old})\setminus \mathcal{S}_{\rm old}\,,\qquad\mathcal{L}_{\rm new} = \mathcal{C}(\mathcal{S}_{\rm new})\setminus \mathcal{S}_{\rm new}.\]

The intuition we follow is to see the two stabilizer codes as two different gauges of the same subsystem code.
The first step, then, is to define a joint subsystem code, $\tilde{C}$, whose gauge group, $\tilde{\mathcal{G}}$, is generated by both $\mathcal{S}_{\rm old}$ and $\mathcal{S}_{\rm new}$,
\[\tilde{\mathcal{G}} = \langle\mathcal{S}_{\rm old}, \mathcal{S}_{\rm new}\rangle.\]
The generated group, $\tilde{\mathcal{G}}$, is not necessarily Abelian, since it contains elements of $\mathcal{S}_{\rm old}$ which may anti-commute with some elements of $\mathcal{S}_{\rm new}$.

The stabilizer group, $\tilde{\mathcal{S}}$, defined as in Eq.~\eqref{eq:centerG}, can be characterised as follows:
Elements in the center of $\tilde{\mathcal{G}}$ also have to be in the centralisers of $\mathcal{S}_{\rm old}$ and $\mathcal{S}_{\rm new}$.
Moreover, being in both centralisers and in $\tilde{\mathcal{G}}$ is sufficient to be in the center, or
\[\tilde{\mathcal{S}} = \mathcal{C}(\mathcal{S}_{\rm old})\cap\mathcal{C}(\mathcal{S}_{\rm new})\cap\tilde{\mathcal{G}}.\]
See \autoref{subfig:venn1} for a representation of $\tilde{\mathcal{S}}$ as a Venn diagram.
Note that, in addition to containing $\mathcal{S}_{\rm old}\cap\mathcal{S}_{\rm new}$, $\tilde{\mathcal{S}}$ can also contain some logical operators from either $\mathcal{L}_{\rm old}$ or $\mathcal{L}_{\rm new}$.
This is the case for the merge operation of lattice surgery where the logical $\overline{Z}_1\overline{Z}_2\in\mathcal{L}_{\rm old}$ but also $\overline{Z}_1\overline{Z}_2 \in \mathcal{S}_{\rm new}$ and therefore $\overline{Z}_1\overline{Z}_2 \in \tilde{\mathcal{S}}$.
Similarly, for the split operation $\overline{Z}_1\overline{Z}_2\in\mathcal{L}_{\rm new}$ but also in $\mathcal{S}_{\rm old}$ and therefore in $\tilde{\mathcal{S}}$.

As defined above, this subsystem code $\tilde{C}$ indeed admits $\mathcal{S}_{\rm old}$ and $\mathcal{S}_{\rm new}$ as two distinct Abelian subgroups of $\tilde{\mathcal{G}}$.
Therefore the codes $\mathcal{S}_{\rm old}$ and $\mathcal{S}_{\rm new}$ correspond to fixing two different sets of states for the gauge qubits of $\tilde{\mathcal{G}}$.
However, for this to function as a subsystem code, one would have to be stabilized at all times by $\tilde{\mathcal{S}}$ and thus be able to measure all values of the stabilizers of $\tilde{\mathcal{S}}$.

This is not the necessarily the case when $\tilde{\mathcal{S}}$ contains some elements of $\mathcal{L}_{\rm old}$ or $\mathcal{L}_{\rm new}$, and we have to further modify $\tilde{\mathcal{G}}$ to a gauge group $\mathcal{G}$ whose center is solely
\[{\rm Z}(\mathcal{G})=\mathcal{S} = \mathcal{S}_{\rm old}\cap\mathcal{S}_{\rm new}.\]

How do we obtain $\mathcal{G}$ from $\tilde{\mathcal{G}}$?
This new gauge group, $\mathcal{G}$ will be generated by $\mathcal{S}_{\rm old}$ and $\mathcal{S}_{\rm new}$ in addition to (anti-commuting) conjugate partners of elements in the sets $\mathcal{M}_{\rm prep}=\mathcal{S}_{\rm old}\cap\mathcal{L}_{\rm new}$ and $\mathcal{M}_{\rm meas}=\mathcal{S}_{\rm new}\cap\mathcal{L}_{\rm old}$.
More precisely, one views $\mathcal{M}_{\rm prep}$ as a subset of $\mathcal{L}_{\rm new}$, and for each independent logical operator contained in $\mathcal{M}_{\rm prep}$ adds a chosen conjugated partner within $\mathcal{L}_{\rm new}$.
One operates similarly for $\mathcal{M}_{\rm meas}$ by viewing it as a subset of $\mathcal{L}_{\rm old}$.
If we then consider the center of $\mathcal{G}$, we see that all elements in $\mathcal{M}_{\rm prep}$ and 
$\mathcal{M}_{\rm meas}$ are excluded from it since they anti-commute with some elements in $\mathcal{G}$. This means that the center of $\mathcal{G}$ is reduced to ${\rm Z}(\mathcal{G})=\mathcal{S}_{\rm old}\cap\mathcal{S}_{\rm new}$ as desired.

The names $\mathcal{M}_{\rm prep}$ and $\mathcal{M}_{\rm meas}$ are chosen to represent their respective roles in the deformation procedure. In such a procedure one starts from a system encoded in $C_{\rm old}$, i.e. stabilized by $\mathcal{S}_{\rm old}$, and then one measures the new stabilizers, $\mathcal{S}_{\rm new}$. When $\mathcal{S}_{\rm new}$ contains some elements of $\mathcal{L}_{\rm old}$, then in general these elements will not stabilize the state of the system, since it can be in any logical state at the beginning of the procedure. Measuring these operators will return information about the logical state and cannot return information about errors. Thus, by switching to $\mathcal{S}_{\rm new}$ one also performs a logical measurement of the elements in $\mathcal{M}_{\rm meas}$.

It is also possible for $\mathcal{S}_{\rm old}$ to contain some elements of $\mathcal{L}_{\rm new}$.
In that case, the state of the system is initially stabilized by these elements, and remains so, since we only measure operators commuting with them. In this sense, the deformation procedure will prepare the logical $+1$ state of elements in $\mathcal{M}_{\rm prep}$.

We denote the code underlying the code deformation step as $C$. Its gauge group, $\mathcal{G}$, is represented as a Venn diagram in \autoref{subfig:venn1}.
Thus the deformation operation that transforms $C_{\rm old}$ into $C_{\rm new}$ is realized by switching what gauge to fix of the code $C$: in one gauge one obtains $C_{\rm old}$, the other gauge gives $C_{\rm new}$. Since the deformation step can also transform logical information, what gauge elements are fixed is subtle. Namely, note that in this gauge fixing of $C$ to either code $C_{\rm old}$ or $C_{\rm new}$ the gauge elements in $\mathcal{G}\setminus\tilde{\mathcal{G}}$ will never be fixed. Said differently, only the elements of $\mathcal{L}_g$ which are in the blue-green dotted region in Fig.~\ref{fig:venndiag} will be fixed, one can also view these as elements of $\mathcal{M}_{\rm fix} \equiv \tilde{\mathcal{G}}\setminus \mathcal{S}_{\rm old}$.

\subsection{Fault-Tolerance of Code Deformation}
\label{subsec:ft-cd}

Given an underlying subsystem deformation code $C$, one can ensure the fault-tolerance of a code deformation operation by checking three criteria:
\begin{enumerate}
	\item \textbf{Code distance:} The distance of the subsystem code, $C$, must be large enough for the desired protection. Ideally it matches the distances of $C_{\rm old}$ and $C_{\rm new}$ so the degree of protection is not reduced during the deformation step.\label{enum:crit1}
	\item \textbf{Error correction:} The error correction procedure follows that of the subsystem code $C$ through the code deformation step.\label{enum:crit2}
	\item \textbf{Gauge fixing:} To fix the gauge, one has to use operators exclusively from $\mathcal{L}_g=\mathcal{G}\backslash \mathcal{S}$. \label{enum:crit3}
\end{enumerate}

More specifically, criterion \ref{enum:crit2} means that to perform error correction, one has to reconstruct from the measurements of $\mathcal{S}_{\rm new}$ the syndrome given by $\mathcal{S}$.
Importantly, criteria \ref{enum:crit2} and \ref{enum:crit3} demonstrate that the processes of error correction and that of gauge fixing are two separate processes with different functionality.
Both processes require the application of Pauli operators (in hardware or in software) to make sure that stabilizer measurements are corrected to have outcome $+1$.
The error correction process does this to correct for errors, while the gauge-fixing process does this to move from $C_{\rm old}$ to $C_{\rm new}$.

This description holds for one step of deformation, so that for each step in a sequence of deformations one has to examine the corresponding subsystem code $C$ and its distance. Depending on the sequence, \autoref{subfig:venn2} illustrates why skipping steps could lead to poor distance and poor protection against errors.
This discussion also assumes that stabilizer measurements are perfect; the effect of noisy stabilizer measurements is considered in the following section.

\subsubsection{Noisy Measurements}

When one considers noisy syndrome measurements, one needs to ensure that both the stabilizer outcomes and the state of the gauge qubits can be learned reliably.
For 2D stabilizer codes such as the surface code this is simply done by repeating the measurements.
To process this repeated measurement information for the surface code, one no longer uses the syndrome but the {\em difference syndrome}: the difference syndrome is marked as non-trivial (we say that a \emph{defect} is present) only when the syndrome value changes from the previous round of measurement.
This difference syndrome or defect gives information about both qubit errors as well as measurement errors.

\begin{figure}[htbp]
	\centering
	\includegraphics[width=.9\textwidth]{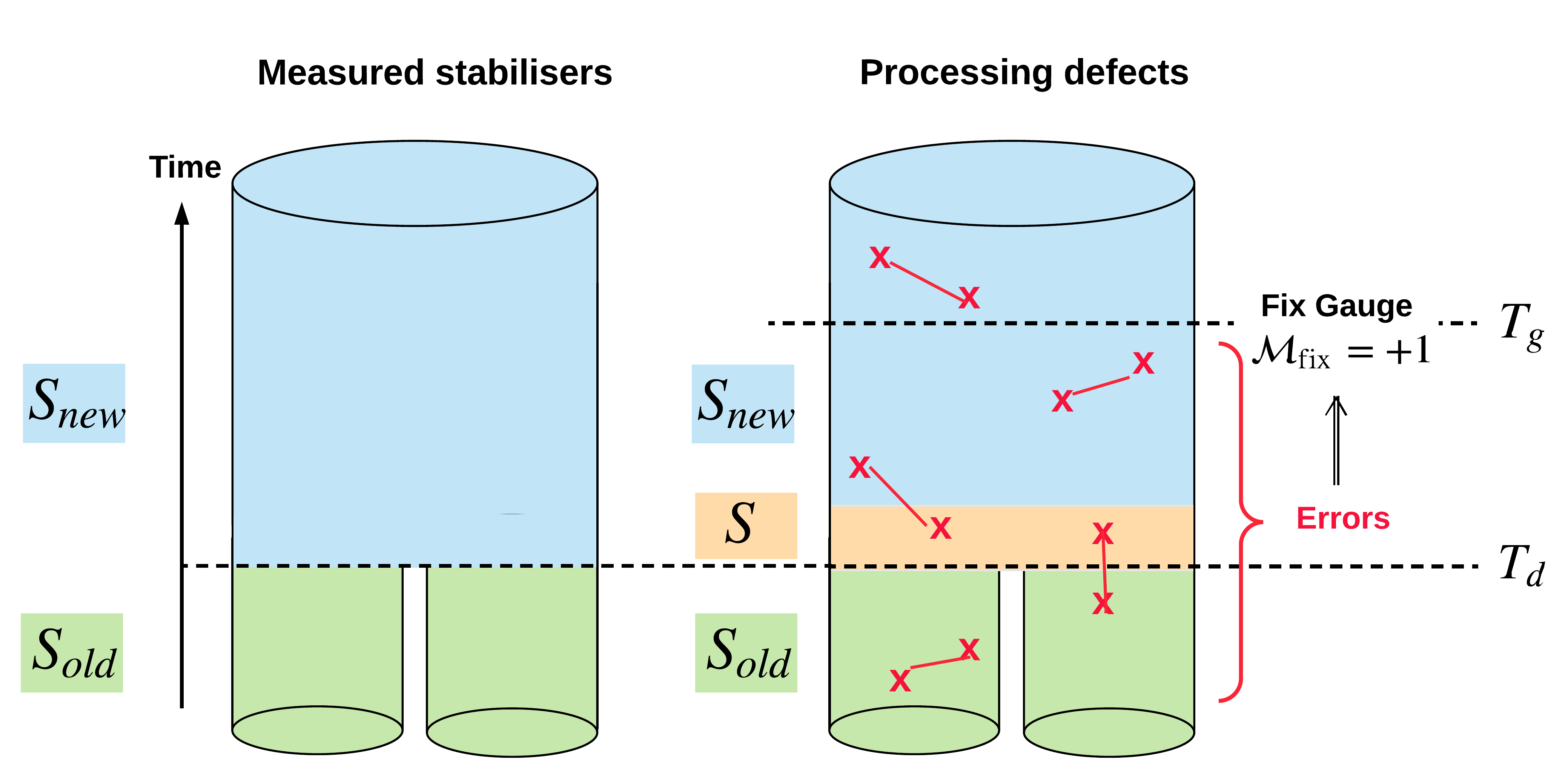}
	\caption{Schematic drawing of a code deformation procedure with repeated noisy measurements, with time increasing upwards. $T_d$ designates the time step at which the code deformation (the switch from measuring the checks of $\mathcal{S}_{\rm old}$ to those of $\mathcal{S}_{\rm new}$) is performed. $T_g$ is the time at which one is confident enough about the state of the gauge qubits, taking into account errors, to fix their states. This means that, after $T_g$, another logical computation can be performed.
	(Right) The first round of measurement of $\mathcal{S}_{\rm new}$ at time $T_d$ does not have a previous value to compare to in order to construct a difference syndrome, i.e. one can only construct defects for $\mathcal{S}$.
	Immediately after this step, one can derive the difference syndrome of the full $\mathcal{S}_{\rm new}$, placing defects accordingly.
	Using defects before and after $T_d$, one processes error information to infer the value of the gauge operators in $\mathcal{M}_{\rm fix}$ at time $T_g$, thus fixing the gauge at $T_g$.}
	\label{fig:decoding_prescription}
\end{figure}

How do we construct the difference syndrome at the code deformation step $T_d$ in Fig.~\ref{fig:decoding_prescription}?
At $T_d$ one obtains the syndrome for the code $\mathcal{S}_{\rm new}$. For those elements in $\mathcal{S}_{\rm new}$ which are in $\mathcal{S}$, we know what this syndrome should have been if no measurement or data errors had occurred since the previous QEC round which measured the stabilizers of $\mathcal{S}_{\rm old}$. Therefore, we can place defects when the found syndrome changes from what it was expected to be based on the last round of measurements with $\mathcal{S}_{\rm old}$.
$\mathcal{S}_{\rm new}$ also contains a subset of elements in $\mathcal{L}_g$, namely the blue-green dotted region $\mathcal{M}_{\rm fix}$ in Fig.~\ref{subfig:venn1}. Some of these elements are also contained in $\mathcal{L}_{\rm old}$ (down-right-dashed area in Fig.~\ref{subfig:venn1}), i.e. they are elements of $\mathcal{M}_{\rm meas}$.
The eigenvalues of these elements in $\mathcal{M}_{\rm meas}$ depends on the logical state and are therefore not a proper syndrome for $S_{\rm old}$. So only after one more round of QEC with 
$\mathcal{S}_{\rm new}$ one can mark whether the syndrome for these elements in $\mathcal{M}_{\rm meas}$ changes, and either place a defect or not.
In addition, the eigenvalues of the gauge operators in the remaining blue-green dotted region take random $\pm 1$ eigenvalues (since they anticommute with some elements in $\mathcal{S}_{\rm old}$): for these checks, like for the elements in $\mathcal{M}_{\rm meas}$, there is no previous record to construct a difference syndrome right away. Again, only after one round of QEC with $\mathcal{S}_{\rm new}$ one can again mark whether the syndrome changed, placing a defect for an element or not. In processing these new syndromes of $\mathcal{S}_{\rm new}$ to do error correction, we should also allow them to be matched with virtual defects placed beyond the past-time boundary $T_d$. For example, a measurement error in the first step when the syndrome is randomly $+1$ or $-1$, followed by many rounds without measurement error, produces a single defect and should be interpreted as the first measurement being incorrect.
In this sense, there is only one layer of time where the defects are those of $\mathcal{S}$ as indicated on the right in Fig.~\ref{fig:decoding_prescription}.

Given all defect syndromes, minimum-weight matching can be used to decode (see Fig.~\ref{fig:decoding_prescription}), to infer some errors as they have occurred in a window of time before and after $T_g$ and $T_d$ (one may use a sliding window as in \cite{dennis2002topological}). Let us then imagine that by matching defects in a window which goes beyond a so-called gauge-fixing time $T_g$, one infers a set of measurement and data errors. These errors are projected forwards to the time-slice $T_g$ and they are used to do three things. One is to correct the value of elements in $M_{\rm meas}$ (if any), so that the logical measurement has been completed and properly interpreted. The second is to determine or fix the gauge, i.e. determine the outcome of elements $\mathcal{M}_{\rm fix}$ in the blue-green dotted region of Fig.~\ref{fig:venndiag}.
As we have argued, these gauge values may be $\pm 1$ at random and hence Pauli gauge-fixing corrections can be added in software to make the outcomes all $+1$ if one wishes to work with the frame where all elements in $\mathcal{S}_{\rm new}$ have $+1$ eigenvalue. These Pauli gauge-fixing corrections are not error corrections and any set of Pauli operators can be chosen as long as they solely fix the values of the elements in $\mathcal{M}_{\rm fix}$.
Thirdly, the projected errors provide the usual update of the Pauli frame for the code $\mathcal{S}$, so together with the gauge-fixing corrections, for the code $\mathcal{S}_{\rm new}$. The whole procedure is represented schematically in Fig.~\ref{fig:decoding_prescription}; at time $T_g$, the code deformation step is finished.

Note that, after $T_d$, the elements in $\mathcal{M}_{\rm prep}$ are no longer measured, but their fixed values before the code deformation now represent logical states prepared by code deformation.

Typically, for 2D stabilizer codes, the time window between $T_g$ and $T_d$ needs be of size $O(d)$ in order to fix the gauge, where $d$ is the distance of code $C$.
In some cases, the measurements contain enough redundant information about the gauge operators so that $T_g$ can be equal to $T_d$ (e.g. in single-shot error correction schemes based on redundancy of the checks).
For example, this is the case when performing the logical measurement of a patch of code by measuring every single qubit in the $Z$ basis.
This is also the case for the logical measurement step of the plain surgery technique explained below.

In the remainder of this section, we apply this formalism to the code deformation and lattice surgery operations discussed earlier.

\subsection{Code Deformation Examples}
\subsubsection{Grow Operations}

\begin{figure}[htb!]
	\centering
	\begin{subfigure}[]{0.4\textwidth}
		\caption{}
		\includegraphics[width=0.8\textwidth]{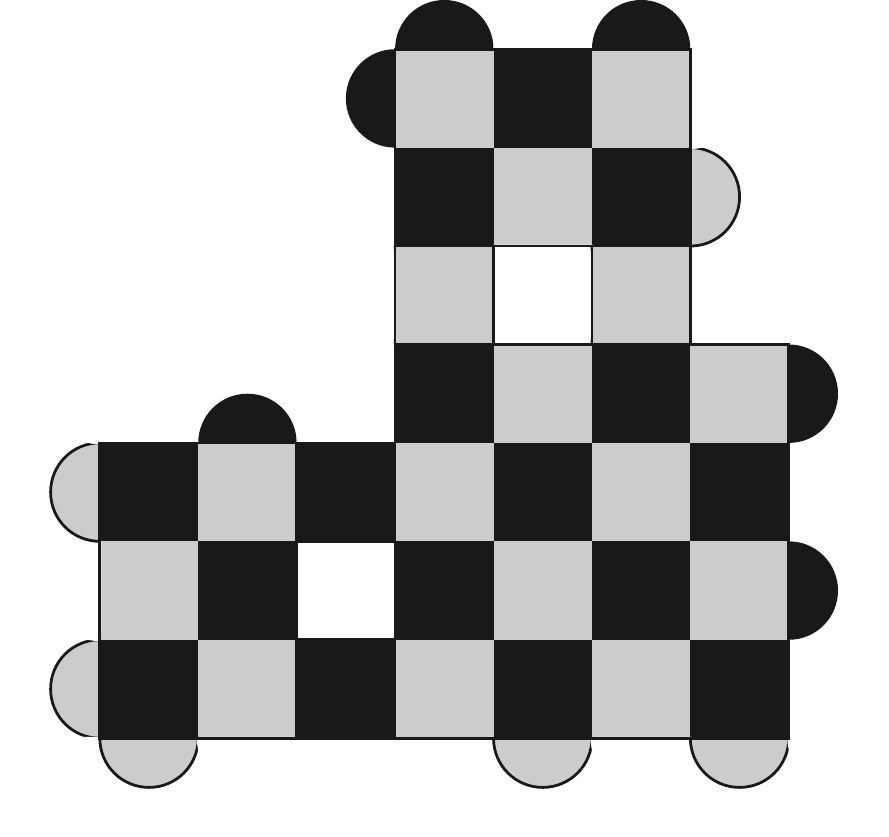}
		\label{fig:grow_eg3}
	\end{subfigure}
	\begin{subfigure}[]{0.4\textwidth}
		\caption{}
		\includegraphics[width=0.8\textwidth]{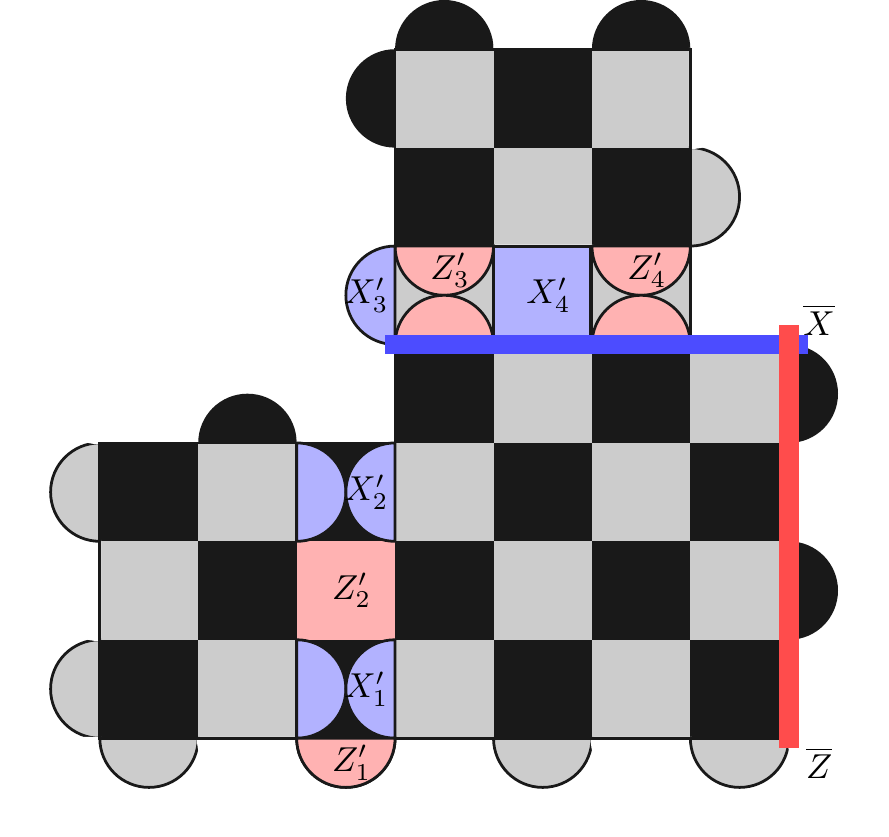}
		\label{fig:grow_eg4}
	\end{subfigure}
	\caption{Description of the subsystem code, $C$, which holds during the first step of the grow operation depicted in \cref{fig:layout_r1,fig:layout_r2}.
		(a) Generators for the stabilizer group, $\mathcal{S}$, of $C$.
		(b) Generators for the whole gauge group $\mathcal{G}$ of $C$. Highlighted in red and blue, respectively, are gauge operators, elements of $\mathcal{L}_g$, of $Z$-type and $X$-type, respectively. The logical operators, $\overline{X},\overline{Z}\in\mathcal{L}_{\rm bare}$, are also represented in brighter colours.}
	\label{fig:growgauge}
\end{figure}

Gauge fixing, when applied to the growing operations of \autoref{fig:h_layout} and \autoref{fig:h_flip}, reveals an underlying subsystem code with a small number of widely-spaced holes and large boundaries, resulting in a high distance.
The stabilizer group, $\mathcal{S}$, as well as the gauge operators, $\mathcal{L}_g$, for the subsystem code $C$ which governs the deformation from \autoref{fig:layout_r1} to \autoref{fig:layout_r2}, are shown in \autoref{fig:growgauge}.

In all figures of this paper, light blue and light red patches individually represent $X$-type and $Z$-type gauge operators, and bright blue and bright red qubit chains are $\overline{X}$ and $\overline{Z}$ operators respectively.
The grow operation is changing the gauge from one in which the gauge operators not overlapping between the initially separate patches are fixed, denoted as $\left \lbrace X_1',\, X_2',\, Z_3',\, Z_4' \right \rbrace$ in \autoref{fig:grow_eg4}, to one in which the overlapping ones are fixed, denoted as $\left \lbrace Z_1',\, Z_2',\, X_3',\, X_4' \right \rbrace$ in \autoref{fig:grow_eg4}.
The distance of $C$ is still $5$, matching the distance of the initial code.

\begin{figure}[htb!]
	\centering
	\begin{subfigure}[]{0.4\textwidth}
		\caption{}
		\includegraphics[width=0.8\textwidth]{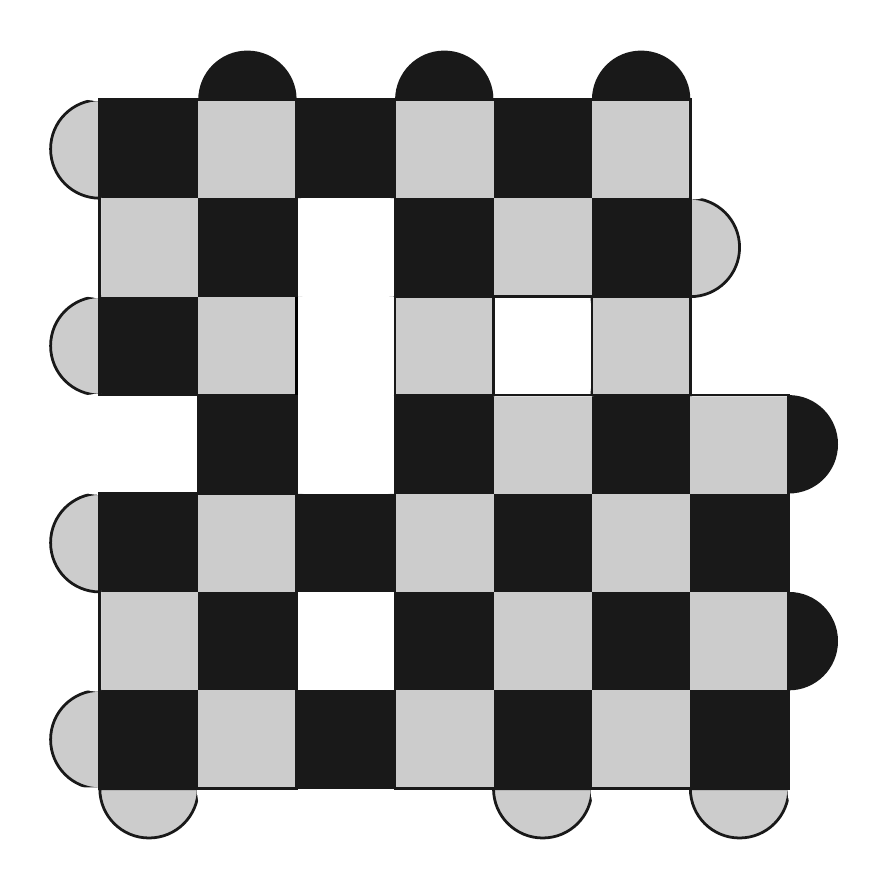}
		\label{fig:nftgrow_eg3}
	\end{subfigure}
	\begin{subfigure}[]{0.4\textwidth}
		\caption{}
		\includegraphics[width=0.8\textwidth]{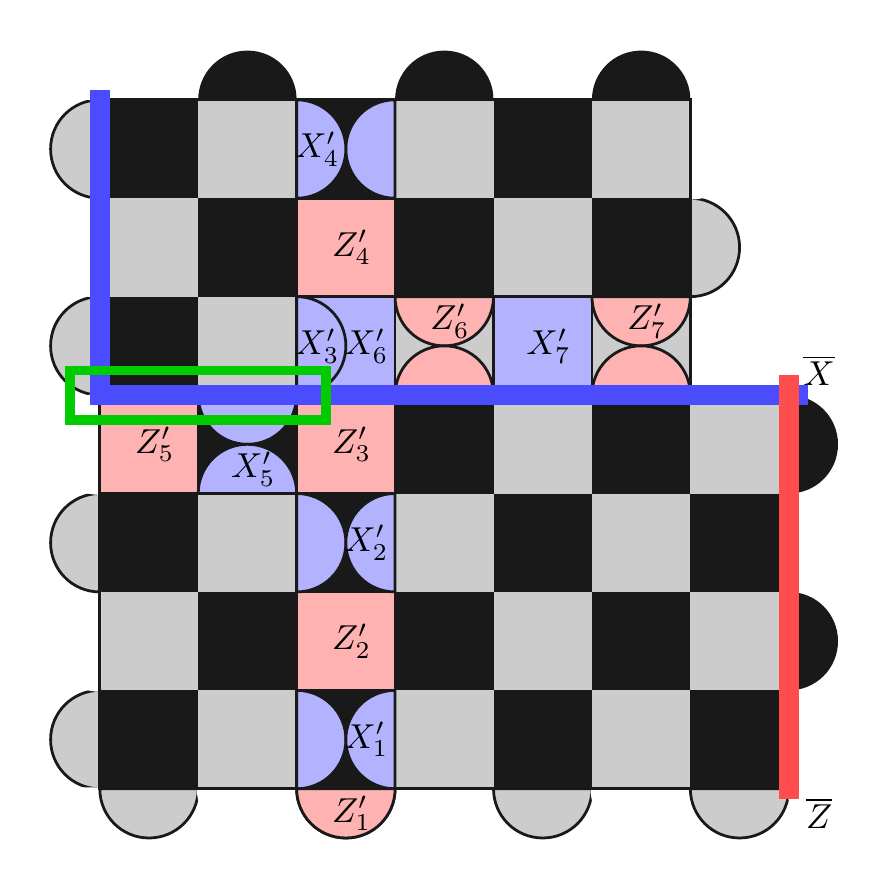}
		\label{fig:nftgrow_eg4}
	\end{subfigure}
	\caption{The operators of the subsystem code for the one-step grow operation from Fig. 1a to Fig. 1c, skipping Fig. 1b: (a) The stabilizers which generate $\mathcal{S}$ and (b) the whole gauge group, $\mathcal{G}$, with highlighted gauge operators and logical operators.}
	\label{fig:nftgrowgauge}
\end{figure}

Now consider what happens if we would go directly from \autoref{fig:layout_r1} to \autoref{fig:layout_r3}.
The stabilizers and the gauge operators for this operation are shown in \autoref{fig:nftgrowgauge}.
Similarly, one fixes the gauge going from separate patches to a single patch.
The distance of the subsystem code for this operation is only $3$.
Indeed one of the minimum-weight dressed logical operators is the $\overline{Z}$ on the qubits in the green box in \autoref{fig:nftgrow_eg4}.
That means that, in order to preserve the code distance, one should perform the intermediary step.

\subsubsection{The merging and splitting operations}

\begin{figure}[htb!]
	\centering
	\begin{subfigure}[]{0.25\textwidth}
		\caption{}
		\includegraphics[width=.9\textwidth]{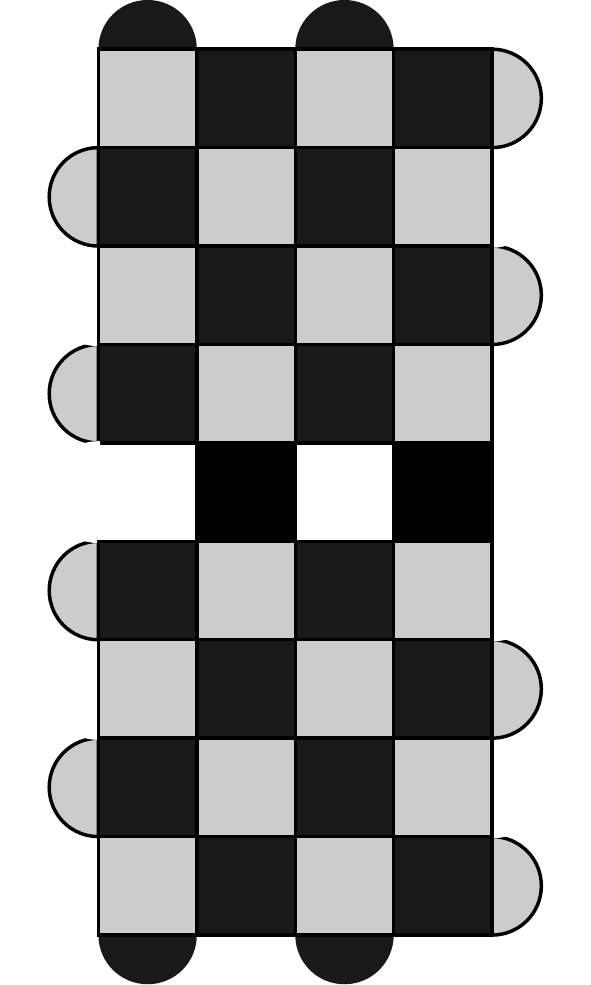}
		\label{fig:cnotgauge_stab}
	\end{subfigure}
	\begin{subfigure}[]{0.25\textwidth}
		\caption{}
		\includegraphics[width=.9\textwidth]{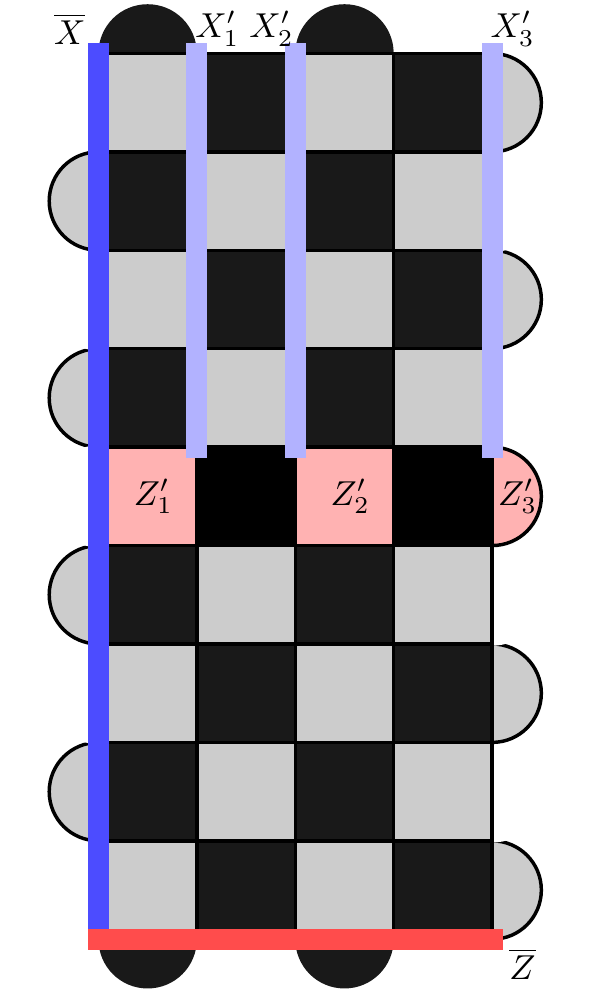}
		\label{fig:cnotgauge_split}
	\end{subfigure}
	\begin{subfigure}[]{0.22\textwidth}
	\caption{}
	\includegraphics[width=.8\textwidth]{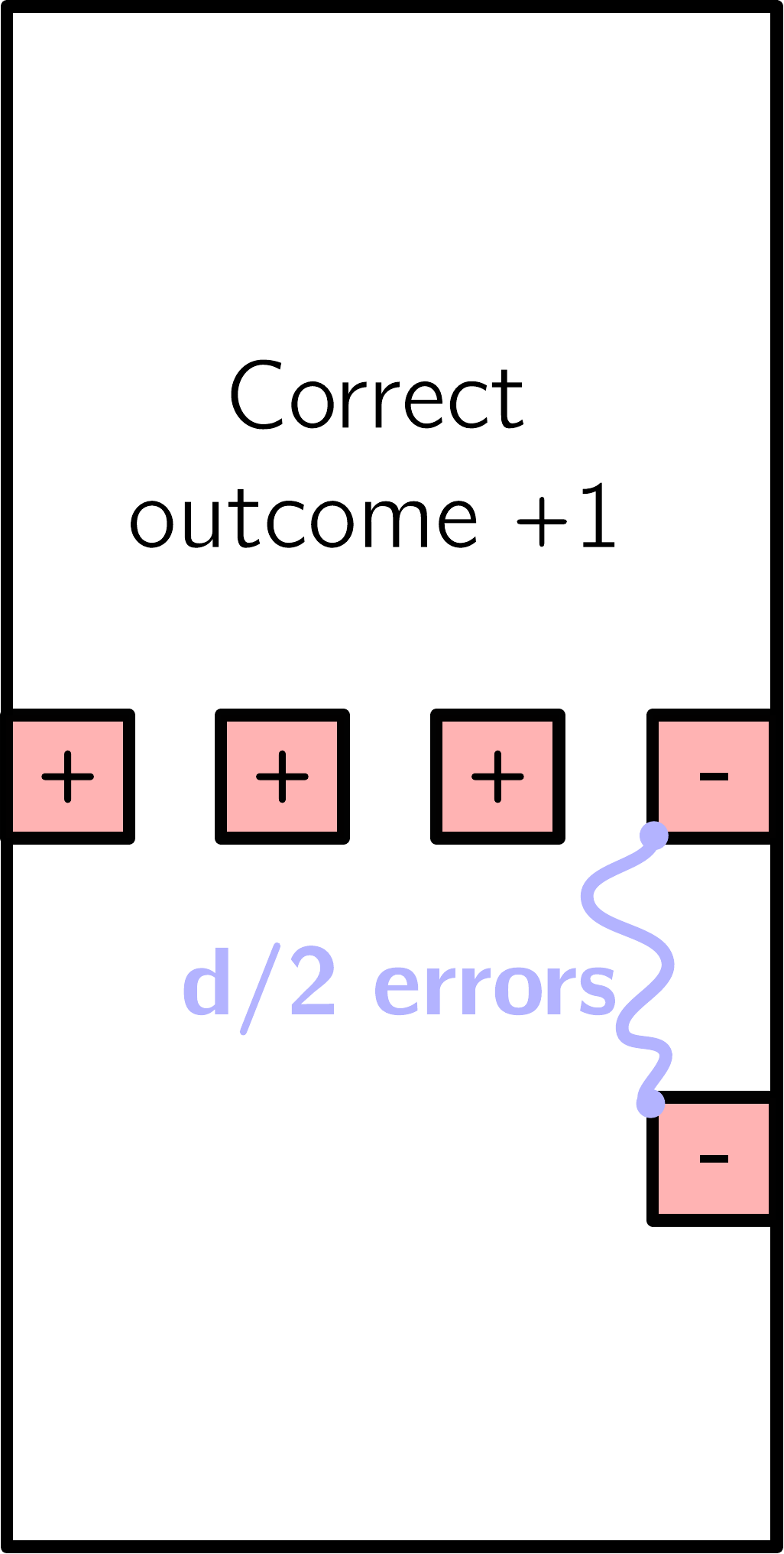}
	\label{fig:expl1}
	\end{subfigure}
	\begin{subfigure}[]{0.22\textwidth}
	\caption{}
	\includegraphics[width=.8\textwidth]{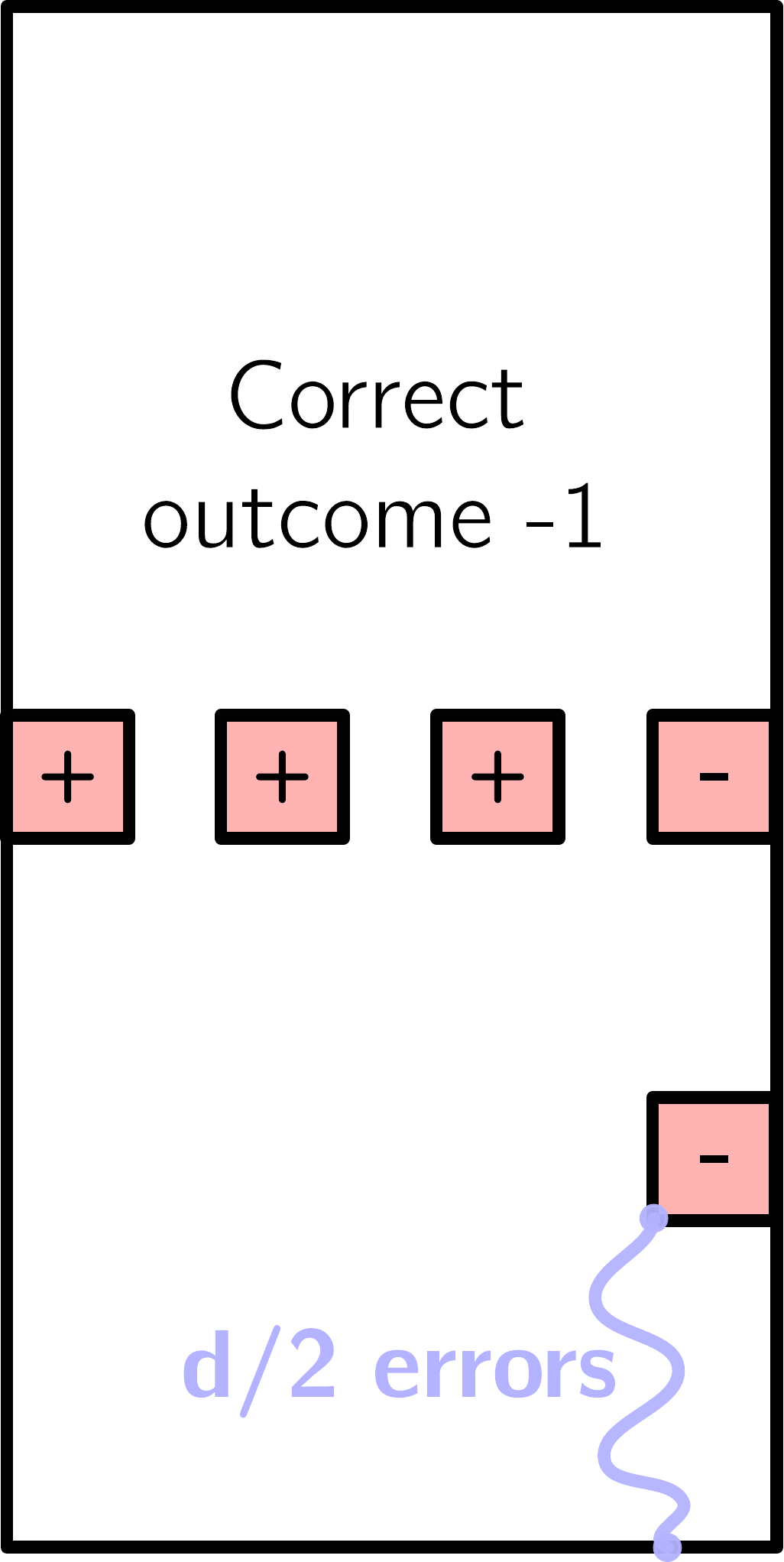}
	\label{fig:expl2}
	\end{subfigure}

	\caption{The operators of the subsystem code, $C$, for the joint measurement $\overline{Z}\overline{Z}$.
		(a) The generators of stabilizer group $\mathcal{S}$.
		(b) The highlighted operators are either gauge operators in $\mathcal{L}_g$ or logical operators in $\mathcal{L}_{\rm bare}$.
		We start in the gauge where the products $X_1'X_2'$ and $X_2'X_3'$ are fixed, and end in the gauge where $Z_1'$, $Z_2'$, and $Z_3'$ are fixed. 
		The distance of the subsystem code is 5, since one can construct a logical $\overline{X}$ with this weight by multiplying it with $X$ gauge operators.
		(\subref{fig:expl1}) and (\subref{fig:expl2}) Two different scenarios with errors of weight $\nicefrac{d}{2}$ with the same observed measurements.}
	\label{fig:cnotgauge}
\end{figure}

In this section, we interpret the joint measurement of $\overline{Z}\overline{Z}$ by lattice surgery in \autoref{fig:sscmerged} as gauge fixing.
The stabilizer group $\mathcal{S}$ is generated by all the stabilizers in \autoref{fig:cnotgauge_stab}. 
The gauge operators, $ \mathcal{L}_{g}$, of the gauge group are given by three representatives of the logical $X$ of the top patch and the intermediary $Z$ plaquettes that anti-commute with them.
They are denoted as $\left \langle {X}'_{1}, {Z}'_{1}, {X}'_{2}, {Z}'_{2}, {X}'_{3}, {Z}'_{3}  \right \rangle$ in \autoref{fig:cnotgauge_split}.
Representatives of the bare logical operators, $\overline{X},\overline{Z}\in\mathcal{L}_{\rm bare}$, are the logical $Z$ of the bottom patch and the logical $X$ of the merged patch (joining the very top to the very bottom), see \autoref{fig:cnotgauge_split}. 
The merge and split operations are realised by fixing some gauge operators of $ \mathcal{L}_{g}$, resulting in new codes $C_{\rm merged}$ or $C_{\rm split}$, respectively.
Note that the weight of $\overline{X}$ of the subsystem code, $C$, is only $d$ and not $2d$ which is the distance for $X$ of the merged code.
Indeed, by using the gauge operators like $X_1^\prime$ and stabilizers, one can construct a dressed logical $X$ of weight $d$.
Another way of seeing this is by realizing that one cannot distinguish between two errors of weight $\nicefrac{d}{2}$ depicted in \autoref{fig:expl1} and \autoref{fig:expl2}. In the first one, the logical measurement outcome is $-1$ and there is a string of $\nicefrac{d}{2}$ $X$-errors from the bottom to the middle of the bottom patch. In the second one the logical measurement outcome is $+1$ and there is a string of $\nicefrac{d}{2}$ $X$-errors from the middle of the bottom patch and the middle (changing the observed logical measurement outcome to $-1$).
Note also that when performing the splitting operation, one wants to correct the $-1$ outcomes for some of the intermediary $X$ stabilizers.
They are gauge operators equivalent to, say $X_1^\prime X_2^\prime$.
They have to be corrected using the $Z$ gauge operators, say $Z_1^\prime$ in this case.
Otherwise one would introduce a logical $Z$ error.

\subsubsection{Plain surgery}

\begin{figure}[htb!]
	\centering
	\begin{subfigure}[]{0.24\textwidth}
		\caption{}
		\includegraphics[width=\textwidth]{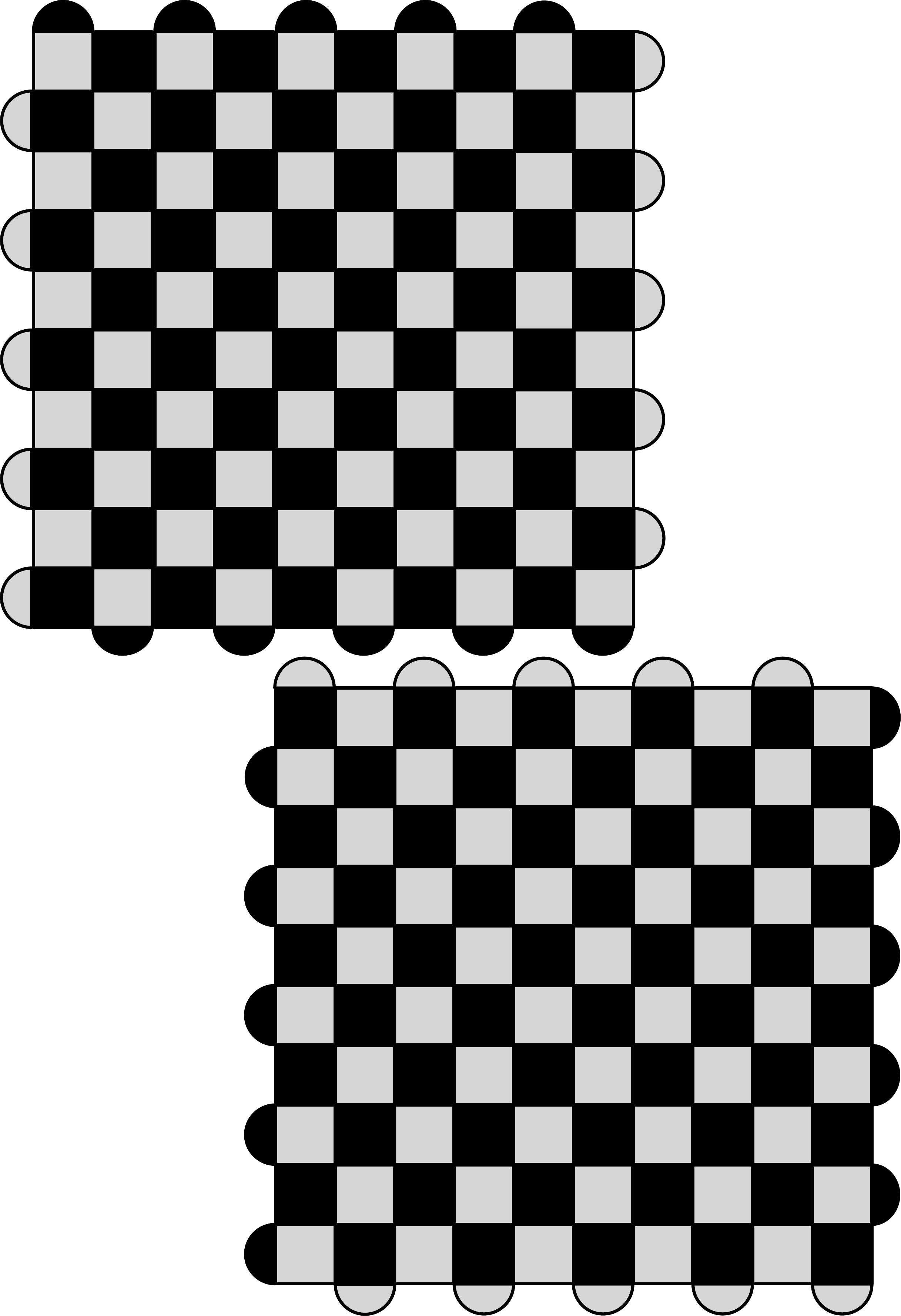}
		\label{fig:plainsurgery_eg1}
	\end{subfigure}
	\begin{subfigure}[]{0.24\textwidth}
		\caption{}
		\includegraphics[width=\textwidth]{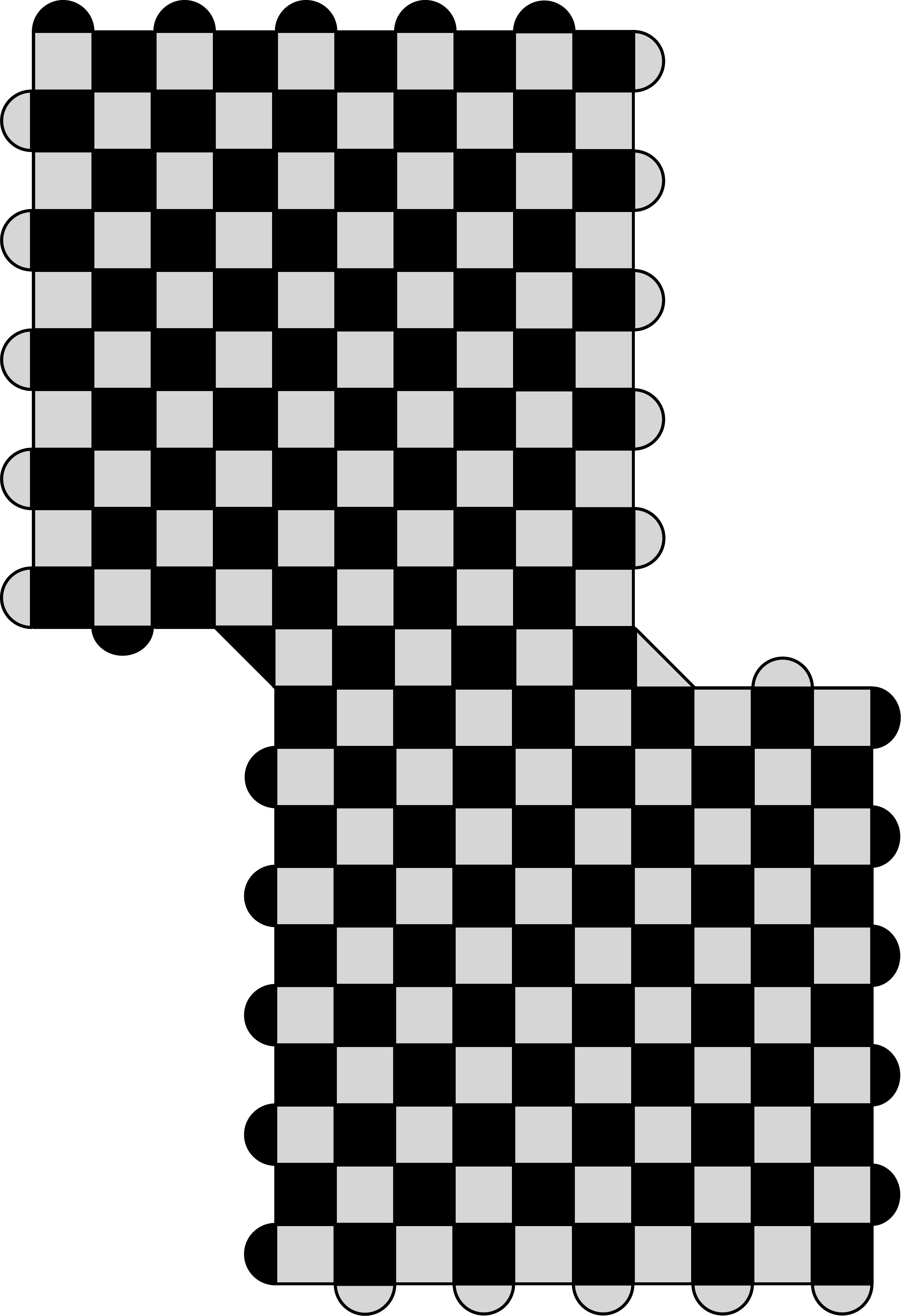}
		\label{fig:plainsurgery_eg2}
	\end{subfigure}
	\begin{subfigure}[]{0.24\textwidth}
		\caption{}
		\includegraphics[width=\textwidth]{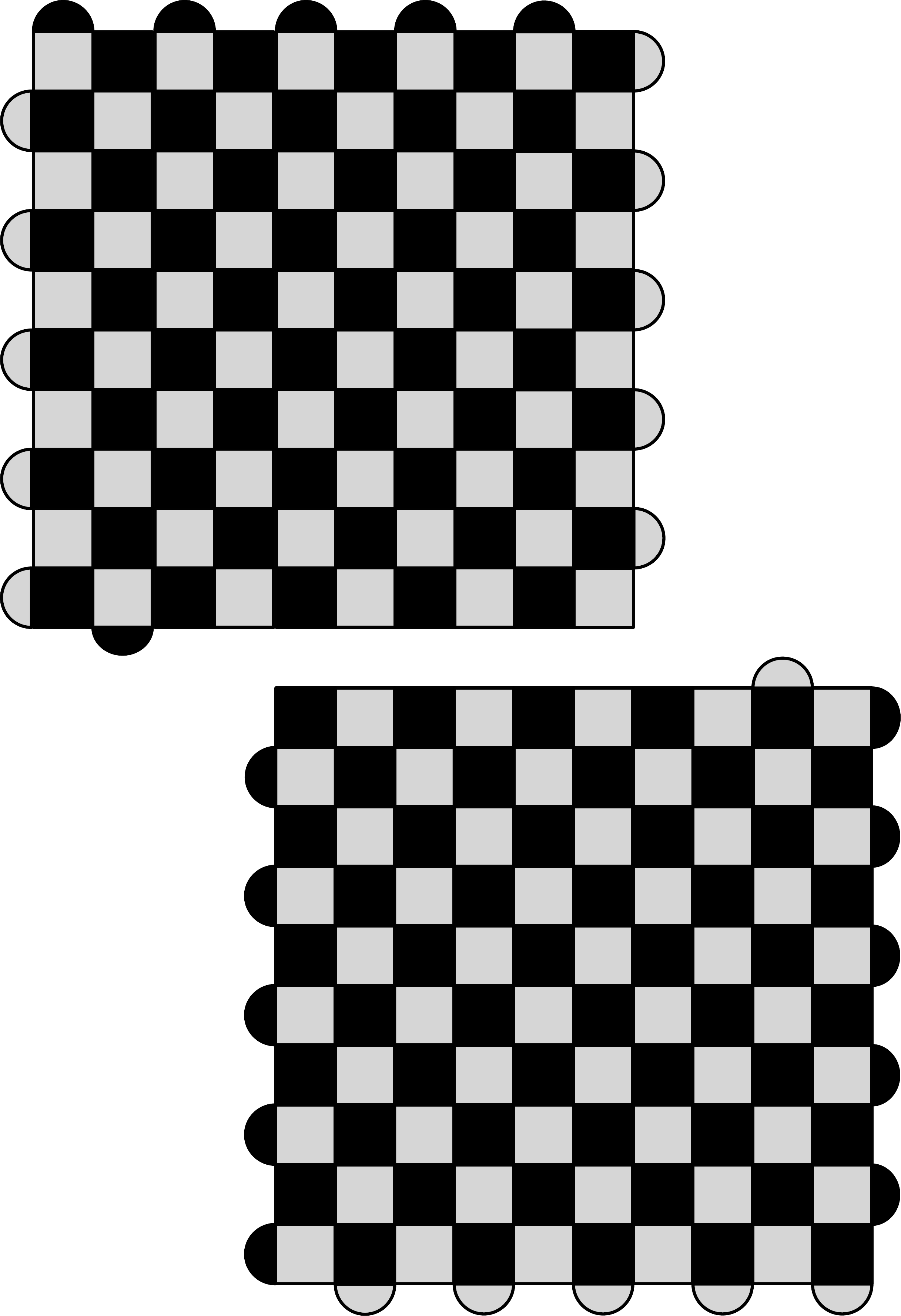}
		\label{fig:plainsurgery_eg3}
	\end{subfigure}
	\begin{subfigure}[]{0.24\textwidth}
		\caption{}
		\includegraphics[width=\textwidth]{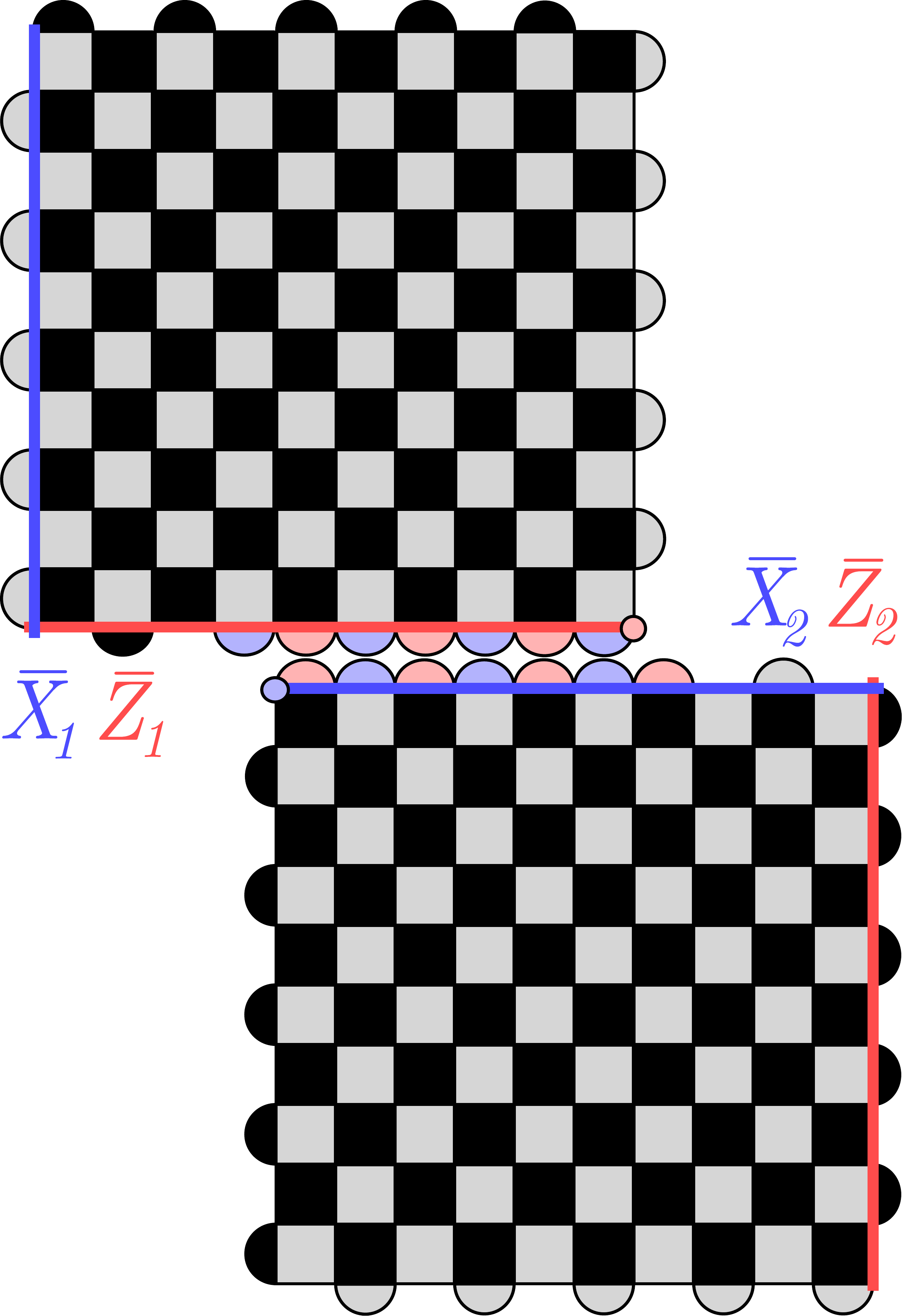}
		\label{fig:plainsurgery_eg4}
	\end{subfigure}
	\caption{(\subref{fig:plainsurgery_eg1}) and (\subref{fig:plainsurgery_eg2}) The qubit layouts before and after the plain merge operation.
		The number of logical qubits is kept constant during this merge operation.
		(\subref{fig:plainsurgery_eg3}) The stabilizers of the subsystem code.
		(\subref{fig:plainsurgery_eg4}) The gauge operators and logical operators of the subsystem code.
		One can see that the distance is guaranteed by the offset between the two blocks.
		The distance of the separate surface codes is 11, and the distance of the subsystem code is 4.}
	\label{fig:plainsurgery}
\end{figure}

We now introduce a new technique with the same goal as lattice surgery, namely performing joint measurements of logical operators, but following a different procedure.
The difference between lattice surgery and the new procedure, \emph{plain surgery}, will be that the logical measurement is performed with redundancy, so that this part of the protocol can be made more robust to noise, at the cost of qubit overhead.

The idea is to separate the merging and logical measurement of lattice surgery into two distinct steps.
The first step deforms the two separated blocks into a single code block where the joint logical operators can be measured redundantly.
Since this step merges the codes, but leaves the logical information unchanged, we call it a \emph{plain} merge.
In the second step, we measure the desired logical operator destructively, similar to the standard logical measurement of a surface code block.
A final deformation step can be used to return to the original code space. 

The layout for the plain merge operation is shown in \autoref{fig:plainsurgery_eg1}.
The patches are placed with an overlap of approximately $\nicefrac{2d}{3}$, the $X$-boundary of one facing the $Z$-boundary of the other.
Then they are merged into a single patch with 3 $X$-boundaries and 3 $Z$-boundaries, so two logical qubits.
Logical operators far away from the interface are left unchanged, and the logical information is untouched.
When looking at the subsystem code for this deformation, shown in \autoref{fig:plainsurgery_eg4}, one can see that the distance is guaranteed by the offset between the two patches.
 
\begin{figure}[htb!]
	\centering
	\begin{subfigure}[]{0.24\textwidth}
		\caption{}
		\includegraphics[width=\textwidth]{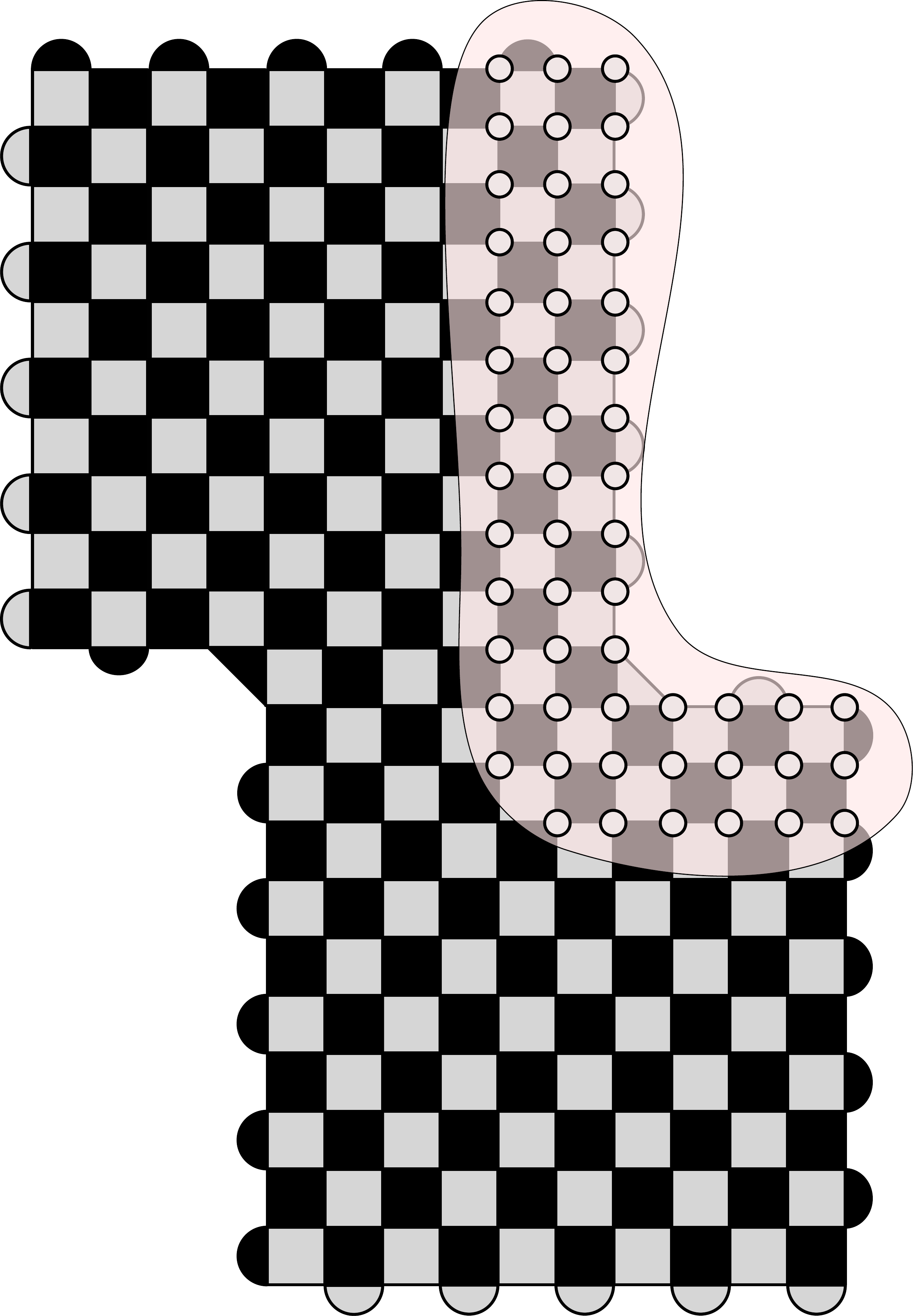}
		\label{fig:plainsurgerymeasure_eg1}
	\end{subfigure}\qquad
	\begin{subfigure}[]{0.24\textwidth}
		\caption{}
		\includegraphics[width=\textwidth]{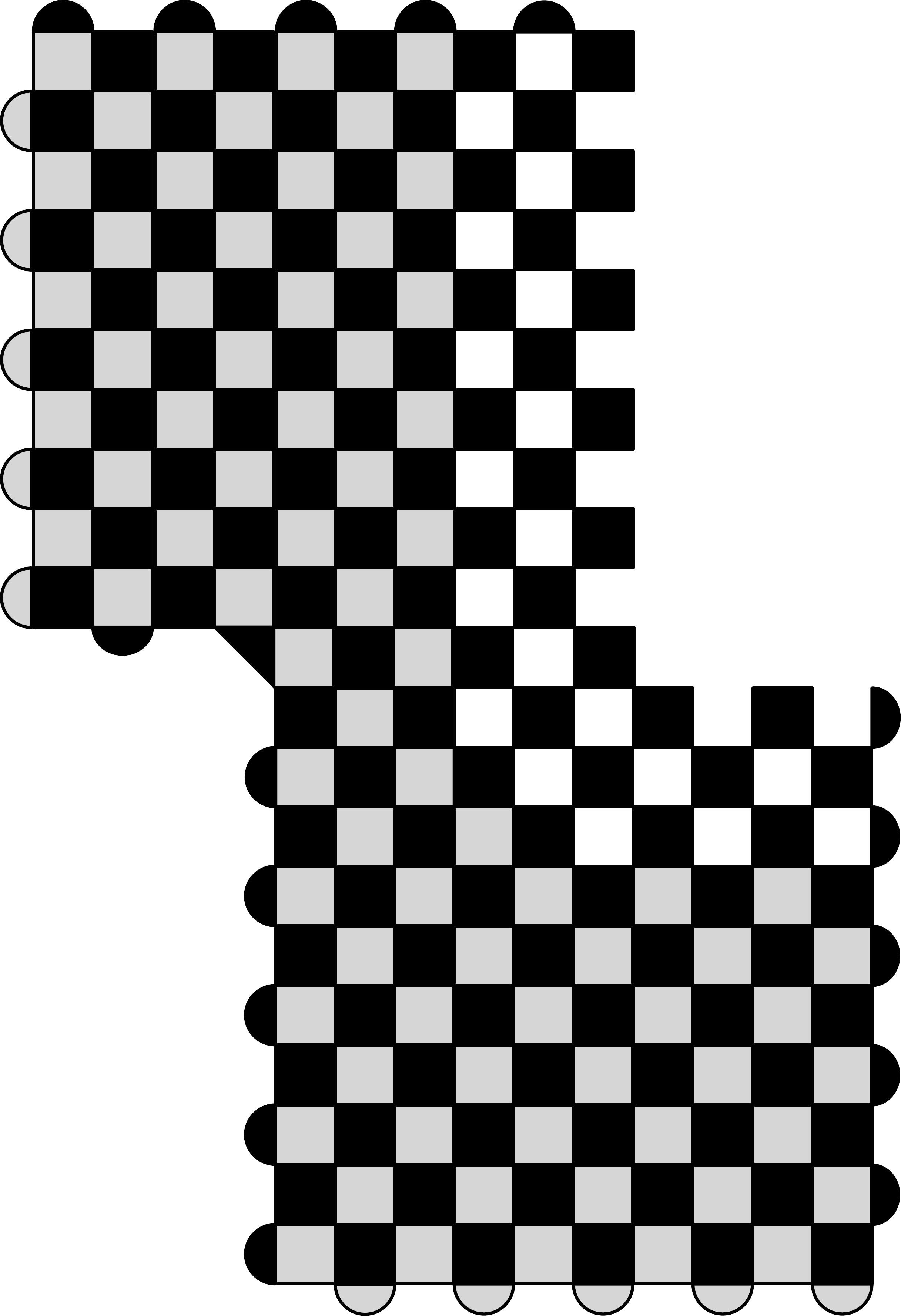}
		\label{fig:plainsurgerymeasure_eg2}
	\end{subfigure}\qquad
	\begin{subfigure}[]{0.24\textwidth}
		\caption{}
		\includegraphics[width=\textwidth]{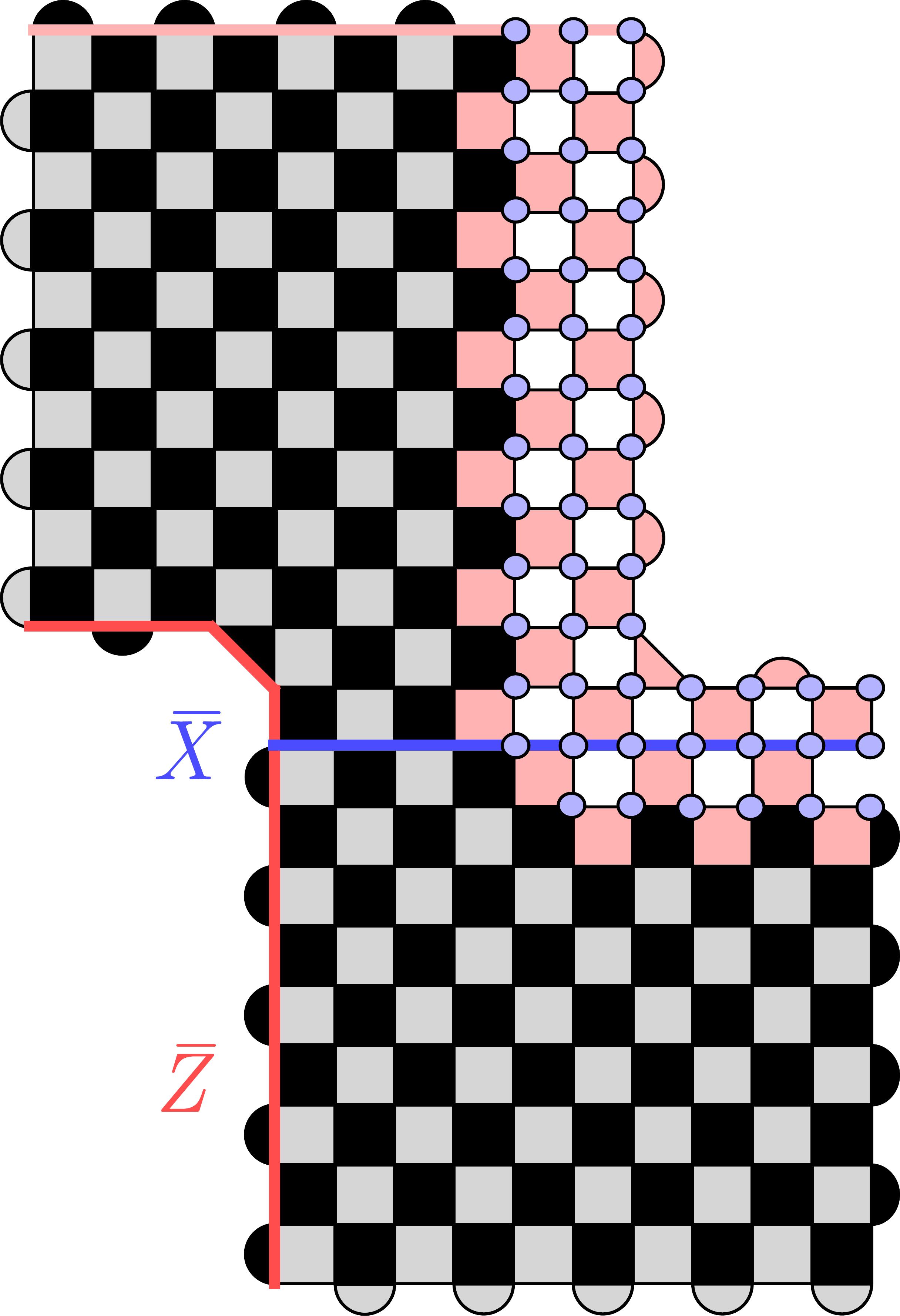}
		\label{fig:plainsurgerymeasure_eg3}
	\end{subfigure}
	\caption{(\subref{fig:plainsurgerymeasure_eg1}) The layout where the qubits in the region highlighted are each to be measured in the $X$ basis.
		(\subref{fig:plainsurgerymeasure_eg2}) The stabilizers of the underlying subsystem code $C$.
		(\subref{fig:plainsurgerymeasure_eg3}) The gauge operators (in pink) and logical operators of the code.
		One can see that the distance is guaranteed by the amount of overlap between the two blocks.
		The distance of the subsystem code is 4.}
	\label{fig:plainsurgerymeasure}
\end{figure}

Then, in this new code, the logical operator $\overline{X}_1\overline{X}_2$ is given by a string starting from the top boundary of the top patch and ending on the right boundary of the bottom patch.
So, by measuring qubits in the $X$ basis in a region away from the third $X$-boundary, one can learn $\overline{X}_1\overline{X}_2$ but not $\overline{X}_1$ or $\overline{X}_2$.
This measurement procedure is depicted in \autoref{fig:plainsurgerymeasure}.
One can check that the associated subsystem code has a distance of at least half the overlap between the patches, $\sim \nicefrac{d}{3}$.
The amount of redundancy in the measurement is also $\sim \nicefrac{d}{3}$, which makes this procedure costly in qubit overhead but as we show in the next section, it offers a better threshold than the standard lattice surgery technique.

\section{Numerics}
\label{sec:Numerics}

To numerically evaluate the fault-tolerance of quantum computation on rotated planar surface codes, we simulate logical measurement, rotation, logical \cnot, and plain surgery, using the Gottesman-Knill formalism \cite{gottesman1998heisenberg}.
These simulations are carried out using two different error models, the \emph{phenomenological} model and the \emph{circuit-based} model.
The phenomenological error model inserts independent $X$ and $Z$ errors on data qubits with equal probability $p$, and measurements output the wrong classical value with probability $p$.
The circuit error model inserts errors with probability $p$ after each operation of the error correction circuit as follows: each single-qubit gate is followed by a $X$, $Y$, or $Z$ with probability $\nicefrac{p}{3}$, each two-qubit gate is followed by an element of $\{I,X,Y,Z\}^{\bigotimes 2}\backslash \{II\}$ with probability $\nicefrac{p}{15}$, and each measurement returns the wrong result with probability $p$.
In this work, except when stated otherwise, the initial logical qubits are prepared without errors when simulating these logical operations.

\begin{figure}[htbp]
	\centering
	\includegraphics[width=.9\textwidth]{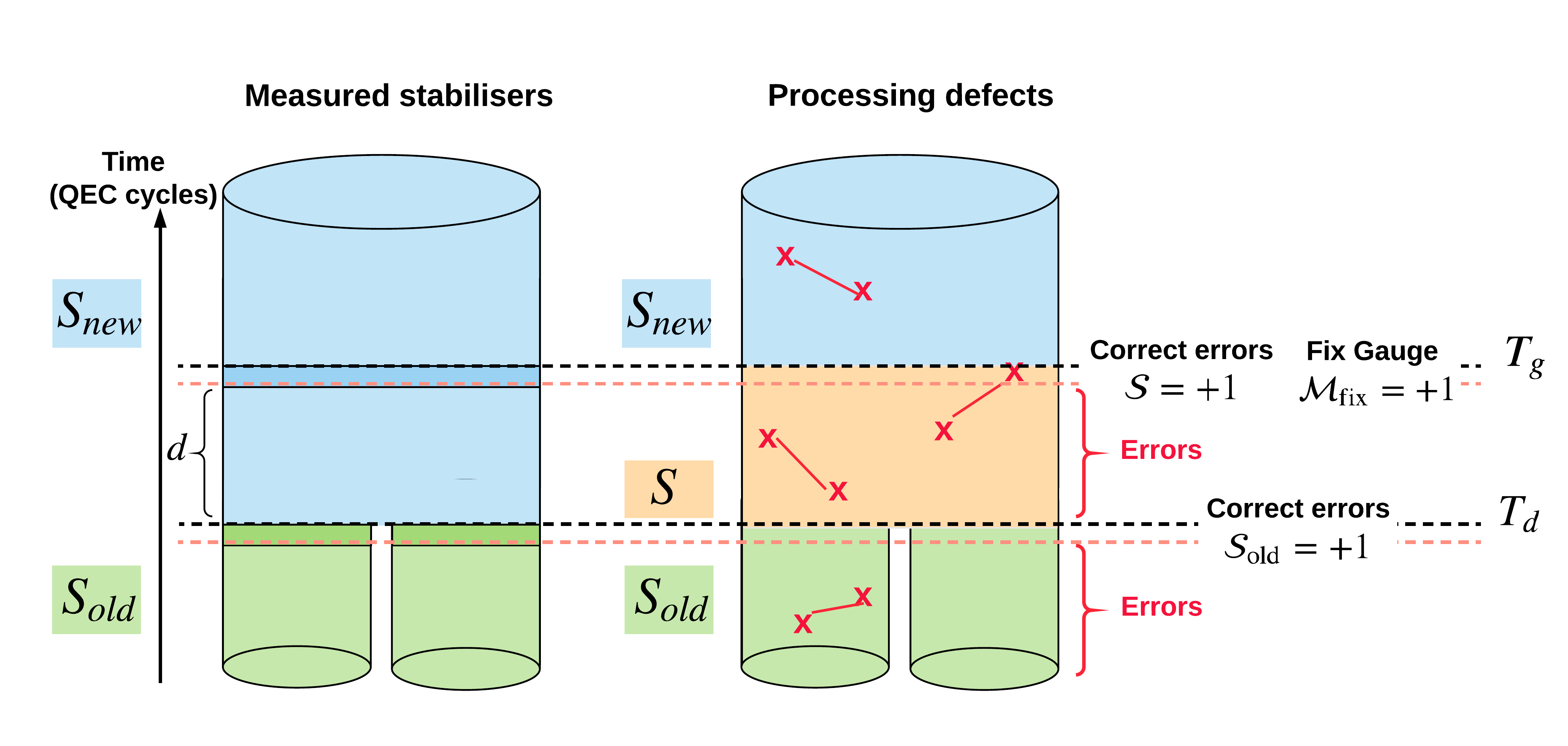}
	\caption{The simulated version of a code deformation procedure in \autoref{fig:decoding_prescription}.
	A perfect round (a small time window from red to black dashed lines) is inserted after each block of noisy $d$ rounds of stabilizer measurements.
	One processes the defects for $\mathcal{S}_{\rm old}$ and corrects errors before the code deformation step $T_d$.
	Then the defects for $\mathcal{S}$ are constructed at time $T_d$ to time $T_g$ and the `defects' for $\mathcal{M}_{\rm fix}$ are constructed one round of measurement later. 
	At time $T_g$, one processes error information to infer the value of the gauge operators and then fixes the gauge.}
	\label{fig:decoding_simulation}
\end{figure}

\begin{figure}[htb!]
\centering
\begin{subfigure}[h]{0.47\textwidth}
\caption{}
\includegraphics[width=\textwidth]{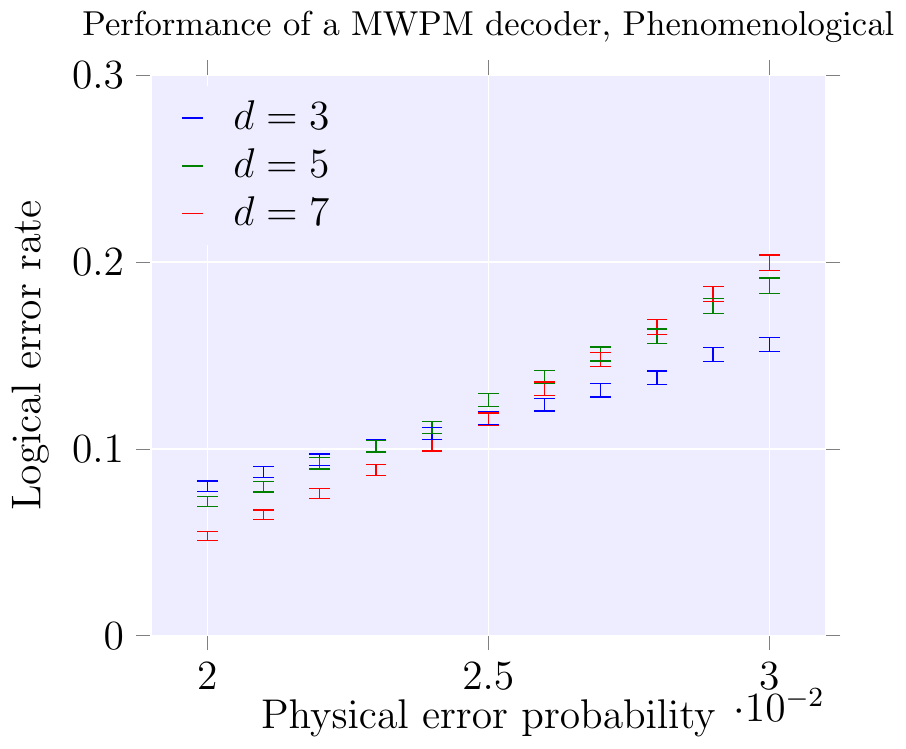}
\label{fig:ler_dec_pq}
\end{subfigure}
\begin{subfigure}[h]{0.45\textwidth}
\caption{}
\includegraphics[width=\textwidth]{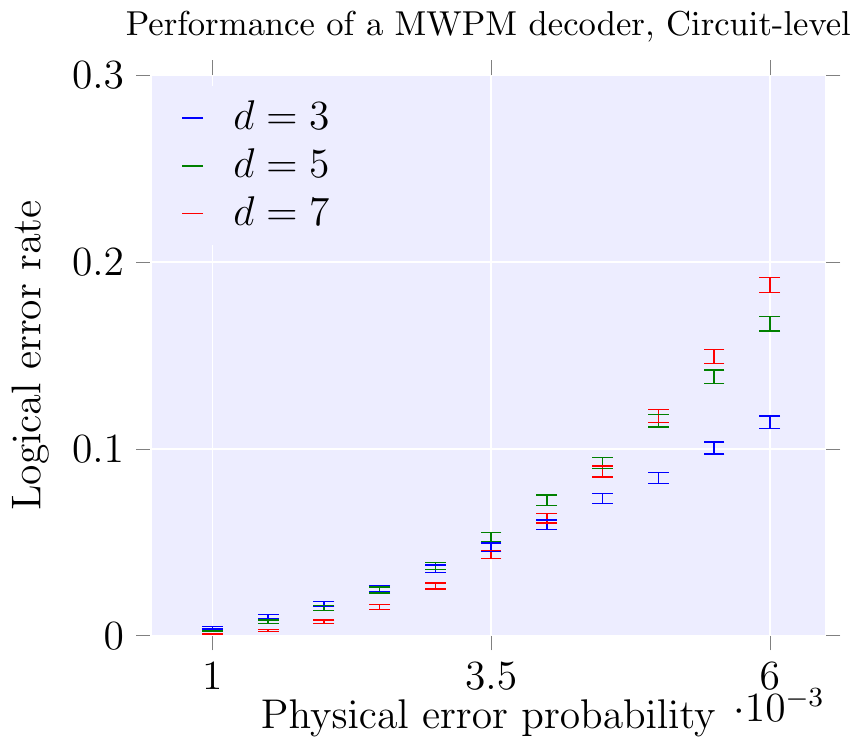}
\label{fig:ler_dec_cm}
\end{subfigure}
\caption{Numerical simulations of a fault-tolerant memory operation with the phenomenological error model near its threshold ($\sim 2.75\%$ (a)) and the circuit-level error model near its threshold ($\sim 0.5\%$ (b)).}
\label{fig:ler_dec}
\end{figure}

In section \ref{subsec:ft-cd}, we have introduced how to construct defects (difference syndromes) for a code deformation step and how to process these defects to infer errors and fix gauge operators (\autoref{fig:decoding_prescription}).
For a realistic implementation of logical operations, a decoder will infer errors in a time window which may include $T_{d}$ or $T_{g}$, by processing the defects within the window.
This means the decoder should be able to match defects across time boundaries, e.g., the defects before and after code deformation time $T_{d}$.
In addition, it needs to construct matching graphs with edges whose endpoints are on different lattices, e.g., defects of $\mathcal{S}_{\rm new}$ may be matched to virtual defects beyond the past-time boundary $T_{d}$.
However, such a decoder is difficult to implement.
In our simulations, we insert perfect measurement rounds after blocks of $d$ rounds of measurement (\autoref{fig:decoding_simulation}) for ease of implementation, where $d$ is the distance of the underlying subsystem code. 
A decoder using the minimum-weight perfect matching (MWPM) algorithm is used and its performance for a fault-tolerant memory operation, that is, $d$ noisy QEC cycles followed by $1$ noiseless cycle, is shown in \autoref{fig:ler_dec}.
For each operation (except for plain surgery), $10^{5}$ ($10^{4}$) iterations were run per point and confidence intervals at $99.9\%$ are plotted in the figures.

\begin{figure}[hbt!]
\centering
\includegraphics[width=0.45\textwidth]{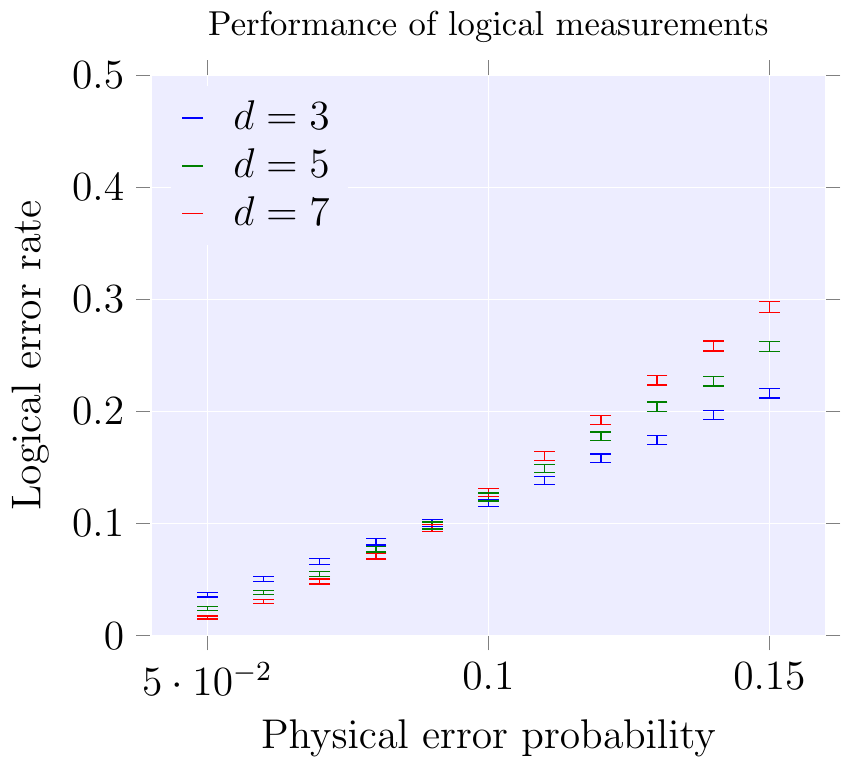}
\caption{Numerical simulations of a transversal $M_{\overline{Z}}$ measurement near its threshold ($\sim 10\%$).}
\label{fig:ler_msmt}
\end{figure}

\begin{figure}[htb!]
\centering
\begin{subfigure}[h]{0.45\textwidth}
\caption{}
\includegraphics[width=\textwidth]{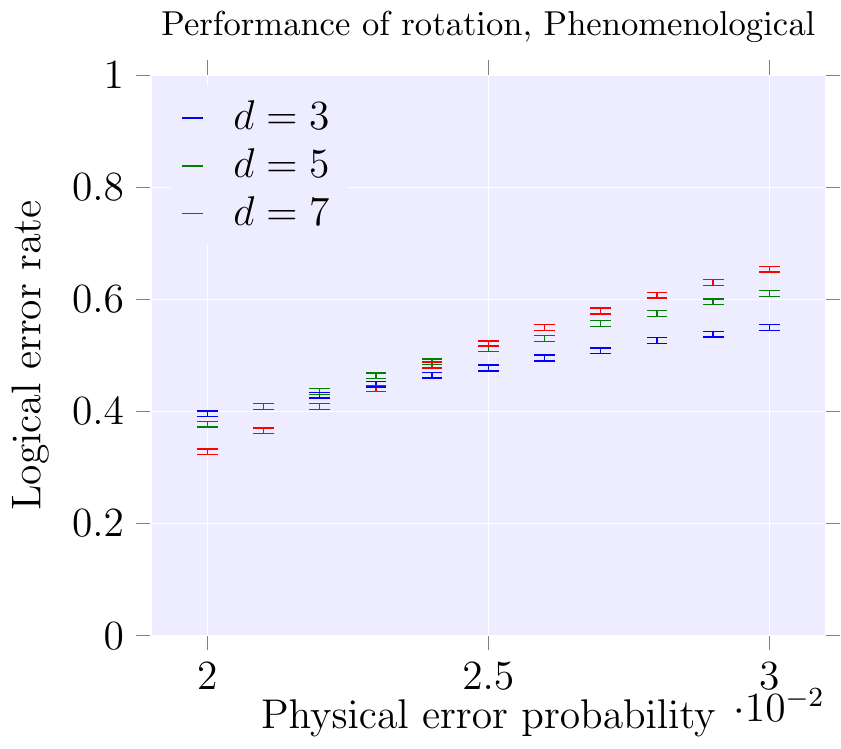}
\label{fig:ler_hpq}
\end{subfigure}
\begin{subfigure}[h]{0.45\textwidth}
\caption{}
\includegraphics[width=\textwidth]{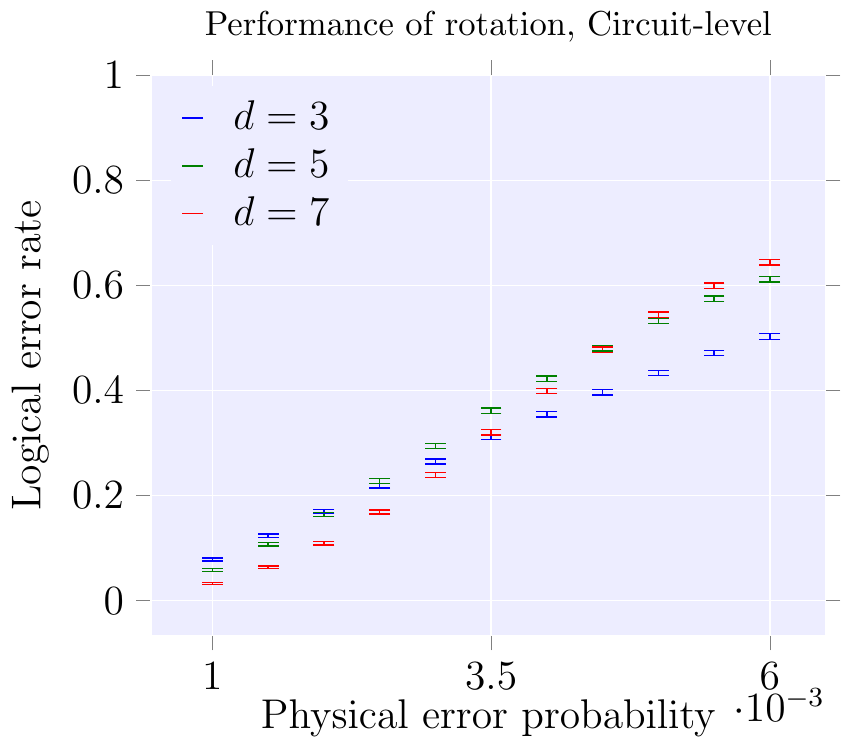}
\label{fig:ler_hcm}
\end{subfigure}
\caption{Numerical simulations of the rotation procedure in \autoref{fig:h_layout} without a final flip operation. 
(a) and (b) The logical error rates of the rotation procedure with phenomenological error model (The error threshold is around $\sim 2.5\%$) and circuit error model (The error threshold is around $\sim 0.45\%$), respectively.}
\label{fig:ler_h}
\end{figure}

\begin{figure}[hbt!]
\centering
\begin{subfigure}[htb!]{0.48\textwidth}
\caption{}
\includegraphics[width=\textwidth]{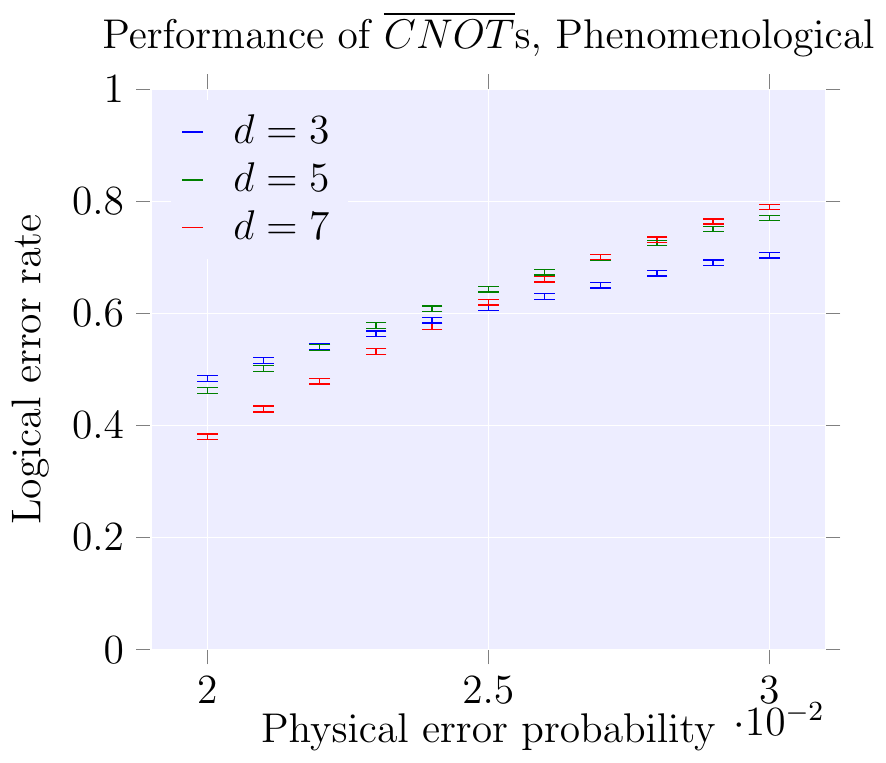}
\label{fig:ler_cnotpq}
\end{subfigure}
\begin{subfigure}[h]{0.45\textwidth}
\caption{}
\includegraphics[width=\textwidth]{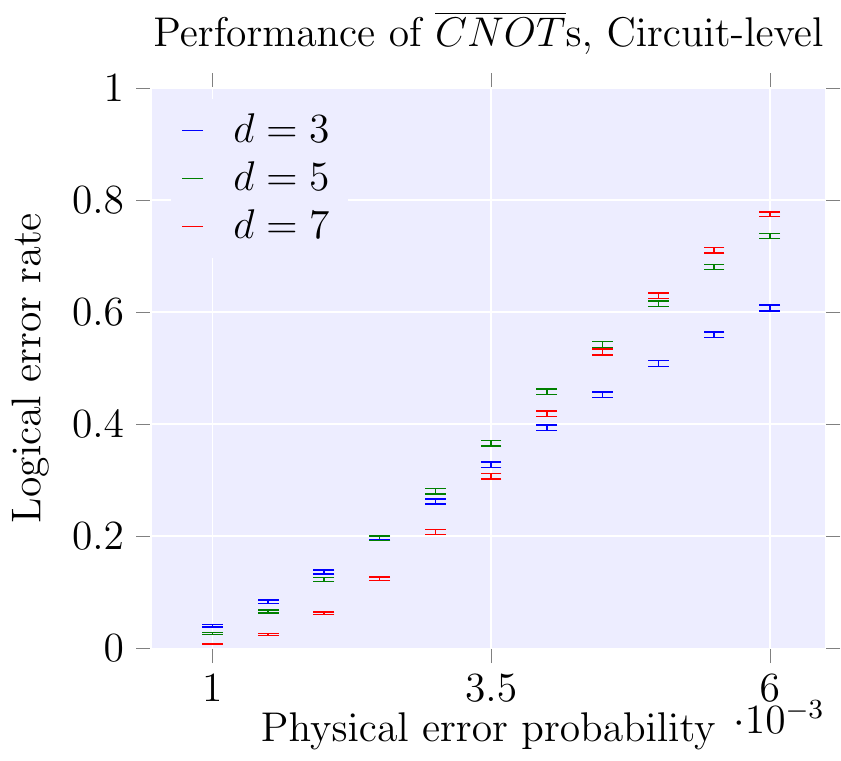}
\label{fig:ler_cnotcm}
\end{subfigure}
\caption{Numerical simulations of a measurement-based \cnot~gate by lattice surgery (The top circuit in \autoref{fig:cnotcircuit}). 
(a) Total error rates for \cnot~gates with the phenomenological error model near the threshold ($\sim 2.7\%$).
(b) Total error rates for \cnot~gates with the circuit-level error model near the threshold ($\sim 0.45\%$ ).}
\label{fig:ler_cnots}
\end{figure}

\begin{figure}[htb!]
\centering
\begin{subfigure}[h]{0.45\textwidth}
\caption{}
\includegraphics[width=\textwidth]{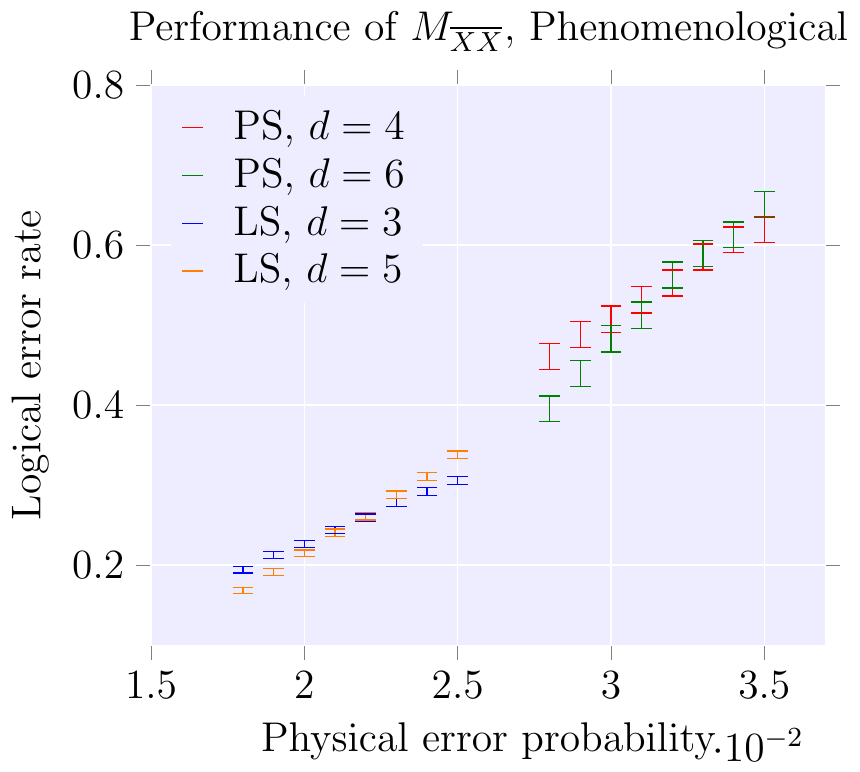}
\label{fig:ler_mxxpq}
\end{subfigure}
\begin{subfigure}[h]{0.455\textwidth}
\caption{}
\includegraphics[width=\textwidth]{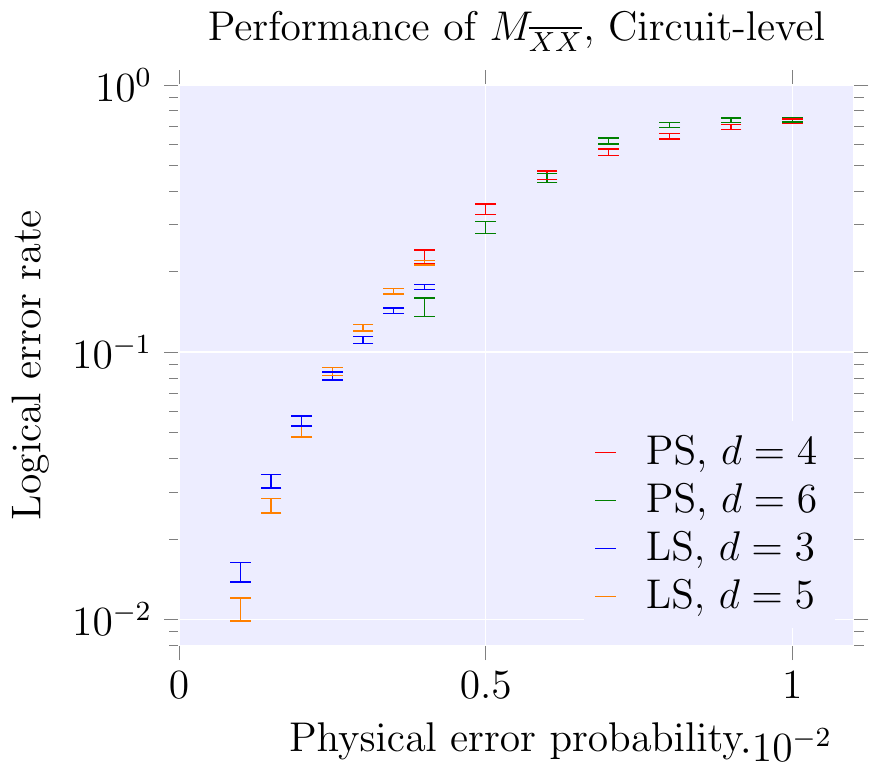}
\label{fig:ler_mxxcm}
\end{subfigure}
\caption{Numerical comparison of the $M_{\overline{X}\overline{X}}$ joint measurements by lattice surgery (LS) and plain surgery (PS), near the points where the two lowest-distance implementations of the two protocols produce the same logical error rate.
The logical error rates of $M_{\overline{X}\overline{X}}$ with the (a) phenomenological error model ((b) circuit-level error model) by LS with a crossing between the $d=3$ and $d=5$ near the physical error probability $\sim 2.2\%$ ($\sim 0.25\%$)) and by PS with a crossing between the $d=4$ and $d=6$ near the physical error probability $\sim 3.2\%$ ($\sim 0.65\%$).}
\label{fig:ler_mxx}
\end{figure}

\textbf{Single-qubit operations:}
Transversal operations (preparation, Pauli gates, measurement) are usually realised by performing qubit-wise physical operations.
They are intrinsically fault-tolerant and their logical error rates will be only slightly higher than a logical identity gate (memory).
Notably, a transversal $M_{\overline{Z}}$ ($M_{\overline{X}}$) measurement does not require quantum error correction cycles (i.e., $T_d = T_g$) since error syndromes of $Z$($X$)-stabilizers can be reconstructed from the measurement outcomes of data qubits, this is also the case for the logical measurement step of plain surgery. 
For instance, one can measure all the data qubits in the $Z$ basis to realise a $M_{\overline{Z}}$ on a planar surface code. 
Afterwards, one can compute the $Z$-syndromes by multiplying the outcomes of corresponding data qubits of each $Z$-stabilizer and then correct the $X$ errors and deduce the value of $\overline{Z}$.  
The performance of a $M_{\overline{Z}}$ measurement for planar surface codes is shown in \autoref{fig:ler_msmt}.
In this simulation, we first prepare a logical qubit in state $\ket{\overline{0}}$ without errors and then perform a $M_{\overline{Z}}$ measurement on it with physical measurement error probability $p$. 
We further numerically simulate the proposed rotating procedure (\autoref{fig:h_layout}) and show the results in \autoref{fig:ler_h}.
For the phenomenological error model, the error threshold of a rotation is slightly lower than the threshold of quantum memory.
For the circuit-level error model, its threshold is similar to that of quantum memory.

\textbf{Two-qubit operations:}
We also simulate the measurement-based \cnot~circuits in \autoref{fig:cnotcircuit} where the split operations of the first joint measurements are parallelised with the merge operations of the second joint measurements (see the decomposed circuits in Appendix \ref{sec:disparity}).
The overall error rates and the error thresholds for a \cnot~gate by lattice surgery are shown in \autoref{fig:ler_cnots}. 
For each error model, the error threshold of \cnot~gates is similar to the threshold of quantum memory.  
Moreover, logical errors propagate through the measurement-based \cnot~circuits, leading to a disparity of logical error rates on control and target qubits, which is demonstrated numerically in Appendix \ref{sec:disparity}. 
In addition, we compare the joint $M_{\overline{X}\overline{X}}$ measurement using lattice surgery with the measurement using plain surgery. 
\autoref{fig:ler_mxx} shows that plain surgery achieves a higher error threshold than lattice surgery, but with higher logical error rates as a consequence of the increased lattice size required to achieve a given code distance.

\section{Discussion \& Conclusion}
\label{sec:Conclusion}

We have illustrated how to describe current measurement-based operations in 2D topological quantum computing using the gauge fixing technique. 
We have shown that, by using the formalism of gauge fixing, the fault tolerance analysis of these code deformation and lattice surgery protocols is considerably simplified, their error correction and gauge fixing schemes also become clear. 
Furthermore, we numerically examined this method with examples on planar surface codes, including some well-known operations such as lattice-surgery-based \cnot~gates and some novel protocols such as lattice rotation and plain surgery.
Although this gauge fixing formalism does not provide direct guidlines on how to design code deformation protocols for a desired logical operation, it does provide an easy way to check the fault-tolerance of protocols and search for new ones via iterations of trial and error.

Moreover, this formalism applies not only to 2D topological codes, but more generally to any stabilizer code.
In the general case (non-topological codes), the analysis of fault-tolerance in the presence of measurement errors becomes more involved, in particular with respect to how much repetition is really needed, see for example \cite{Earlsingleshot,Grospellier}.
We leave for future work how to obtain general and simple criteria for fault-tolerance.

\begin{acknowledgments}
The authors would like to thank Benjamin Brown for enlightening discussions.
LLL acknowledges funding from the China Scholarship Council.
BMT and CV acknowledge support by the European Research Council (EQEC, ERC Consolidator Grant No: 682726).
BMT, KB and CGA acknowledge support from the QuantERA ERA-NET Co-fund in Quantum Technologies implemented within the European Union’s Horizon 2020 Programme (for the QCDA consortium).
KB and CGA acknowledge support from the Intel Corporation.
\end{acknowledgments}

\bibliographystyle{unsrt}
\bibliography{code_deformation_is_gauge_fixing}

\begin{thebibliography}{10}

\bibitem{shor1994algorithms}
Peter~W Shor.
\newblock Algorithms for quantum computation: Discrete logarithms and
  factoring.
\newblock In {\em Foundations of Computer Science, 1994 Proceedings., 35th
  Annual Symposium on}, pages 124--134. IEEE, 1994.

\bibitem{jordan2011quantum}
Stephen Jordan.
\newblock Quantum algorithm zoo.
\newblock \url{http://math.nist.gov/quantum/zoo/}, 2011.

\bibitem{riste2015detecting}
Diego Rist{\`e}, Stefano Poletto, M-Z Huang, Alessandro Bruno, Visa Vesterinen,
  O-P Saira, and Leonardo DiCarlo.
\newblock Detecting bit-flip errors in a logical qubit using stabilizer
  measurements.
\newblock {\em Nature communications}, 6:6983, 2015.

\bibitem{kelly2015state}
Julian Kelly, R~Barends, AG~Fowler, A~Megrant, E~Jeffrey, TC~White, D~Sank,
  JY~Mutus, B~Campbell, Yu~Chen, et~al.
\newblock State preservation by repetitive error detection in a superconducting
  quantum circuit.
\newblock {\em Nature}, 519(7541):66--69, 2015.

\bibitem{steane1996error}
Andrew~M Steane.
\newblock Error correcting codes in quantum theory.
\newblock {\em Phys. Rev. Lett.}, 77(5):793, 1996.

\bibitem{knill1996concatenated}
Emanuel Knill and Raymond Laflamme.
\newblock Concatenated quantum codes.
\newblock {\em arXiv:9608012}, 1996.

\bibitem{bacon2006operator}
Dave Bacon.
\newblock Operator quantum error-correcting subsystems for self-correcting
  quantum memories.
\newblock {\em Phys. Rev. A}, 73(1):012340, 2006.

\bibitem{fowler2012surface}
Austin~G Fowler, Matteo Mariantoni, John~M Martinis, and Andrew~N Cleland.
\newblock Surface codes: Towards practical large-scale quantum computation.
\newblock {\em Phys. Rev. A}, 86(3):032324, 2012.

\bibitem{kitaev2003fault}
A~Yu Kitaev.
\newblock Fault-tolerant quantum computation by anyons.
\newblock {\em Annals of Physics}, 303(1):2--30, 2003.

\bibitem{wang2011surface}
David~S Wang, Austin~G Fowler, and Lloyd~CL Hollenberg.
\newblock Surface code quantum computing with error rates over {1\%}.
\newblock {\em Phys. Rev. A}, 83:020302, 2011.

\bibitem{bombin2009quantum}
H{\'e}ctor Bomb{\'\i}n and Miguel~A Martin-Delgado.
\newblock Quantum measurements and gates by code deformation.
\newblock {\em Journal of Physics A: Mathematical and Theoretical},
  42(9):095302, 2009.

\bibitem{horsman2012surface}
Clare Horsman, Austin~G Fowler, Simon Devitt, and Rodney Van~Meter.
\newblock Surface code quantum computing by lattice surgery.
\newblock {\em New Journal of Physics}, 14(12):123011, 2012.

\bibitem{paetznick2013universal}
Adam Paetznick and Ben~W Reichardt.
\newblock Universal fault-tolerant quantum computation with only transversal
  gates and error correction.
\newblock {\em Phys. Rev. Lett.}, 111(9):090505, 2013.

\bibitem{Bombin2011Cliff}
H~Bombin.
\newblock Clifford gates by code deformation.
\newblock {\em New Journal of Physics}, 13(4):043005, 2011.

\bibitem{landahl2014quantum}
Andrew~J Landahl and Ciaran Ryan-Anderson.
\newblock Quantum computing by color-code lattice surgery.
\newblock {\em arXiv:1407.5103}, 2014.

\bibitem{Bravyi16deform}
Sergey Bravyi.
\newblock Fault-tolerant quantum computing by code deformation.
\newblock {\em QIP Tutorial}, 2016.

\bibitem{Nautrup2016FTi}
Hendrik Poulsen~Nautrup, Nicolai Friis, and Hans~J. Briegel.
\newblock Fault-tolerant interface between quantum memories and quantum
  processors.
\newblock {\em Nature Communications}, 8(1):1321, 2017.

\bibitem{Brown2017poking}
Benjamin~J. Brown, Katharina Laubscher, Markus~S. Kesselring, and James~R.
  Wootton.
\newblock Poking holes and cutting corners to achieve clifford gates with the
  surface code.
\newblock {\em Phys. Rev. X}, 7:021029, May 2017.

\bibitem{Litinski2018latticesurgery}
Daniel Litinski and Felix~von Oppen.
\newblock Lattice {S}urgery with a {T}wist: {S}implifying {C}lifford {G}ates of
  {S}urface {C}odes.
\newblock {\em {Quantum}}, 2:62, May 2018.

\bibitem{Fowler2018lowoverhead}
A.~G. {Fowler} and C.~{Gidney}.
\newblock {Low overhead quantum computation using lattice surgery}.
\newblock {\em arXiv:1808.06709}, August 2018.

\bibitem{Vasmer20183DSurf}
M.~{Vasmer} and D.~E. {Browne}.
\newblock {Universal Quantum Computing with 3D Surface Codes}.
\newblock {\em arXiv:1801.04255}, January 2018.

\bibitem{dennis2002topological}
Eric Dennis, Alexei Kitaev, Andrew Landahl, and John Preskill.
\newblock Topological quantum memory.
\newblock {\em Journal of Mathematical Physics}, 43(9):4452--4505, 2002.

\bibitem{Rauss2007holebraiding}
Robert Raussendorf and Jim Harrington.
\newblock Fault-tolerant quantum computation with high threshold in two
  dimensions.
\newblock {\em Phys. Rev. Lett.}, 98:190504, May 2007.

\bibitem{Bombin2010twists}
H{\'e}ctor Bomb{\'\i}n.
\newblock Topological order with a twist: Ising anyons from an abelian model.
\newblock {\em Phys. Rev. Lett.}, 105:030403, Jul 2010.

\bibitem{gottesman1998heisenberg}
Daniel Gottesman.
\newblock The heisenberg representation of quantum computers.
\newblock {\em arXiv:9807006}, 1998.

\bibitem{gottesman1998fault}
Daniel Gottesman.
\newblock Fault-tolerant quantum computation with higher-dimensional systems.
\newblock {\em arXiv:9802007}, 1998.

\bibitem{beaudrap2017zx}
Niel de~Beaudrap and Dominic Horsman.
\newblock The zx calculus is a language for surface code lattice surgery.
\newblock {\em arXiv:1704.08670}, 2017.

\bibitem{poulin2005stabilizer}
David Poulin.
\newblock Stabilizer formalism for operator quantum error correction.
\newblock {\em Phys. Rev. Lett.}, 95(23):230504, 2005.

\bibitem{Earlsingleshot}
E.~T. {Campbell}.
\newblock {A theory of single-shot error correction for adversarial noise}.
\newblock {\em ArXiv e-prints}, May 2018.

\bibitem{Grospellier}
O.~{Fawzi}, A.~{Grospellier}, and A.~{Leverrier}.
\newblock {Constant overhead quantum fault-tolerance with quantum expander
  codes}.
\newblock {\em ArXiv e-prints}, August 2018.

\bibitem{anderson2014fault}
Jonas~T Anderson, Guillaume Duclos-Cianci, and David Poulin.
\newblock Fault-tolerant conversion between the steane and reed-muller quantum
  codes.
\newblock {\em Phys. Rev. Lett.}, 113(8):080501, 2014.

\bibitem{Colladay2017rewiring}
Kristina~Renee Colladay and Erich Mueller.
\newblock Rewiring stabilizer codes.
\newblock {\em New Journal of Physics}, 2018.

\end{thebibliography}
\newpage
\appendix
\section{Algebraic proof of the correctness of the merge and split operations}
\label{sec:proofs}
In this appendix, we denote the set of physical qubits as $\mathsf{Q}$. For any subset of $k$ qubits, $s=\{j_1,\dots,j_k\}\subset\mathsf{Q}$, we denote the operator composed of a Pauli $Z$ resp. $X$ on each qubit in $s$ as $Z(s)$, resp. $X(s)$, i.e. 
\[Z(s) = Z_{j_1}\otimes\cdots\otimes Z_{j_k},\qquad X(s) = X_{j_1}\otimes\cdots\otimes X_{j_k}.\]
\subsection{Merge operation}

The setting for the merge operation is drawn in \autoref{fig:ssclayout}.
The starting code, $C_{\rm split}$, with stabilizer $\mathcal{S}_{\rm split}$, consists of two adjacent $L\times L$ patches of rotated surface code with the opposite boundaries being supports for their $\overline{Z}$ operators.
We label the upper logical qubit as $1$ and the lower qubit as $2$.
The new code, $C_{\rm merged}$, with stabilizer $\mathcal{S}_{\rm merged}$, consists of only one $2L\times L$ patch of rotated surface code.

We define the subsystem code, $C$, and its gauge group, $\mathcal{G}$, as specified in \autoref{sec:Unification}, see \autoref{fig:cnotgauge}.
Notably, we exclude from the center of $\tilde{\mathcal{G}}$ the logical operator $\overline{Z}_1\overline{Z}_2\in\mathcal{S}_{\rm merged}$.
We therefore add $\overline{X}_1$ to $\tilde{\mathcal{G}}$ to form $\mathcal{G}$, and so have $\overline{X}_1\in\mathcal{L}_g$.
Call $\mathcal{I}$ the set of intermediary plaquettes (red plaquettes in \autoref{fig:ssclayout}) to be measured to perform the merge operation.
For $p\in\mathcal{I}$ we have $Z(p)\in\mathcal{L}_g$, these are the gauge operators to be fixed by the merge operation.
For each $p\in\mathcal{I}$, one measures the operator $Z(p)$ and let its outcome be $m_p$.

To explain the action of the merge operation at the logical level, we first prove that this operation transforms code states of the two original $L\times L$ patches of surface code into code states of the $2L\times L$ patch surface code with some $X$ errors.
To accomplish this, we use the standard prescription from the Gottesman-Knill theorem \cite{gottesman1998heisenberg}.
It is straightforward to see that the original $Z$ checks stay unchanged, and the newly-measured checks, the $p\in \mathcal{I}$, are added, with sign $m_p$.
The original $X$ checks all commute with the new intermediary $Z$ checks except for the two-body boundary checks between the two patches, which are also part of $\mathcal{L}_g$.
Those boundary checks can be merged in pairs in order to commute with the new $Z$ checks.
The situation is then the same as depicted in \autoref{fig:sscmerged}.

The product of all measurement outcomes gives the desired outcome for the $\overline{Z}_1\overline{Z}_2$ measurement, we denote it as
\[m_{L} = \prod_{p\in\mathcal{I}}m_p.\]
Then one fixes the gauge by applying the conjugate $X$-gauge operators to the $Z(p)$ with $m_p=-1$.
Let's call $c_{m_{L}}$ the set of qubits involved in this fixing operation.
Note that when $m_{L}=+1$ then the correction is equivalent to a stabilizer in $\mathcal{S}_{\rm split}$ whereas when $m_{L}=-1$, the correction is equivalent to $\overline{X}_1$.
Then, the full merge operation at the physical qubit level is easily written as
\[X(c_{m_{L}})\cdot\left(\prod_{p\in\mathcal{I}}\frac{\id+(-1)^{m_p}Z(p)}{2}\right) = \left(\prod_{p\in\mathcal{I}}\frac{\id+Z(p)}{2}\right)\cdot X(c_{m_{L}}).\]
Due to the definition of $X(c_{m_{L}})$, commuting it through the $Z$ projections eliminates the $(-1)^{m_p}$ terms.

To determine the logical operation realised by this procedure, we use encoding isometries of $C_{\rm split}$ and $C_{\rm merged}$, called $E_{\rm split}$ and $E_{\rm merged}$, respectively.
These isometries map unencoded logical states to code states in the full physical Hilbert space.
Since $C_{\rm split}$ contains two logical qubits and $C_{\rm merged}$ contains only one, the isometries have the following signatures:
\[E_{\rm split}:\mathbb{C}^2\otimes\mathbb{C}^2\rightarrow\mathbb{C}^{2\mathsf{Q}},\qquad E_{\rm merged}:\mathbb{C}^2\rightarrow\mathbb{C}^{2\mathsf{Q}}.\]
Let $\tilde{M}_{m_{L}}$ be the operation on the logical level, which can be expressed as
\begin{align}
\tilde{M}_{m_{L}} &:\mathbb{C}^2\otimes\mathbb{C}^2\rightarrow\mathbb{C}^2,\nonumber\\ 
\tilde{M}_{m_{L}} &= \left(E_{\rm merged}\right)^\dagger\cdot \left(\prod_{p\in\mathcal{I}}\frac{\id+Z(p)}{2}\right)\cdot X(c_{m_{L}})\cdot E_{\rm split}.\label{eq:tildeM}
\end{align}

An important fact about encoding isometries $E$ is that, if $S$ is a stabilizer of the code and $\overline{L}$ a representative for the logical operator $L$, then
\begin{align}
S\cdot E &= E,\\
\overline{L}\cdot E &= E\cdot L,
\end{align}
where $L$ is the corresponding physical operator.
This means that $\tilde{M}_{m_{L}}$, defined in \autoref{eq:tildeM}, simplifies to
\begin{equation}
\tilde{M}_{m_{L}} = \left(E_{\rm merged}\right)^\dagger\cdot E_{\rm split}\cdot X_1^{(1-m_{L})/2}.
\label{eq:tildeMsimp}
\end{equation}
To show this, we use the fact that for all $p\in\mathcal{I}$, $Z(p)$ is a stabilizer of $C_{\rm merged}$ and the correction $X(c_{+})$ is in $\mathcal{S}_{\rm split}$ whereas $X(c_{-})$ is a representative of $\overline{X}_1$ in $C_{\rm split}$.

To show that the operation $\tilde{M}_{m_{L}}$ is equal to $M_{m_{L}}$, as defined in Eq.~\eqref{eq:MSp} and Eq.~\eqref{eq:MSm}, one can analyse how $\tilde{M}_{m_{L}}$ acts on the computational basis, i.e. we track how it transforms the stabilizers of those states.
For example, the state $\ket{00}$ is stabilized by $Z_1$ and $Z_2$, this means that
\begin{align*}
\tilde{M}_+\ket{00} &= \left(E_{\rm merged}\right)^\dagger\cdot E_{\rm split}\ket{00} \\
& = \left(E_{\rm merged}\right)^\dagger\cdot E_{\rm split}\cdot Z_1\ket{00}\\
& = \left(E_{\rm merged}\right)^\dagger\cdot \overline{Z}_1\cdot E_{\rm split}\ket{00}\\
& = Z\cdot\left(E_{\rm merged}\right)^\dagger\cdot E_{\rm split}\ket{00}\\
& = Z\cdot\tilde{M}_+\ket{00},
\end{align*}
and therefore $\tilde{M}_+\ket{00}$ is stabilized by $Z$.
Here, we have used the properties of the encoding isometries and the fact that a representative $\overline{Z}_1$ for $C_{\rm split}$ is also a representative $\overline{Z}$ for $C_{\rm merged}$.
Doing the same with the other stabilizer, $Z_2$, also yields $Z$ as a stabilizer (so $Z_1 Z_2$ yields the identity).
One can also verify that $\tilde{M}_+\ket{00}$ is not stabilized by $-Z$ by reversing the previous equalities and therefore $\langle Z\rangle$ is the full stabilizer group of $\tilde{M}_+\ket{00}$.
Looking now at $\tilde{M}_-\ket{00}$ one can see that $Z_2$ also yields $Z$ but $Z_1$ will yield $-Z$, indeed 
\begin{align*}
\tilde{M}_-\ket{00} &= \left(E_{\rm merged}\right)^\dagger\cdot E_{\rm split}\cdot X_1\ket{00} \\
& =  \left(E_{\rm merged}\right)^\dagger\cdot E_{\rm split}\cdot X_1\cdot Z_1\ket{00}\\
& = -\left(E_{\rm merged}\right)^\dagger\cdot \overline{Z}_1\cdot E_{\rm split}\cdot  X_1\ket{00}\\
& = -Z\cdot\left(E_{\rm merged}\right)^\dagger\cdot E_{\rm split}\cdot X_1\ket{00}\\
& = -Z\cdot\tilde{M}_-\ket{00}.
\end{align*}
Hence, $\tilde{M}_-\ket{00}$ is both stabilized by $Z$ and $-Z$, and is therefore the null vector.
In other words, the state $\ket{00}$ will never give an outcome $-1$ for $m_{L}$, which is what we expect.
\begin{table}[!ht]
\centering
\begin{tabular}{c c || c c  c c }
	 &	 & \multicolumn{2}{c}{$\tilde{M}_+$} & \multicolumn{2}{c}{$\tilde{M}_-$} \\
$\mathcal{S}$ &	State	& $\mathcal{S}$ & State	& $\mathcal{S}$ & State\\\hline\hline
$\langle Z_1, Z_2\rangle$ & $\ket{00}$ & $\langle Z \rangle$ & $\ket0$ & $\langle Z, -Z \rangle$ & $0$ \\
$\langle Z_1, -Z_2\rangle$ & $\ket{01}$ & $\langle Z, -Z \rangle$ & $0$ & $\langle Z \rangle$ & $\ket0$ \\
$\langle -Z_1, Z_2\rangle$ & $\ket{10}$ & $\langle -Z, Z \rangle$ & $0$ & $\langle -Z \rangle$ & $\ket1$ \\
$\langle -Z_1, -Z_2\rangle$ & $\ket{11}$ & $\langle -Z \rangle$ & $\ket1$ & $\langle -Z, Z \rangle$ & $0$ \\
\end{tabular}
\caption{How $\tilde{M}_\pm$ transforms the computational basis states characterised by their stabilizer group.}
\label{tab:stabMpm}
\end{table}
The full results (shown in \autoref{tab:stabMpm}) indicate that
\begin{align*}
\tilde{M}_+ &= \alpha_+ \ket{0}\bra{00} + \beta_+ \ket{1}\bra{11}\\
\tilde{M}_- &= \alpha_- \ket{0}\bra{01} + \beta_- \ket{1}\bra{10},
\end{align*}
for some non-zero complex numbers $\alpha_\pm$ and $\beta_\pm$.
To complete the proof, we verify that there are no relative phases or amplitude differences between $\alpha_\pm$ and $\beta_\pm$.
To see that, one can look at the action of $\tilde{M}_{m_{L}}$ on the Bell states.
For $\tilde{M}_+$ we look at the Bell state $\left (\ket{00}+\ket{11}\right )/\sqrt{2}$, stabilized by $\langle X_1X_2, Z_1Z_2\rangle$ and for $\tilde{M}_-$ the Bell state $\left (\ket{01}+\ket{10}\right )/\sqrt{2}$ stabilized by $\langle X_1 X_2, -Z_1Z_2\rangle$.
The important fact is that a representative $\overline{X}_1\overline{X}_2$  for $C_{\rm split}$ is also a representative of $\overline{X}$ for $C_{\rm merged}$.
That is to say 
\begin{align*}
\tilde{M}_+\frac{\ket{00}+\ket{11}}{\sqrt{2}} &= \gamma_+\frac{\ket{0}+\ket{1}}{\sqrt{2}}\\
\tilde{M}_-\frac{\ket{01}+\ket{10}}{\sqrt{2}} &= \gamma_-\frac{\ket{0}+\ket{1}}{\sqrt{2}},
\end{align*}
for some non-zero complex numbers $\gamma_\pm$.
By linearity of $\tilde{M}_{m_{L}}$ we can conclude that $\alpha_+ = \beta_+ = \gamma_+$ and that $\alpha_- = \beta_- = \gamma_-$.
In conclusion, we have shown that $\tilde{M}_{m_{L}} \propto M_{m_{L}}$, meaning that it performs the desired logical operation.

\subsection{Split operation}
For the $Z$-split operation one reverses the roles of $C_{\rm split}$ and $C_{\rm merged}$.
The starting point is the same as shown in \autoref{fig:sscmerged}, without $\pm$ terms in the middle.
Then, in order to split the patch, one has to split each four-body $X$ stabilizer in the middle row into a pair of two-body $X$ stabilizers.
Those stabilizers are shown with $\pm$ signs on \autoref{fig:sscsplit}.
They commute with everything except for the central row of $Z$-plaquettes.
One can see that measuring them will remove those $Z$-plaquettes from the stabilizer group, but keep the product of all those plaquettes, the logical $\overline{Z}_1\overline{Z}_2$ of the two separate patches.
Note that it is sufficient to measure only the top (or bottom) row of two-body $X$-checks as the bottom (or top) one is then the product of those and the previous four-body $X$-checks.
This also means that the outcomes of those two-body checks are perfectly correlated between facing pairs.
Letting $\mathcal{I}$ be the set of the top row of those checks and $m_p = \pm1$ the measurement outcome of the two-body plaquette $p$, the operation performed is then
\[\prod_{p\in\mathcal{I}}\frac{\id+(-1)^{m_p}X(p)}{2}.\] 
Then, to correct to standard surface codes with no remaining minus signs, one has to apply some of the previous $Z$-plaquettes that were removed from the stabilizer, correcting the correlated facing $X$-checks.
Labeling the set of qubits affected by the correction $c$, one has
\[Z(c)\cdot\prod_{p\in\mathcal{I}}\frac{\id+(-1)^{m_p}X(p)}{2} = \prod_{p\in\mathcal{I}}\frac{\id+X(p)}{2}\cdot Z(c).\]
This operation corresponds to $S_+$, defined in Eq.~\eqref{eq:MSp}. If one wants to implement $S_-$, defined in Eq.~\eqref{eq:MSm}, then one has to additionally apply a logical representative of $X$ on the first patch, $\overline{X}_1$.
The choice of one or the other version is conditioned by the previous $m_{L}$ outcome that we received during the merging step.
Then, to show that this performs the correct logical operation, we analyse
\[\tilde{S}_{m_{L}} = \left (E_{\rm split}\right )^\dagger\cdot\overline{X}_1^{\frac{1-m_{L}}{2}}\cdot\prod_{p\in\mathcal{I}}\frac{\id+X(p)}{2}\cdot Z(c)\cdot E_{\rm merged},\]
which, using the properties of the encoding isometries, $\tilde{S}_{m_{L}}$ simplifies to
\begin{equation}
\tilde{S}_{m_{L}} = X_1^{\frac{1-m_{L}}{2}}\cdot\left (E_{\rm split}\right )^\dagger\cdot E_{\rm merged}.\label{eq:tildeSsimp}
\end{equation}
At this point, recalling Eq.~\eqref{eq:tildeMsimp}, we can see that 
\[\tilde{S}_\pm = \left (\tilde{M}_\pm \right)^\dagger = \left (M_\pm\right )^\dagger = S_\pm,\]
which concludes the proof of correctness for the split operation.
Note that it was crucial to apply the intermediary $Z$-plaquettes (in $\mathcal{L}_g$) as the correction. If we had instead applied a string of $Z$-flips between the faulty $X$-plaquettes, the correction would not be absorbed in the encoding map of $C_{\rm merged}$ and moreover would anti-commute with any representative $\overline{X}$ of $C_{\rm merged}$ or $\overline{X}_1\overline{X}_2$ of  $C_{\rm split}$ and therefore flip the phase between the $\ket0$ and $\ket1$ states.

\section{Example: Code Conversion as Gauge Fixing}
\label{app:conv}
\begin{figure}[htp!]
\centering
\hfill
\includegraphics[width=0.18\textwidth]{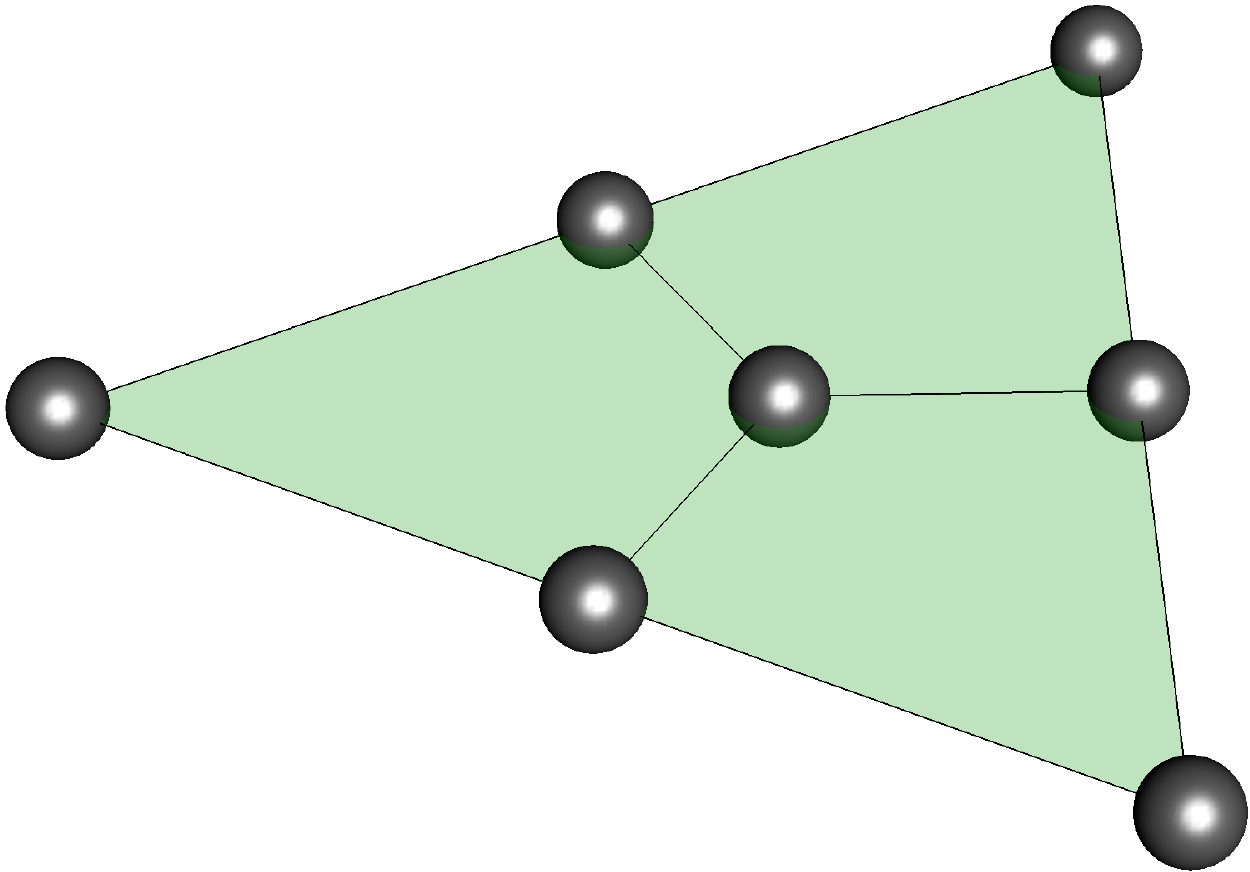}
\hfill
\includegraphics[width=0.18\textwidth]{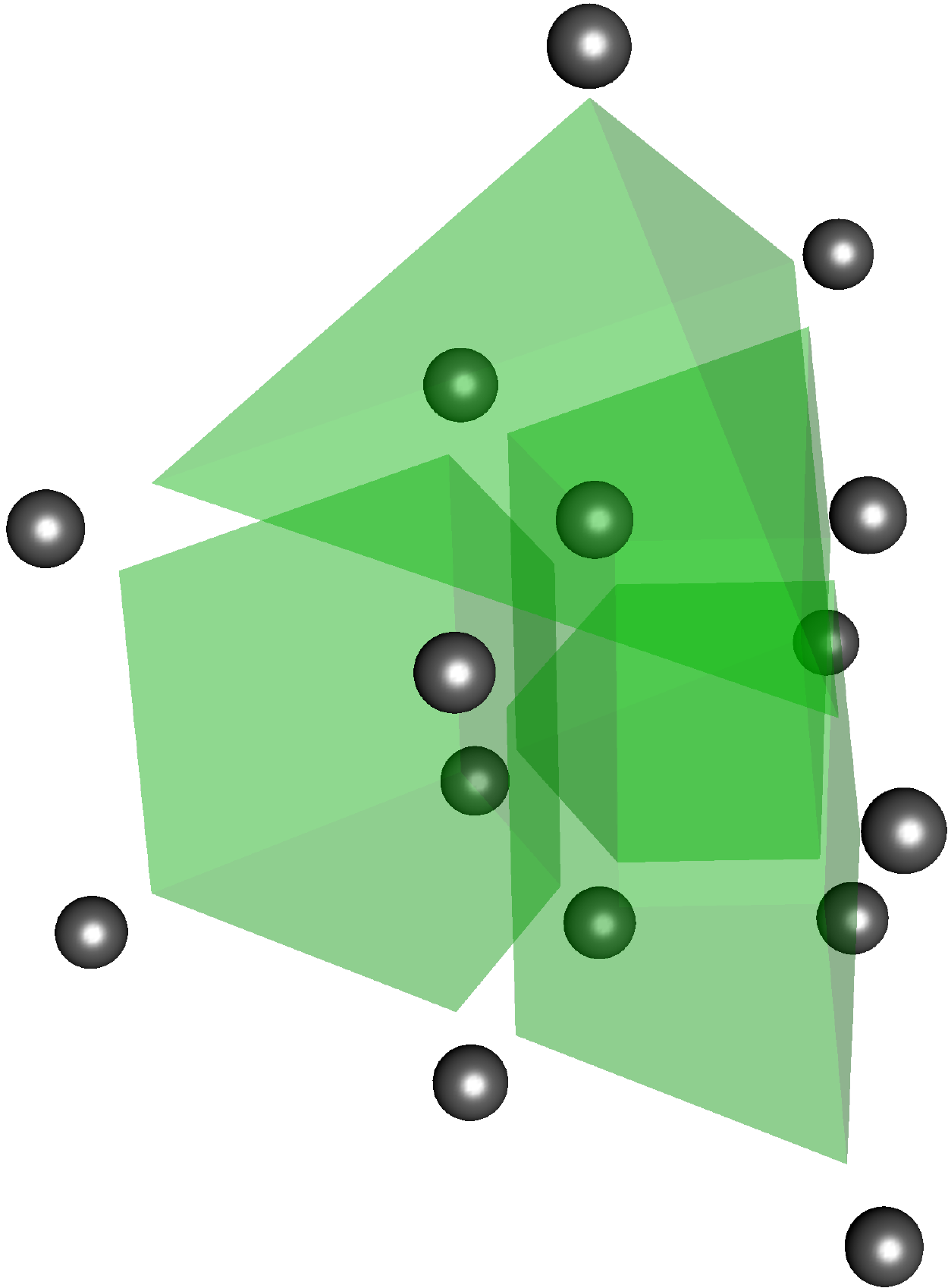}
\hfill
\includegraphics[width=0.18\textwidth]{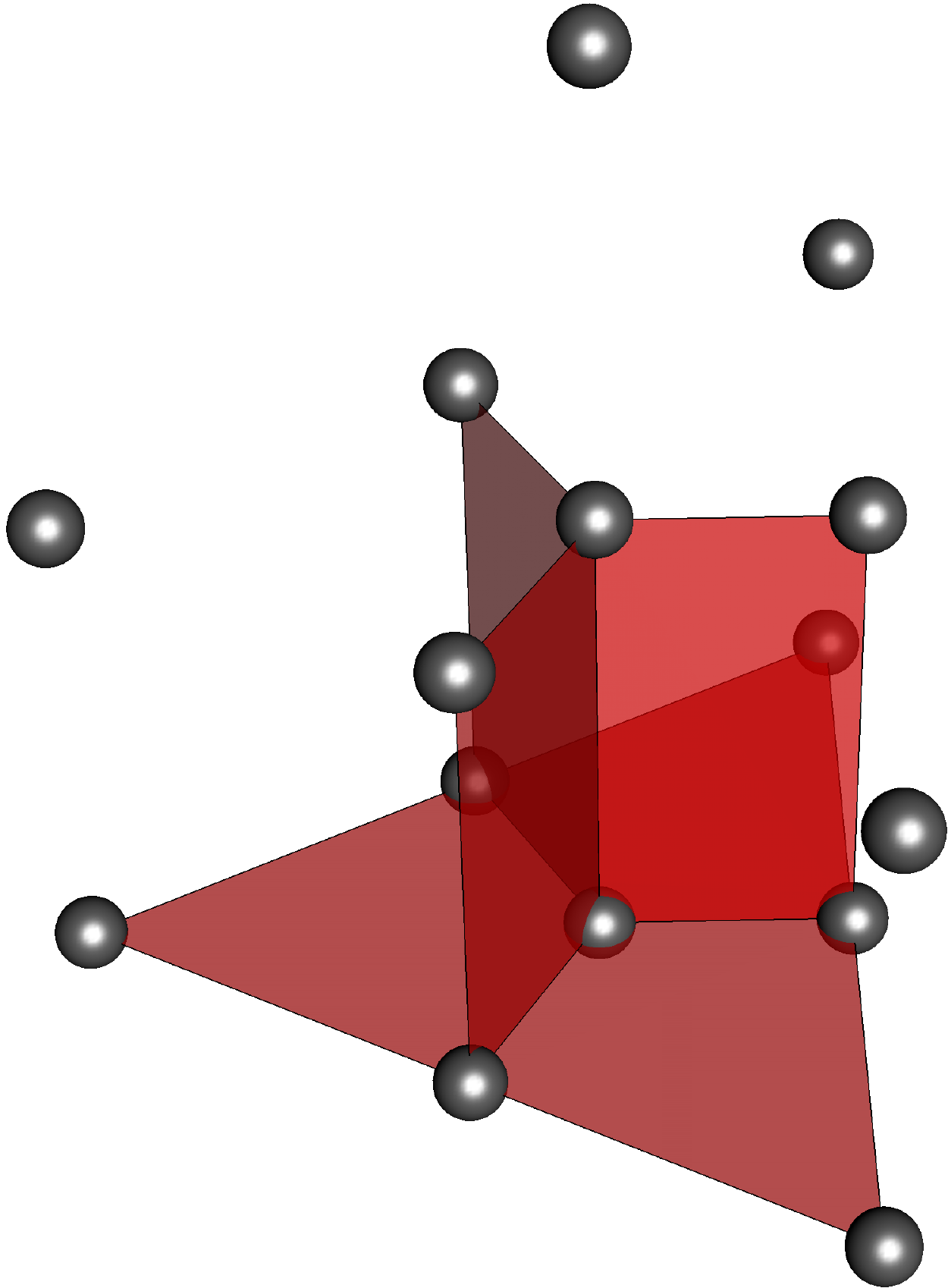}
\hfill
\hfill

\hfill
\makebox[0.2\textwidth]{Steane stabilizers}
\hfill
\makebox[0.2\textwidth]{Reed-Muller stabilizers}
\hfill
\makebox[0.2\textwidth]{Reed-Muller Gauge Operators}
\hfill
\hfill
\caption{stabilizers of the Steane and Reed-Muller codes, and $Z$ gauge operators of the Reed-Muller code.
Red tinting on a face or volume indicates the presence of a $Z$ operator on the vertices which make up that face or volume.
For example, there are six Reed-Muller gauge operators of the form $Z^{\otimes 4}$, supported on the red-tinted quadrilaterals seen on the right. 
Green tinting indicates the presence of both an $X$ and a $Z$ stabilizer operator.}
\label{fig:steanereedmullerstabs}
\end{figure}

\begin{figure}[htp!]
\centering
\includegraphics[width=0.18\textwidth]{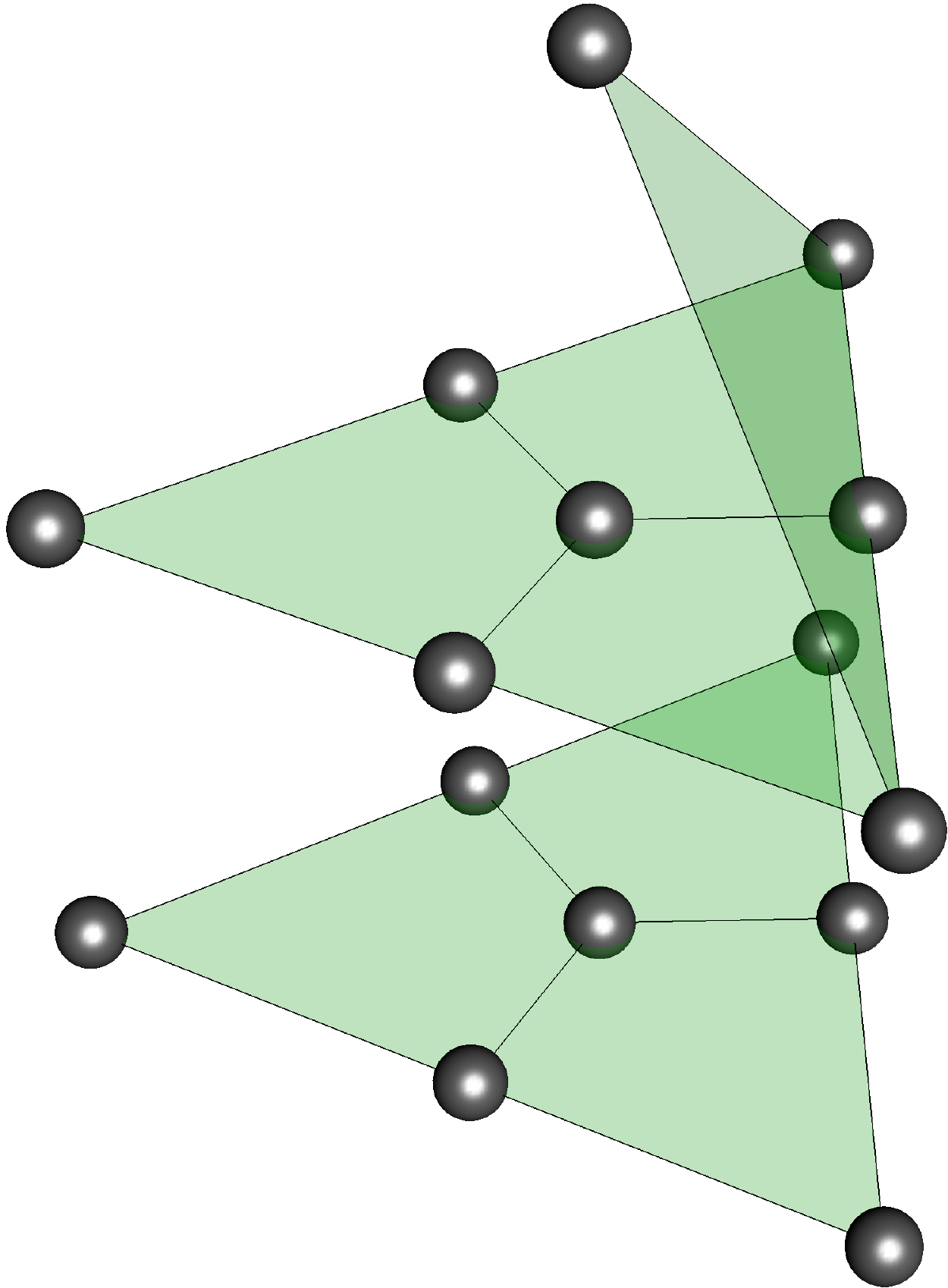}
\hfill
\includegraphics[width=0.18\textwidth]{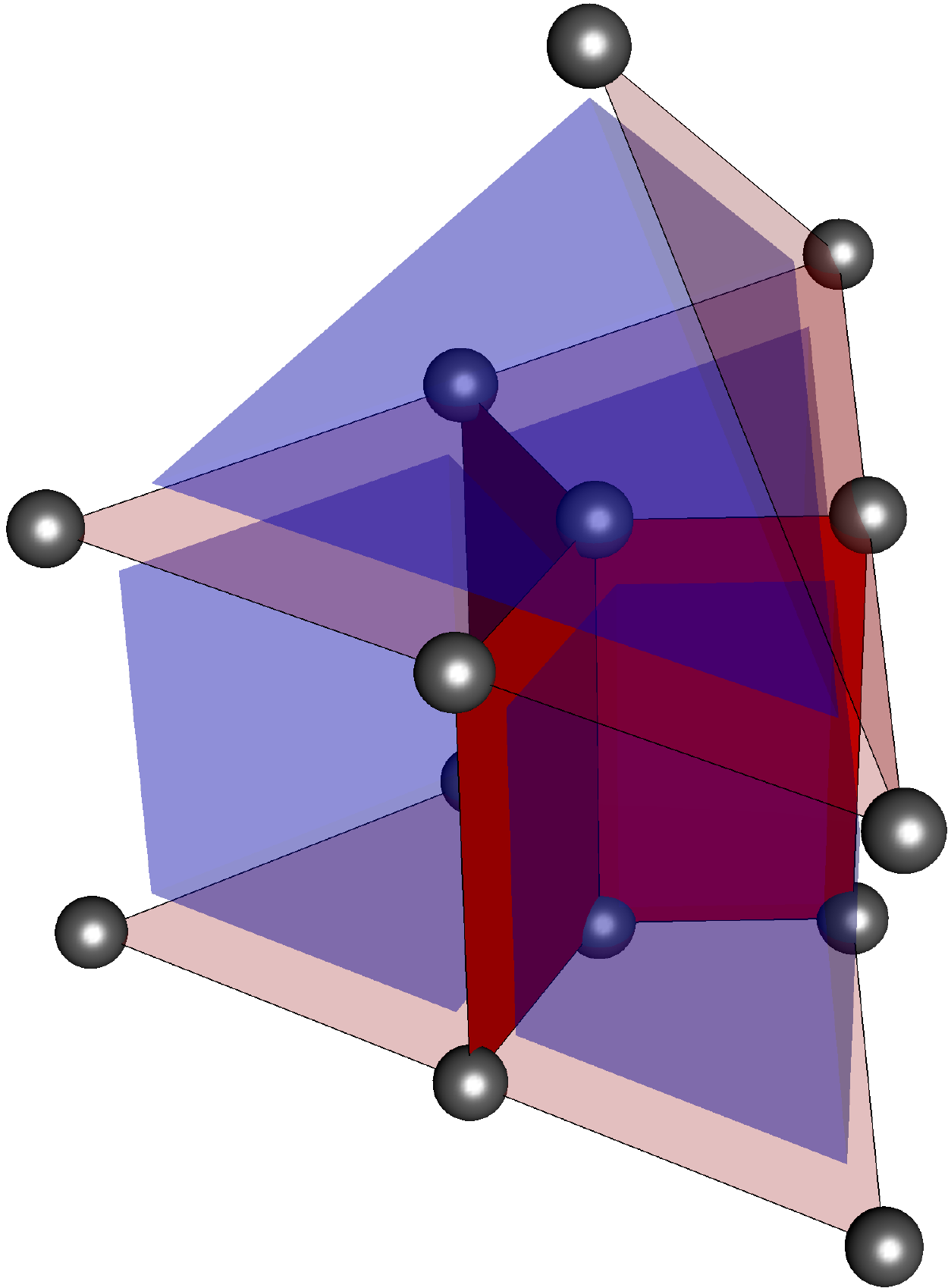}
\hfill
\includegraphics[width=0.18\textwidth]{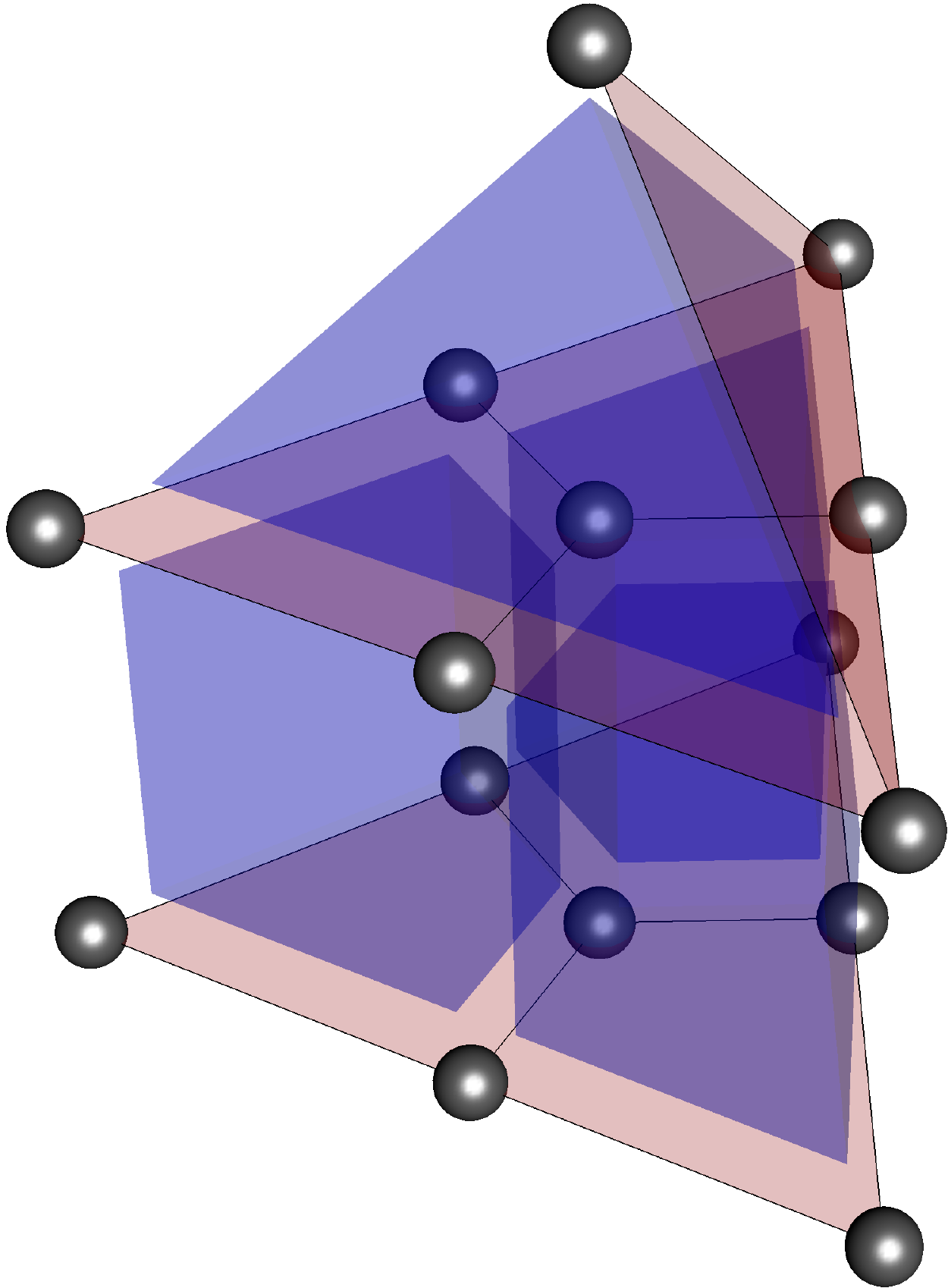}
\hfill
\includegraphics[width=0.18\textwidth]{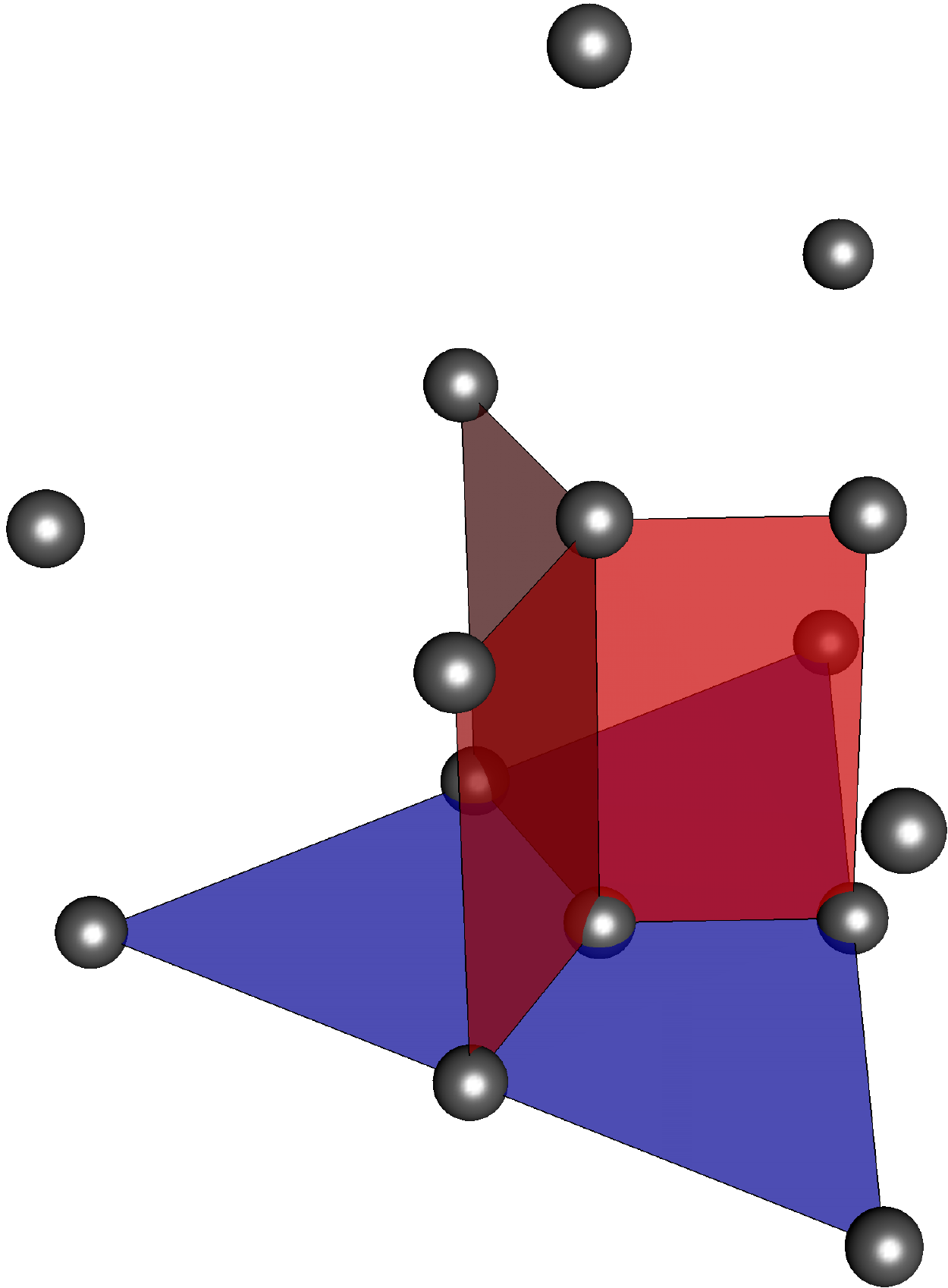}

\includegraphics[width=0.18\textwidth]{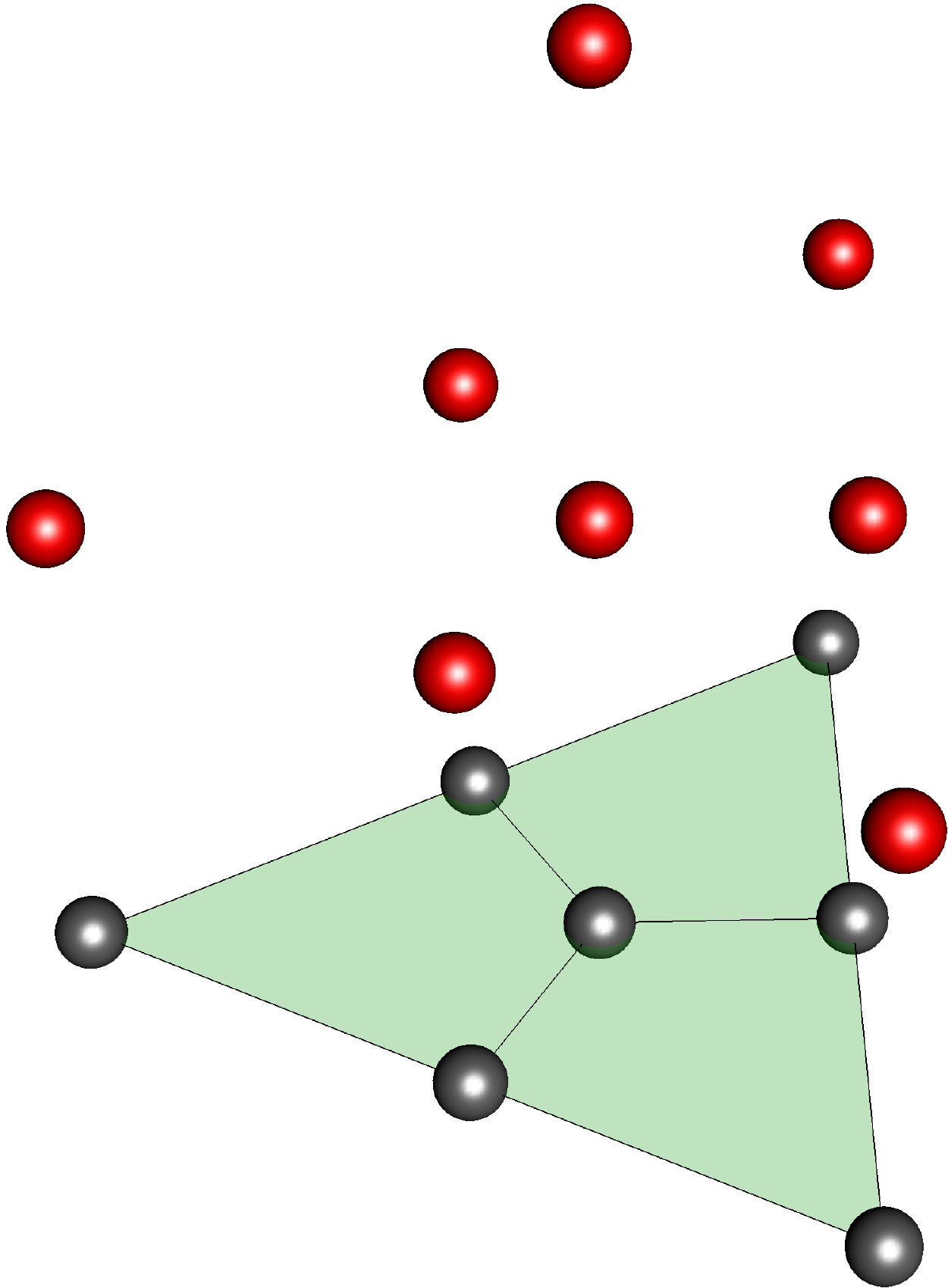}
\hfill
\includegraphics[width=0.18\textwidth]{ReedMullerChecks.pdf}
\hfill
\includegraphics[width=0.18\textwidth]{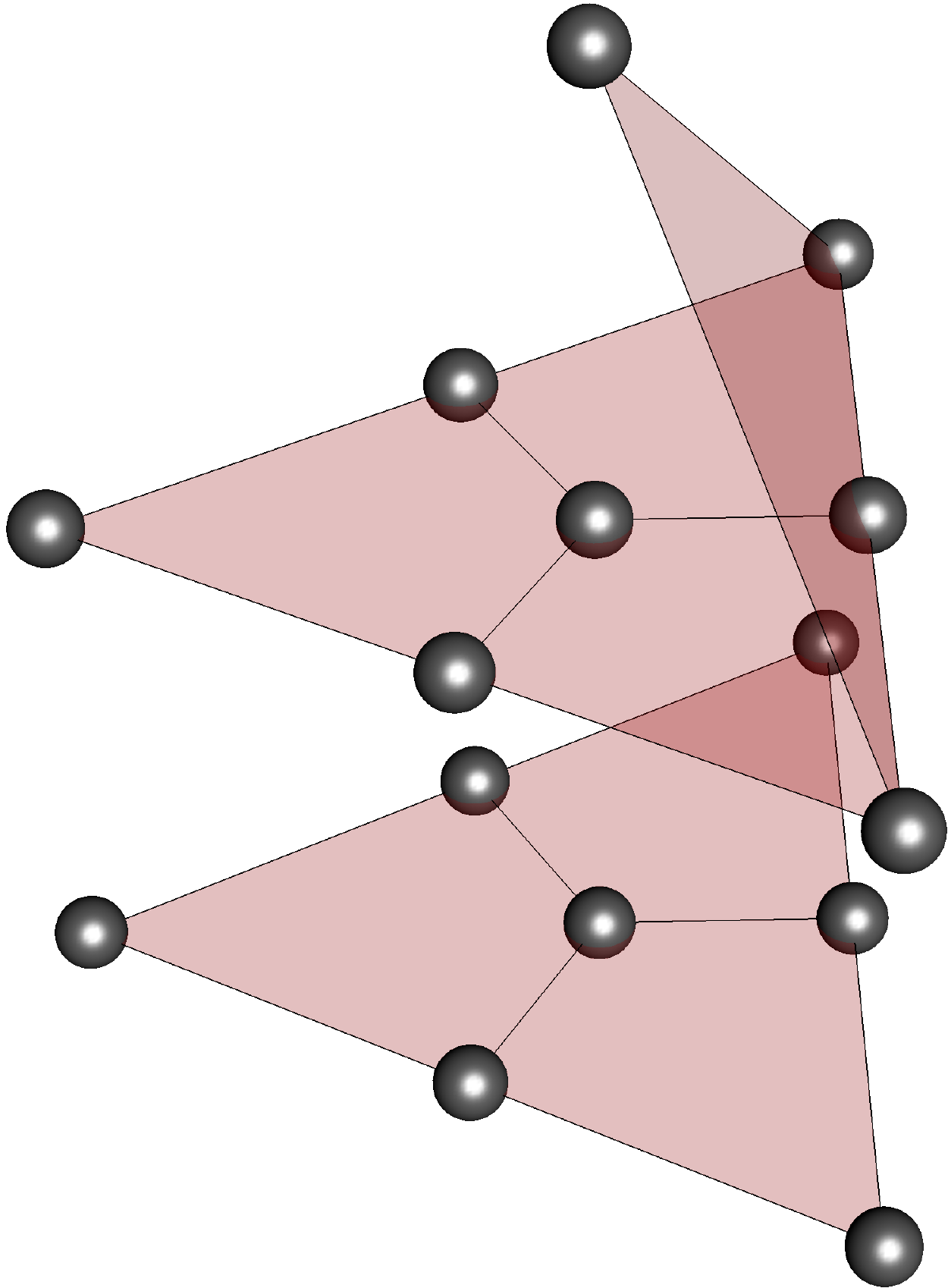}
\hfill
\includegraphics[width=0.18\textwidth]{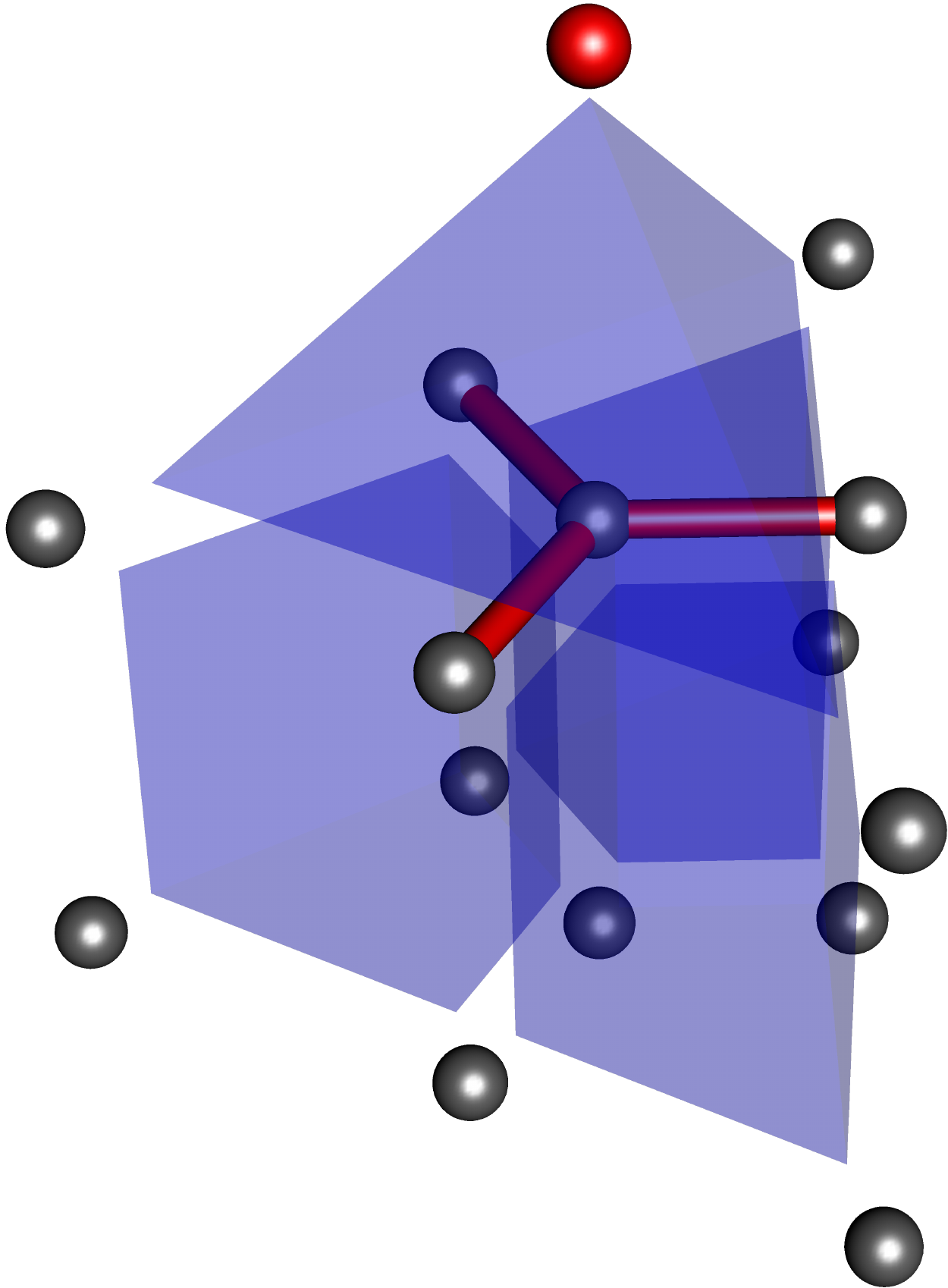}

\makebox[0.2\textwidth]{Initial Checks ($\mathcal{S}_{\rm old}$)}
\hfill
\makebox[0.2\textwidth]{Final Checks ($\mathcal{S}_{\rm new}$)}
\hfill
\makebox[0.2\textwidth]{stabilizers ($\mathcal{S}$)}
\hfill
\makebox[0.2\textwidth]{Gauge Operators ($\mathcal{L}_g$)}
\caption{
Comparison between Steane-to-Reed-Muller conversion schemes from \cite{anderson2014fault} (top) and \cite{Colladay2017rewiring} (bottom).
Red and green tinting match \autoref{fig:steanereedmullerstabs}, blue tinting indicates an $X$ operator supported on the vertices of the tinted face or volume.
Tinted vertices/edges indicate weight-one/two operators supported on the tinted vertex/edge. 
In the Anderson scheme, the subsystem code which applies during the code deformation is made explicit; it is the distance-three Reed-Muller code.
The Colladay scheme, however, does not have any $X$ operators in the relevant stabilizer, $\mathcal{S}$, so the distance of the relevant subsystem code is only 1, see \autoref{sec:Unification}. 
Note: Gauge operators in the top right should also be present in the bottom right, they are not drawn here for clarity.
}
\label{fig:steane_reed_muller}
\end{figure}

To see the utility of gauge fixing for analysing code conversion protocols, we consider two protocols for converting from the $\llbracket 7,\,1,\,3 \rrbracket$ Steane code to the $\llbracket 15,\,7,\,3 \rrbracket$ Reed-Muller code with six gauge $Z$ operators fixed (see \autoref{fig:steanereedmullerstabs} for the stabilizers and gauge operators that define these codes).
The first, from Anderson et al \cite{anderson2014fault}, is based on the realisation that the state $\ket{\psi}_{\textrm{Steane}} \otimes \frac{1}{\sqrt{2}} \left( \ket{0}_{\textrm{Steane}}\ket{0} + \ket{1}_{\textrm{Steane}}\ket{1}\right)$ is a code state of the Reed-Muller code with its horizontal $X$ gauge logical operators fixed, see top-right of \autoref{fig:steane_reed_muller}. 
Conversion from the Steane code to the Reed-Muller code then involves fault-tolerantly preparing the eight-qubit ancilla state and fixing the three appropriate $Z$ gauge operators.
The state is always stabilized by the Reed-Muller stabilizers, whose eigenvalues can be reconstructed from the checks which are measured at every round, preserving the code distance and allowing error correction by syndrome decoding.

The second scheme, from Colladay and Mueller \cite{Colladay2017rewiring}, is not based on gauge fixing, and begins with the eight qubits needed for conversion initialised in the state $\ket{0}^{\otimes 8}$. 
This ensures that the initial checks anticommute with any potential $X$ stabilizer supported on the final eight qubits, so that the only operators whose eigenvalues can be reconstructed from the measured operators are $Z$ operators, preventing the correction of $Z$ errors (see \autoref{fig:steane_reed_muller} for a graphical comparison of these code conversion protocols). 
The difference in fault tolerance between these two protocols which accomplish the same task provides us with a good motive to incorporate subsystem codes into the analysis of code deformation and lattice surgery, considered in the main text.

Examining the Criterion 1 from \autoref{subsec:ft-cd}, one can see that the Anderson scheme has an underlying subsystem code with distance 3, whereas not having any $X$-stabilizers, the Colladay scheme has an underlying subsystem code with distance 1.

\section{Disparity in error rates of CNOT gates}
\label{sec:disparity}

\begin{figure}[!ht]
\centering
\begin{subfigure}[h]{0.45\textwidth}
\caption{}
\includegraphics[width = \textwidth]{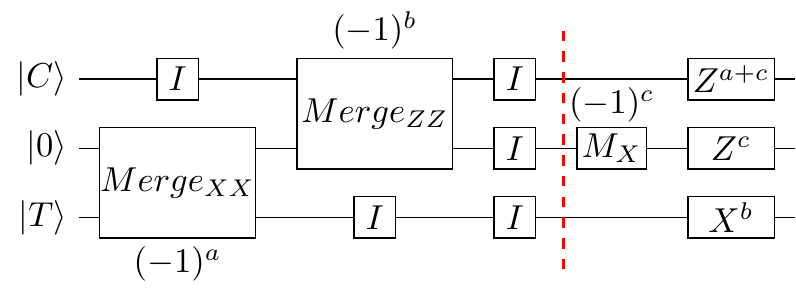}
\label{fig:cnotcircuit1_2}
\end{subfigure}\hspace{5mm}
\begin{subfigure}[h]{0.45\textwidth}
\caption{}
\includegraphics[width = \textwidth]{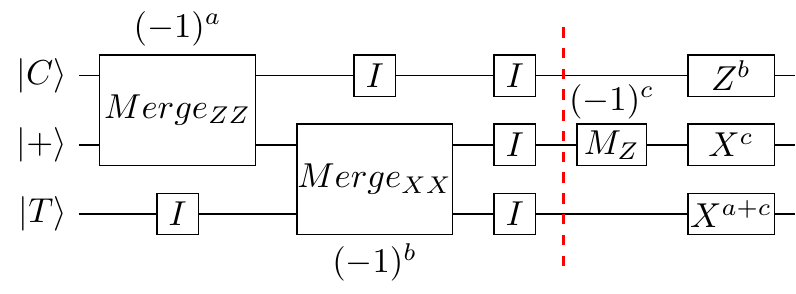}
\label{fig:cnotcircuit2_2}
\end{subfigure}\hspace{10mm}
\caption{The decomposed circuits (a) and (b) of the top and bottom measurement-based \cnot~circuits in \autoref{fig:cnotcircuit}.}
\label{fig:cnotdecompose}
\end{figure}

\begin{figure}[htb!]
\centering
\begin{subfigure}[h]{0.45\textwidth}
\caption{}
\includegraphics[width=\textwidth]{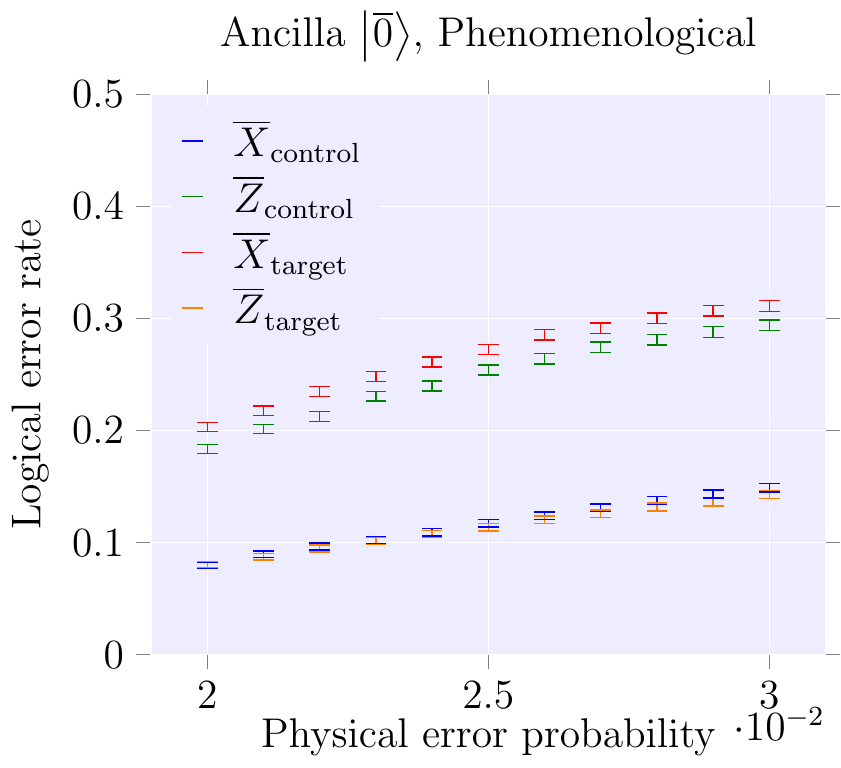}
\label{fig:ler_cnotpq_grd5}
\end{subfigure}
\begin{subfigure}[h]{0.45\textwidth}
\caption{}
\includegraphics[width=\textwidth]{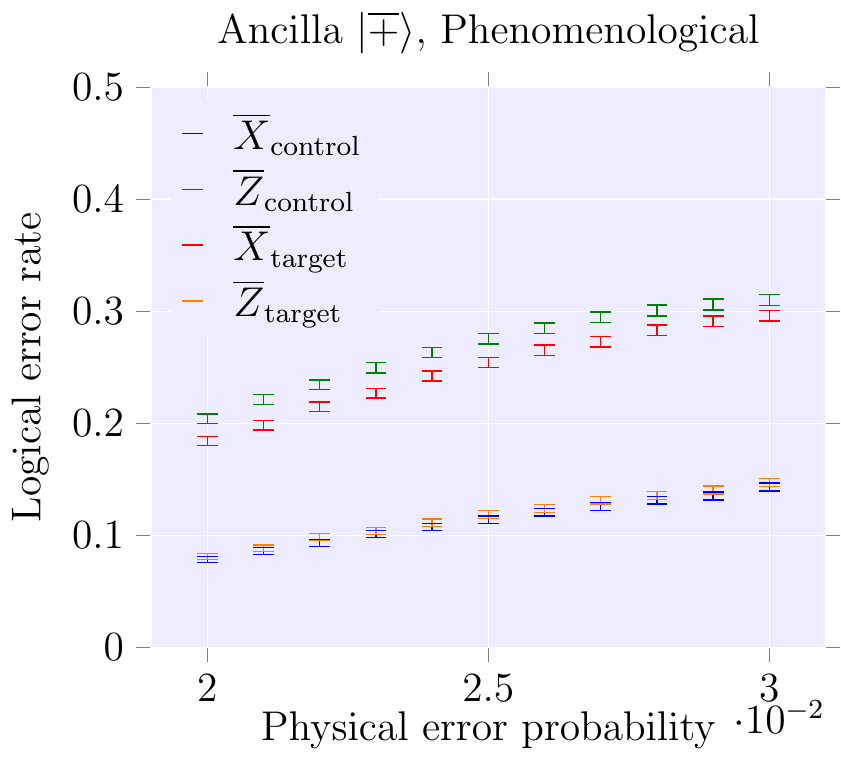}
\label{fig:ler_cnotpq_pls5}
\end{subfigure}
\begin{subfigure}[h]{0.45\textwidth}
\caption{}
\includegraphics[width=\textwidth]{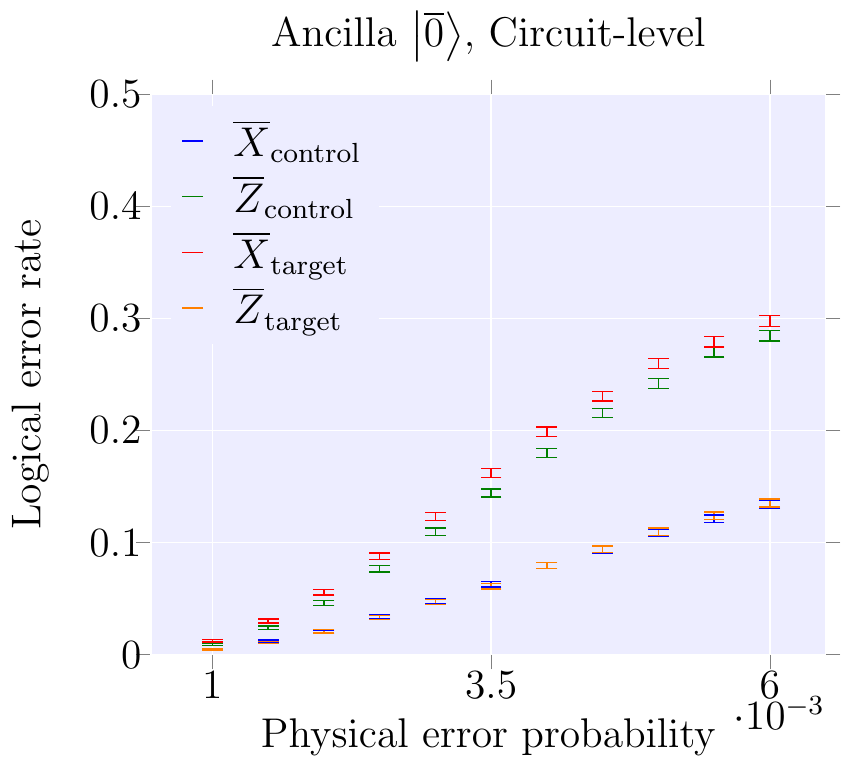}
\label{fig:ler_cnotcm_grd5}
\end{subfigure}
\begin{subfigure}[h]{0.45\textwidth}
\caption{}
\includegraphics[width=\textwidth]{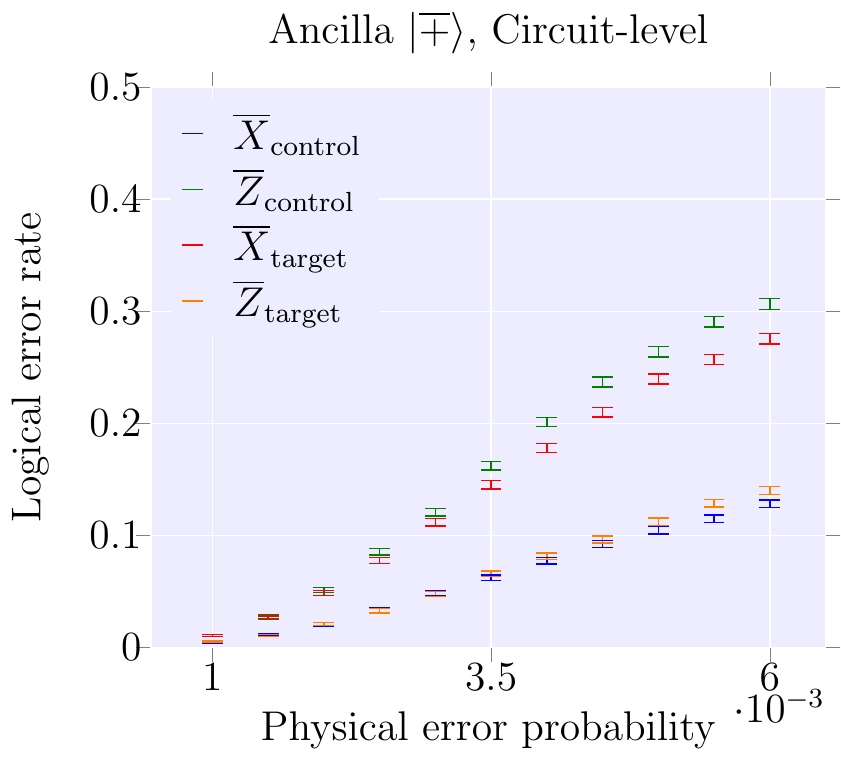}
\label{fig:ler_cnotcm_pls5}
\end{subfigure}
\caption{$\overline{X}$ and $\overline{Z}$ error rates on the control and target qubits for lattice-surgery-based \cnot~operations at distance 5. 
(a) and (b) correspond to the phenomenological error model, (c) and (d) correspond to the circuit-based error model. 
The disparity in error rates is explained by error propagation through the measurement-based circuit implementing the \cnot.}
\label{fig:ler_cnot_qubits}
\end{figure}

A joint measurement is realised by performing a merge and a split operation in sequence.
In our simulation, the circuits in \autoref{fig:cnotcircuit} are decomposed into the ones in \autoref{fig:cnotdecompose}.
\autoref{fig:ler_cnot_qubits} shows that the rates of $\overline{X}$/$\overline{Z}$ errors on the control and target qubits are different for the rotated surface code with $d=5$. 
This disparity can be explained using a toy model to account for propagation of logical errors through measurement-controlled corrections. 

In this toy model, identity gates result in an $\overline{X}$ or $\overline{Z}$ error with probability $p$ ($\overline{Y}$ errors are assumed to occur with probability $\sim p^2$, since the minimum-weight $\overline{Y}$ operator has weight $2d-1$ in the surface code).
The merge operations are modeled as ideal joint measurements, followed by an error of the form $\overline{X \otimes \id}$, $\overline{\id \otimes X}$, $\overline{Z \otimes \id}$, or $\overline{\id \otimes Z}$, each occurring with probability $p$, since these are the likeliest logical errors.
If a logical Pauli error occurs, it propagates forward through the circuit, changing the measured eigenvalue for any measurement operator with which it anticommutes. 
For example, if an $\overline{X \otimes \id}$ error occurs after the $M_{XX}$ operation in \cref{fig:cnotcircuit} (in which the ancilla begins in the $\ket{0}$ state), the measured value $b$ will be mapped to $1 - b$, causing an $\overline{X}$ operator to be incorrectly applied to the target qubit at the end of the $\overline{\cnot}$.
It is easy to confirm that there are 7 such first-order errors which result in an $\overline{X}$ error on the target qubit, 6 errors which result in a $\overline{Z}$ error on the control qubit, and 3 errors which result in the other logical errors shown in \cref{fig:ler_cnotpq_grd5,fig:ler_cnotcm_grd5} (a similar analysis holds for the error rates shown in \cref{fig:ler_cnotpq_pls5,fig:ler_cnotcm_pls5}).
The biased logical error rates predicted by this simplified model are in good agreement with the logical error rates observed in simulation, shown in \cref{fig:ler_cnot_qubits}.
Preventing this bias from growing during the execution of a long algorithm, by appropriate selection of decomposition for \cnot{s}, is likely an important step in the design of high-performance fault-tolerant circuits for quantum computation.

\end{document}